\documentclass{aa}  

\usepackage{color}
\usepackage{float}
\usepackage{soul}
\usepackage{dblfloatfix}
\usepackage{graphicx}
\usepackage{longtable}
\usepackage{multicol}
\usepackage{supertabular}

\usepackage{txfonts}
\usepackage{natbib}
\newcommand{\cahk}{Ca\,{\sc ii}\,H\&K}
\newcommand{\kms}{\,kms$^{-1}$}
\usepackage[dvipsnames]{xcolor}%
\begin{document}

\title{A star under multiple influences}
   \subtitle{Magnetic activity in V815\,Her, a compact 2+2 hierarchical system\thanks{Based on data obtained with the STELLA robotic observatory in Tenerife, an AIP facility jointly operated by AIP and the Instituto de Astrofisica de Canarias.}}

%   \subtitle{Subtitle}

\author{Zs.~K\H{o}v\'ari
          \inst{1,2}
        \and
          K.~G.~Strassmeier
          \inst{3,4}
        \and
          L.~Kriskovics
          \inst{1,2}%\thanks{Bolyai J\'anos Research Fellow}
        \and
          K.~Ol\'ah
          \inst{1,2}
        \and
        T.~Borkovits
            \inst{5,6,7}
        \and
          \'A.~Radv\'anyi
          \inst{8}
        \and
          T.~Granzer
          \inst{3} 
        \and
          B.~Seli
          \inst{1,2}
        \and
          K.~Vida
          \inst{1,2}
        \and
          M.~Weber
          \inst{3}
         %\samethanks
  }

   \institute{Konkoly Observatory, HUN-REN Research Centre for Astronomy and Earth Sciences, Konkoly Thege \'ut 15-17., H-1121, Budapest, Hungary\\
              \email{kovari@konkoly.hu}
        \and
        CSFK, MTA Centre of Excellence, Budapest, Konkoly Thege út 15-17., H-1121, Hungary
        \and
        Leibniz-Institute for Astrophysics Potsdam (AIP), An der Sternwarte 16, D-14482 Potsdam, Germany
        \and
        Institute for Physics and Astronomy, University of Potsdam, Karl-Liebknecht-Strasse 24/25, D-14476 Potsdam, Germany
        \and
        Baja Astronomical Observatory of University of Szeged, Szegedi \'ut, Kt. 766, H-6500 Baja, Hungary
       \and
        HUN-REN-SZTE Stellar Astrophysics Research Group, Szegedi út, Kt. 766, H-6500 Baja, Hungary
        \and
        ELTE Gothard Astrophysical Observatory, Szent Imre h. u. 112, H-9700 Szombathely, Hungary
%        \and
%        MTA-ELTE Exoplanet Research Group, Szent Imre h. u. 112, H-9700 Szombathely, Hungary
        \and
        Moholy-Nagy University of Art and Design, Centre for Data Science and Digital Development, Budapest, Zugligeti út 9, H-1121, Hungary
        %            ELTE Eötvös Loránd University, Institute of Physics, Pázmány Péter sétány 1/A, H-1117, Budapest, Hungary
            }

   \date{Received ...; accepted ...}

  \abstract
   {Close binaries with magnetically active components are astrophysical laboratories for studying the effects of binarity on activity. Of particular interest are binary and multiple star systems that contain a solar-type active component with an internal structure similar to the Sun, allowing us to study how the dynamo of a solar-type star would work under different conditions.}
   {We are conducting a comprehensive investigation of V815\,Her using photometric and spectroscopic data to understand the origin of the activity and what influences it in the short and long term.}
   {Using space photometry we performed light curve modeling in order to derive astrophysical and orbital parameters for the eclipsing binary subsystem V815\,Her~B. Using archival photometric data covering a century we carried out a time frequency analysis.% to investigate long-term variability and search for activity cycles of the spotted primary component V815\,Her~Aa.
   Spectral synthesis was applied to determine the basic astrophysical parameters of the rapidly rotating primary using high-resolution STELLA spectra recorded in 2018. }
   {Photometric analysis of archived data revealed multiple cycles on timescales between $\sim$6.5 and $\sim$26 years, some of which may be harmonic. From \emph{TESS} photometry we obtained orbital solution for the V815\,Her~B subsystem. By placing the primary component on the Hertzsprung$-$Russell-diagram, we can deduce an age of $\approx$30\,Myr, in line with the high Li-6707 abundance. The STELLA spectra covering the ~200 day-long observing season enabled to create 19 time-series Doppler images, which revealed a constantly changing spotted surface on a time scale of a few weeks. From the consecutive image pairs we built up the average cross-correlation function map to measure the surface differential rotation of the spotted star, from which we derive a weak solar-type surface shear.}
   {We found evidence that the V815\,Her~B component previously apostrophized as a ’third body’ is actually an eclipsing close binary subsystem of two M dwarfs with a period of 0.5\,d, i.e., V815\,Her is a 2+2 hierarchical quadruple system. The system is apparently young, only a few times ten million years old, consistent with the spotted primary V815\,Her~Aa being a zero-age main sequence star. Spot activity on the primary was found to be vivid. Fast starspot decay suggests that convective-turbulent erosion plays a more significant role in such a rapidly rotating star. The weak surface shear of V815\,Her~Aa due to differential rotation is presumably confined by tidal forces of the close companion V815\,Her~Ab. The slowly increasing photometric cycle of about 6.5 years on average is interpreted as a spot cycle of V815\,Her~Aa, which is probably modulated by the eccentric wide orbit.}


   \maketitle
%
%-------------------------------------------------------------------

\section{Introduction}

Close binaries that contain magnetically active components may be regarded as astrophysical laboratories for studying the effect of binarity on activity. Tidal forces in such systems are supposed to modify the operation of the magnetic dynamo, by altering the flux emergence patterns at the surface \citep{2003A&A...405..291H,2003A&A...405..303H}. Amplified dynamo operation is expected, for instance, when fast rotation is sustained by weakening magnetic braking through angular momentum transport in a close binary system \citep{2018A&A...609A...3S}. Although, other binarity-related effects like early-evolution accretion processes can also result in faster rotation \citep{2007ApJ...665L.155M}. In all probability, the gravitational influence of a close companion can suppress the differential rotation of the convection zone \citep{1982ApJ...253..298S,2007AN....328.1030C} or even force the formation of active longitudes at certain phases fixed to the orbit \citep{2006Ap&SS.304..145O,2021A&A...650A.158K}.
Especially interesting are those binaries and multiple star systems which contain active G-type main-sequence (MS) stars with inner structure (convective zone) similar to that of the Sun, making it possible to investigate how the solar-type dynamo would work under different circumstances.

In accordance with the above, this time we choose the active G dwarf component of the single-lined but in fact a quaternary star system V815\,Her (=HD\,166181) for our study. In the beginning, it was believed that the G star with its enhanced magnetic activity inferred from strong \cahk\ emission is the primary member of a single-lined binary (SB1) system with an orbital period of $\approx$1.81~days \citep{1974A&A....37..191N}. Later using new spectroscopic data taken in 1992 at Kitt Peak National Observatory \citet{1996AJ....111.1356D} found a significant radial velocity difference of 11\kms\ compared to the data of \citet{1974A&A....37..191N}, which was attributed to a possible third body with an orbital period of many years. This idea was warmed up by \citet{2004RMxAC..21...45F} who, using new observations from 2002, calculated long-period orbital elements for the outer companion in the system for the first time with a preliminary period of 6.3\,yr. With additional radial velocity measurements from 2003, this value was soon refined to 2092.2$\pm$5.8\,d \citep{2005AJ....129.1001F}. From their mass estimates for the system components of the long-period orbit \citet{2005AJ....129.1001F} also suggested that the unseen component might also be a close binary. An important step forward in the story was that the outer component has been resolved by direct imaging within the frame of Gemini Deep Planet Survey (GDPS) by \citet[][]{2007ApJ...670.1367L} who confirmed the astrometric solution of \citet{2005AJ....129.1001F} for the third body.
Going even further, in our paper we present space photometry from NASA’s Transiting Exoplanet Survey Satellite (\emph{TESS}) to demonstrate that the suspected third body itself is indeed an eclipsing close binary system of two unseen but very likely M dwarfs, i.e. V815\,Her is in fact a four-star system consisting of two close binaries orbiting each other.

According to photometric observations, it has long been known that the G (most probably G5-G6) active component features cool spots on its surface \citep{1980IBVS.1791....1M}. \citet{1989ApJS...69..141S} reported automatic photoelectric telescope (APT) data from 1984-1986 in $U$, $B$ and $V$ colours and found that the photometric period associated to the rotation of the G star was 1.8195\,d and 1.8356\,d for 1985 and 1986, respectively, i.e. $\approx$1\% longer than the orbital period from \citet[][]{1974A&A....37..191N}. From an extended study using long-term APT data taken in $BV$ between 1984 and 1998 \citet{2000A&A...362..223J} found seasonal photometric period changes attributable to surface differential rotation of the spotted star. In addition, their time-series analysis revealed that one dominant active longitude was present for the full 14 year-long observing period while another weaker one emerged during some of the seasons. This APT data were re-analyzed in \citet{2009ARep...53..941S} to reconstruct temperature inhomogeneities of the stellar surface. Again, two active longitudes were found, separated by about 0.5 rotation phase \citep[cf.][]{2000A&A...362..223J}, showing the so-called flip-flop phenomenon, i.e. when dominance switches quasi-periodically from one active longitude to the other.

Like the photosphere, the chromosphere and corona of the G component show the signatures of strong magnetic activity over a wide range of the electromagnetic spectrum.  In addition to the strong and narrow \cahk\ emission \citep{1974A&A....37..191N,1984ApJS...54..387B} 
$UV$ observations from the \emph{IUE} satellite \citep{1986ApJS...60..551F} also indicate an active chromosphere. The absolute luminosity in the Mg\,{\sc ii}\,k line supports this enhanced activity \citep{2003A&A...408..337C}, however, the lack of hot component in the $UV$ spectra suggests that the companions without contribution should be M-dwarfs or fainter.
Moreover, the G star performs extreme-ultraviolet (EUV) radiation measured during the ROSAT Wide Field Camera All-Sky Survey  \citep{1993MNRAS.260...77P} and by the \emph{Extreme Ultraviolet Explorer} satellite \citep{1994AJ....107..751M}. 
Due to the ROSAT mission, V815\,Her is known for its bright X-ray corona \citep[][]{1993ApJS...86..599D,1993ApJ...413..333D}. Indeed, its coronal luminosity of $L_X$=3.183\,10$^{30}$\,erg \citep{2003AJ....126.1996M} classifies V815\,Her as one of the 100 brightest stellar X-ray sources within 50\,pc. Accordingly, with such an active corona, the star is known as a radio source as well \citep{1992ApJS...82..311D}.

V815\,Her is not associated with any particular moving group \citep[but see][who cited the star as Pleiades-like]{2003A&A...399..983W}. The strong Li line in the optical spectrum \citep[][]{2005AJ....129.1001F} indicates that the G component is indeed young, near the zero-age main sequence (ZAMS).
This observation is consistent with the star being listed in the Two Micron All Sky Survey Point Source Catalog \citep[2MASS/PSC,][]{2009ApJS..184..138H} due to its infrared excess, most probably related to primordial protoplanetary material. 

In our paper, we perform a Doppler imaging study for the G star for the first time with reconstructing 19 time-series Doppler images using high-resolution optical spectra taken during a 9-month long observing run in 2018.
The paper is organized as follows. In Sect.\,\ref{sect_phot_data} we present the analysis of photometric data. Using \emph{TESS} observations we analyze the newly discovered eclipses and give an orbital solution for the unseen, but most likely M+M dwarf binary companion V815\,Her~B in the quadruple system. In Sect.\,\ref{sect_par} we derive precise astrophysical parameters of the active G component for the Doppler imaging study presented in Sect.\,\ref{sect_di}. %Also, using the time-series spot maps we can track the surface changes of the spotted star in a very good time resolution. 
Results are discussed in Sect.\,\ref{sect_disc}, and summarized in Sect.\,\ref{sect_sum}.

%-------------------------------------

\section{Photometric data and analysis}\label{sect_phot_data}

\subsection{New light curve and orbital solution for V815\,Her~B using \emph{TESS} data}\label{sect_TESS}

\emph{TESS} has observed V815\,Her during sectors 26, 40 and 53.\footnote{We note here that there is a 605-day interval between the end of our spectroscopic observations in 2018 and the first data point of the \emph{TESS} sector 26 dataset, so unfortunately there is no overlap.} The sector 26 observations were carried out in 2-min cadence mode, so the light curves produced by the Science Processing Operations Center (SPOC) became directly available, while the Year 3 observations were carried out only in the 600-min cadenced full frame image (FFI) mode. Hence, we downloaded the 2-min cadence Pre-search Data Conditioned Simple Aperture Photometry (PDCSAP) data from the Mikulski Archive for Space Telescopes, while we processed the sector 40 and 53 FFI data with the software package {\sc Fitsh} \citep{2012MNRAS.421.1825P}. V815\,Her (=TIC~320959269) is listed as a \emph{TESS} planet host candidate with a supposed planetary orbit of 0.26\,d. 
However, from a closer inspection, after removing the spot variability of the G star using the 1.8\,d rotational period, it became clear to us that the depths of successive minima are slightly different, i.e., primary and secondary minima alternate. Evidently, instead of a planet orbiting with a period of 0.26\,d we found a close eclipsing binary with a double-length period of 0.52\,d bound to V815\,Her~A in the wide orbit.

\begin{figure}[thb]
    \includegraphics[width=\columnwidth]{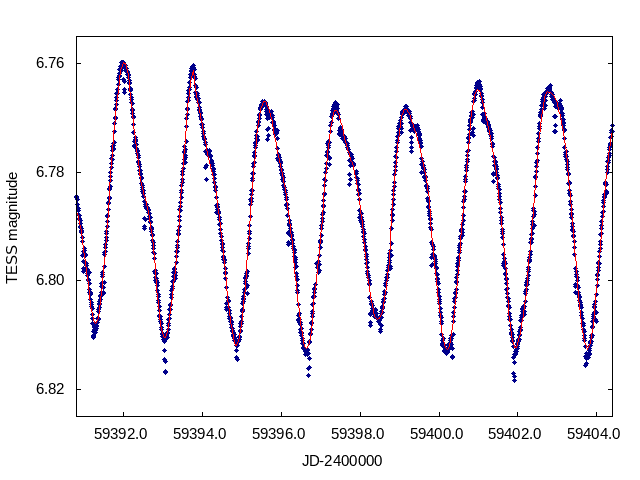}
      \caption{\emph{TESS} sector 40 light curve of V815\,Her (blue dots) fitted with a time-series three-spot model (red line).
      }
      \label{fig_TESS_spotfit}
\end{figure}

To prepare data for the computation of the orbital solution of this new close binary we fitted the spotted light curve of the G star with a time-series three-spot model (see Fig.~\ref{fig_TESS_spotfit}), which could properly follow the ever-changing light variation instead of a constant rotational period and its harmonics. For the fit we used an upgraded version of the SpotModeL code originally presented in \citet{2003AN....324..202R}. For the light curve modeling the fixed parameters used were only crudely approximated (temperature, limb darkening, etc.) since the main goal was only to fit the light curve as smoothly as possible. The data were also cleaned from obvious outliers and a possible flare. The resulting model was subtracted from the original data and binned to 0.005 days (7.2 minutes) which was then used for binary modeling.

\begin{figure}[thb]
    \includegraphics[width=\columnwidth]{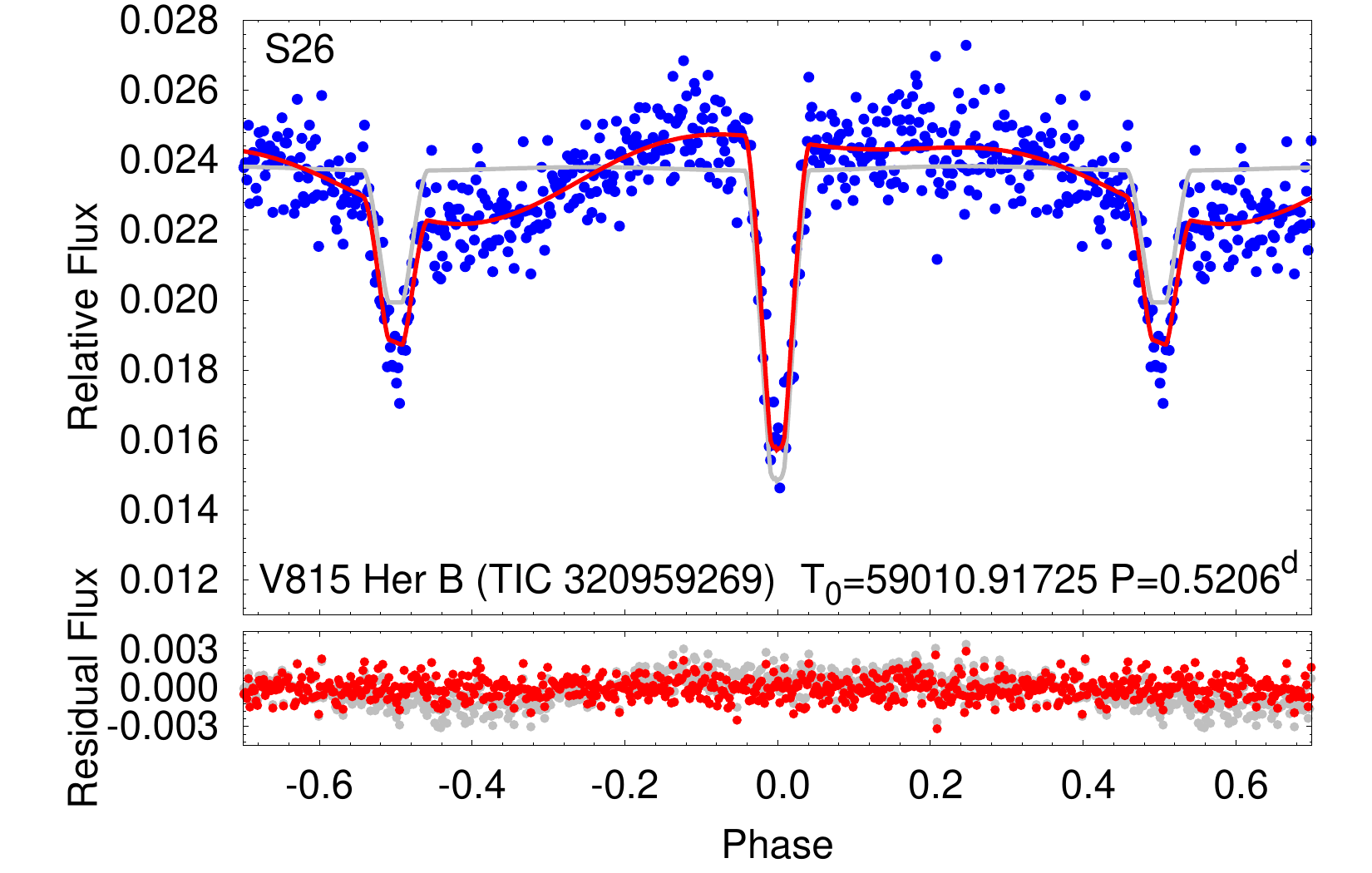}
    \includegraphics[width=\columnwidth]{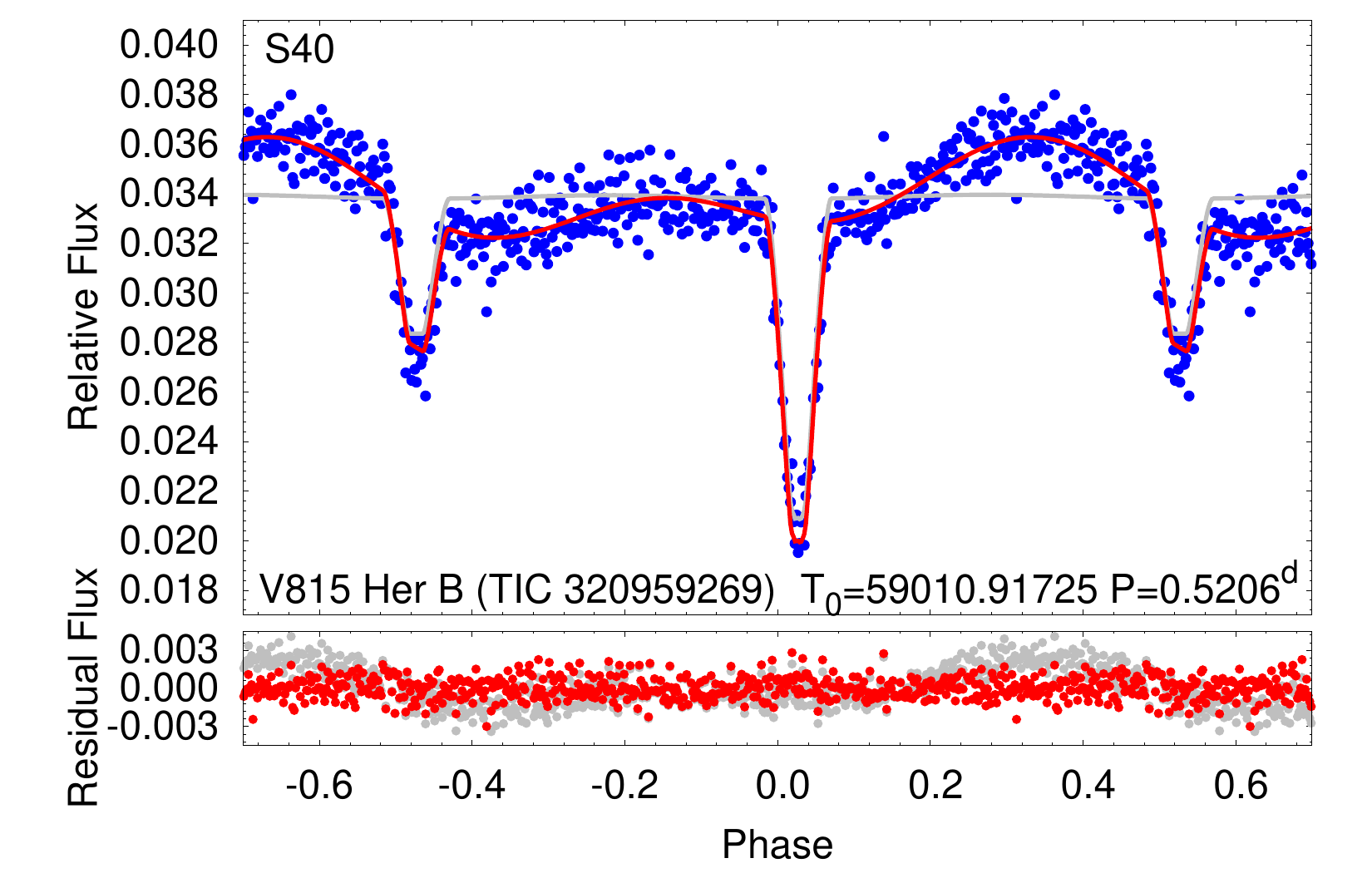}
    \includegraphics[width=\columnwidth]{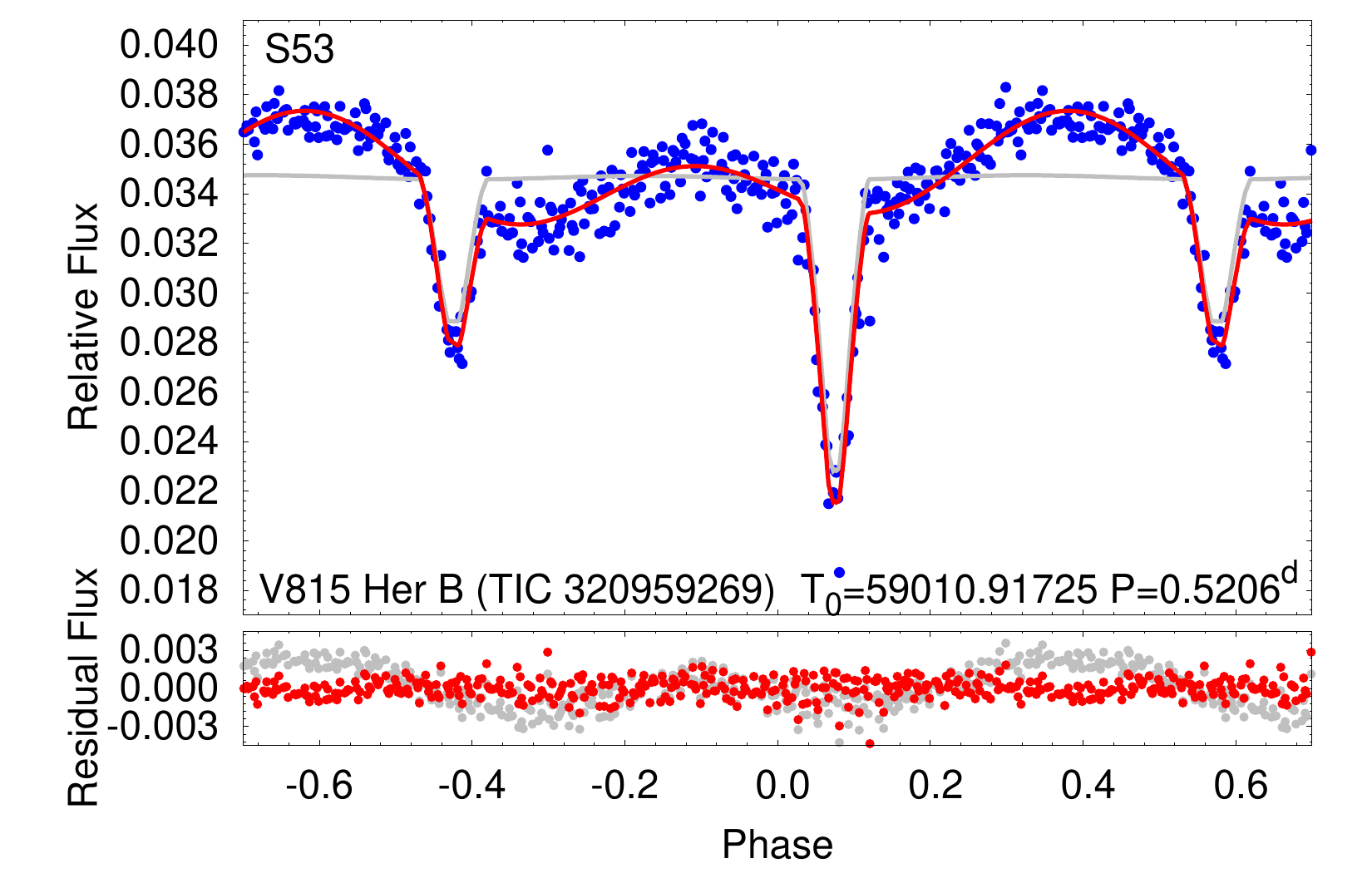}
      \caption{Cleaned and folded \emph{TESS} Sector 26, 40 and 53 light curves (blue dots) for the wide binary companion V815\,Her~B. The original orbital solutions without assuming surface inhomogeneities are drawn with grey lines, while the models drawn with red lines fit also the light curve variations arising from stellar spots. The residuals of both fits are plotted on the bottom parts of all panels. Note, that the eclipses in sectors 40 and 53 (middle and bottom panels) shifted from phase 0.0 and 0.5 indicating timing variations which is discussed in the text in detail. }
        \label{fig_TESS_foldfit}
\end{figure}

For the eclipsing binary (EB) light curve modeling the residual curves of the three \textit{TESS} sectors were converted into the flux domain, and normalized to unity. Then they were phase-folded with the previously determined eclipsing period of $0\fd5206$-day and binned into 500 evenly spaced phase cells. The fluxes in each cell were simply averaged, and these 500 mean flux values were rendered to the mid-phase moments of each cell. In such a manner we obtained evenly spaced, folded, binned, mean light curves separately for sectors 26, 40 and 53, which are plotted in Fig.\,\ref{fig_TESS_foldfit}.

The EB light curve analyses of these prepared data sets were carried out with software package {\sc Lightcurvefactory} \citep[see][and further references therein]{2019MNRAS.483.1934B,2020MNRAS.493.5005B}. We analysed the data of the three sectors separately. During our MCMC-based parameter search process seven light curve parameters were adjusted, as follows: the total duration of an eclipse (which is a direct observable, and closely relates to the sum of the fractional radii of the two stars); the $R_\mathrm{Bb}/R_\mathrm{Ba}$ ratio of the stellar radii; the $T_\mathrm{Ba}$ effective temperature of the primary; the logarithm of the ratio of the effective temperatures $\log(T_\mathrm{Bb}/T_\mathrm{Ba})$; the ``third'' light $\ell$, which allows us to take into account the light contamination caused by the much brighter binary V815\,Her~A; the EB's inclination $i_\mathrm{B}$ and, finally, the $T_0$ time of the reference epoch.\footnote{Note, while the light curve folding was applied in the phase domain, as {\sc Lightcurvefactory} operates in the time domain, the phase values were nominally converted into the time domain. Thus, the zero epoch $T_0$ was chosen as the mid-time of the very first primary eclipse observed with \textit{TESS}.} Regarding the other orbital elements, the folded mean light curve in Fig.\,\ref{fig_TESS_foldfit} reveals that the secondary eclipses occur just at (or, at least, very close to) phase 0.5, i.e. in the mid-times between the consecutive primary eclipses and, moreover, their durations look similar than that of the primary ones. Thus, we assume a circular EB orbit and, hence, keep the eccentricity fixed to $e_\mathrm{B}=0$.

The adjustment of the effective temperature of the primary  $T_\mathrm{Ba}$ needs some further explanation. In general, in the case of a single photometric band EB light curve, the temperature ratio of the two stars is well determined, while the effective temperatures of each star, themselves, play only a minor role in forming the light curve and hence, they are ill-determined. So, the usual treatment is to take one of the temperatures from some outer sources (e.g. spectroscopy or SED analysis), and to keep it fixed.
However, in the present case, $T_\mathrm{Ba}$ was treated as a free parameter in order to check the physical reliability and consistency of our solution.
This was done as follows. We know that the brighter binary should be a very young object. We reasonably assume that the two binaries are coeval, and thus, the two stars of binary B locate also on (or, very near to) the zero-age main sequence. Hence, we expect that the empirical mass--luminosity and mass--radius relations of \citet{1996MNRAS.281..257T} are applicable for these stars. Therefore, with the combined and inverted use of these relations, we can compute the stellar masses and the radii, as well, from the effective temperatures. Then we can compare the total mass of binary B, obtained from our analysis to the minimum mass inferred from the RV solution of \citet{2005AJ....129.1001F}. Moreover, we can compare the stellar radii, calculated directly through the $M-R$ relations of \citet{1996MNRAS.281..257T} to that calculated from the combination of the fractional radii (which are direct outputs of the light curve analysis) and the masses, inferred again from the \citet{1996MNRAS.281..257T} relations.

Finally, we also note, that a first inspection of the folded, mean light curve of binary B reveals that despite the averaging over $\sim$25 orbital cycles, the out-of-eclipse light variations did not average out perfectly. One possible origin of these out-of-eclipse light variations might be the surface brightness inhomogeneities (e.g. stellar spots), which, for the (almost) synchronized stellar rotations may have `survived' the averaging for the orbital period. We model these variations together with the EB light curve modeling simultaneously in a purely mathematical way, i.e. independent of their real origin, as follows. In each trial step, after the removal of the model EB light curve from the input light curve data, the residual is modelled with harmonic functions of two fixed frequencies (namely, the twice of the orbital frequency and, the orbital frequency itself), of which the four (plus one) coefficients are obtained via matrix inversion. Then, this mathematical model of the residual light curve is added to the EB model light curve and the actual $\chi^2$ value is calculated for this mixed model curve.

\begin{table}
 %\centering
\caption{Adjusted and derived astrophysical and orbital parameters for V815\,Her~B.}
 \label{tab:lcfit_V815HerB}
\begin{tabular}{@{}lll}
\hline
\hline\noalign{\smallskip}
\multicolumn{3}{c}{Orbital elements} \\
\hline\noalign{\smallskip}
  $P$ [d] & \multicolumn{2}{c}{$0.5206$}  \\
  $a$ [$R_\odot$] & \multicolumn{2}{c}{$2.33_{-0.24}^{+0.29}$}\\
  $e$ & \multicolumn{2}{c}{$0$}\\
  $\omega$ [$^{\circ}$] & \multicolumn{2}{c}{...} \\ 
  $i$ [$^{\circ}$] & \multicolumn{2}{c}{$89.17_{-1.45}^{+0.69}$}  \\
  $T_0-2400000$ [d] & \multicolumn{2}{c}{$59010.9173_{-0.0002}^{+0.0002}$}\smallskip  \\% 59011.04717
  \hline\noalign{\smallskip}
  mass ratio $(q=m_\mathrm{sec}/m_\mathrm{pri})$ & \multicolumn{2}{c}{$0.48_{-0.08}^{+0.12}$}\smallskip \\
  \hline\noalign{\smallskip}
\multicolumn{3}{c}{Stellar parameters} \\
\hline\noalign{\smallskip}
   & Ba & Bb \\
  \hline\noalign{\smallskip}
 \multicolumn{3}{c}{\smallskip Relative quantities} \\
  \hline\noalign{\smallskip}
 fractional radius [$R/a$] & $0.1603_{-0.0057}^{+0.0046}$ & $0.1102_{-0.0060}^{+0.0053}$\smallskip \\
 temperature relative to $(T_\mathrm{eff})_\mathrm{Ba}$ & $1$ & $0.8736_{-0.0293}^{+0.0222}$\smallskip \\
 fractional flux (in \textit{TESS}-band) & $0.0179_{-0.0021}^{+0.0020}$ & $0.0038_{-0.0003}^{+0.0003}$\smallskip \\
 \hline\noalign{\smallskip}
 \multicolumn{3}{c}{Physical quantities} \\
 \hline\noalign{\smallskip}
 $m$ [$M_\odot$] & $0.444_{-0.158}^{+0.198}$ & $0.185_{-0.022}^{+0.053}$\smallskip \\
 $R$ [$R_\odot$] & $0.377_{-0.051}^{+0.047}$ & $0.256_{-0.023}^{+0.026}$\smallskip \\
 $T_\mathrm{eff}$ [K] & $3702_{-174}^{+369}$ & $3245_{-91}^{+174}$\smallskip \\
 $L_\mathrm{bol}$ [$L_\odot$] & $0.024_{-0.010}^{+0.019}$ & $0.007_{-0.002}^{+0.003}$\smallskip \\
 $M_\mathrm{bol}$ [mag] & $8.77_{-0.64}^{+0.54}$ & $10.20_{-0.41}^{+0.34}$\smallskip \\
 $M_V           $ [mag] & $10.47_{-1.31}^{+1.04}$ & $13.20_{-0.41}^{+0.34}$\smallskip \\
 $\log g$ (in cgs) & $4.94_{-0.08}^{+0.05}$ & $4.91_{-0.06}^{+0.07}$\smallskip \\
 \hline\noalign{\smallskip}
 extra flux (in \textit{TESS}-band) & \multicolumn{2}{c}{$0.978_{-0.002}^{+0.002}$}\smallskip \\
 \hline\noalign{\smallskip}
\end{tabular}
%\textit{Notes. }{}
\end{table}

In Table~\ref{tab:lcfit_V815HerB} we tabulate the median values and the $1\sigma$ statistical uncertainties of the parameters (including several derived, physical ones) that were obtained from our analysis for the sector 26 data. We also display the synthetic model fits (both with and without the mathematically modelled out-of-eclipse light variations) derived from the best-fit solution in Fig.~\ref{fig_TESS_foldfit}. We note that in case of the analysis of sector 40 and 53 light curves we obtained mostly similar parameters within the tabulated 1-$\sigma$ uncertainties and, hence, we do not tabulate these results separately. The two exceptions are the $\ell$ extra flux and the $T_0$ epoch of the primary minimum. Regarding the former one, in sectors 40 and 53 we obtained smaller $\ell$ parameters with $\sim$5$-$10\%, which we explain with a smaller level of contaminated fluxes in the \textit{TESS} pixels. What physically more interesting is, the shift of the phase of the primary eclipses, which can also be seen in the consecutive panels of Fig.~\ref{fig_TESS_foldfit}. In order to understand the nature and origin of this effect, we carried out a detailed analysis of the times of eclipsing minima.

\subsection{Eclipse timing variation analysis of V815\,Her~B}
\label{TESS_ETV}

\begin{figure}
    \centering
    \includegraphics[width=1.0\linewidth]{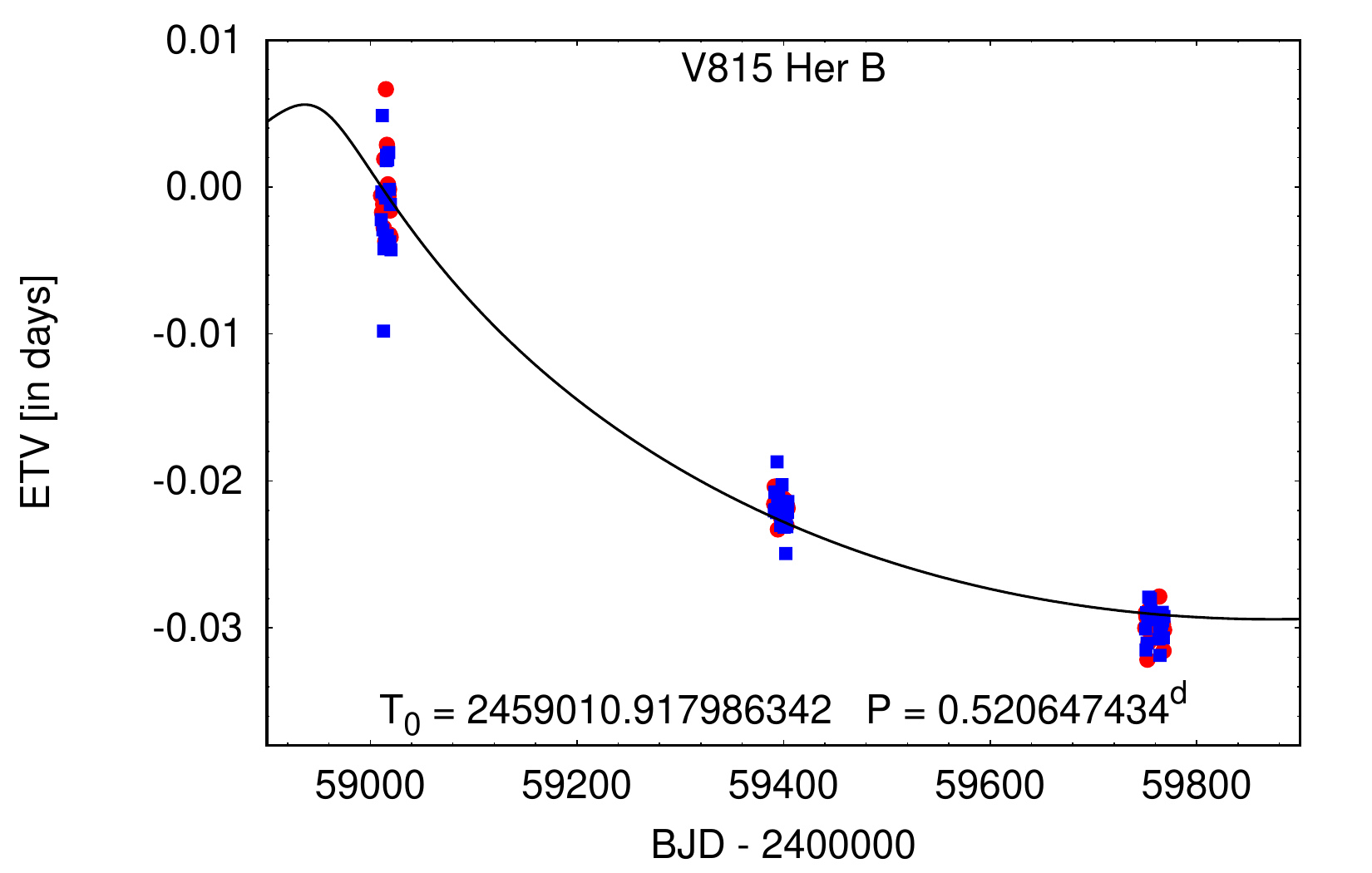}
     \caption{Eclipse timing variations of V815\,Her~B, calculated from the mid-minima times of primary and secondary eclipses (red dots and blue squares, respectively), observed with \textit{TESS}. The black line represents the theoretical LTTE curve derived from the former third-body RV solution of \citet{2005AJ....129.1001F}. See text for details.}
   \label{fig:etv}
\end{figure}

We determined mid-eclipse times for all the observed primary and secondary eclipses of V815\,Her~B for the available \textit{TESS} cleaned time-series (see in Sect.\,\ref{sect_TESS}). The individual times of minima are tabulated in Table~\ref{Tab:V815_Her_B_ToM} of Appendix A. We derived an eclipse timing variation (ETV) curve using the linear ephemeris
\begin{equation}
    \mathrm{MIN}_\mathrm{I}=2459010.917986 + 0.520647 \times E,
\end{equation}
where the zero epoch is given in barycentric Julian Days (BJD). We plot the obtained ETV curve in Fig.~\ref{fig:etv}. As one can see, the ETV curve displays clearly a non-linear behaviour, which reveals that the observed eclipsing period is varying.

As a next step, we demonstrate that this variation can be explained by the light-travel time effect (LTTE) caused by the revolution of the eclipsing binary V815\,Her~B around its brighter counterpart V815\,Her~A, and the parameters of this LTTE orbit is in accordance with the former RV solution of \citet{2005AJ....129.1001F}. This indicates, that this currently discovered EB is identical with the third component suspected formerly through RV measurements. In order to demonstrate this fact, we derive LTTE solution from the outer RV orbit of \citet[][their Table~2]{2005AJ....129.1001F}. While doing so, one should be cautious about a few things. First, the argument of pericenter of the outer orbit ($\omega_\mathrm{A}$) given in the RV solution refers to the orbit of component A, while in the LTTE solution of component B, the argument of pericenter of component B ($\omega_\mathrm{B}=\omega_\mathrm{A}+180^\mathrm{o}$) is present. Moreover, the projected semi-major axis $a_\mathrm{A}\sin i_\mathrm{out}$ of the orbit of component A around the center of mass of the whole quadruple system, which was deduced from the amplitude of the RV curve, should also be transformed to the projected semi-major axis of component B ($a_\mathrm{B}\sin i_\mathrm{out}=m_\mathrm{A}/m_\mathrm{B}\times a_\mathrm{A}\sin i_\mathrm{out}$). For this step, we must know the mass ratio of the two binaries. For the formerly known, brighter component we accept $m_\mathrm{A}=m_\mathrm{Aa}+m_\mathrm{Ab}=1.27\,M\sun$ \citep{2005AJ....129.1001F}, while our solution gives the mass of component B to be $m_\mathrm{B}=m_\mathrm{Ba}+m_\mathrm{Bb}=0.63\,M\sun$, which results in an outer mass ratio of $m_\mathrm{A}/m_\mathrm{B}=2.02$. Taking into account these constraints, and plotting the theoretical ETV curve against the measured ones, one can see a nice agreement (see Fig.~\ref{fig:etv}), which, in our readings demonstrate nicely that the spectroscopic multiplicity of V815\,Her is confirmed independently through \textit{TESS} photometry, and the third spectroscopic component itself is a close, eclipsing binary, forming such a way a close quadruple system with a 2+2 hierarchy.

For comparison, the periods and geometrical parameters of the wide (V815\,Her~AB) and the two close orbits (V815\,Her~A and B) in the 2+2 hierarchical system are given in Table\,\ref{Tab:twoplustwo}.

\begin{table}
 %\centering
\caption{Comparison of the periods and geometry of the wide (AB) and the two close orbits (A and B) in the 2+2 quadruple system V815\,Her.}
 \label{Tab:twoplustwo}
\begin{tabular}{@{}ccc}
\hline
\hline\noalign{\smallskip}
 & AB\tablefootmark{a} &   \\
\hline\noalign{\smallskip}
 &  $P_{\rm AB}$=2092.2\,d &  \\
 & $a_{\rm AB}$=236.6$\times$10$^6$\,km & \\
 & $i_{\rm AB}$=78.4$^{\circ}$ & \\
 & $e_{\rm AB}$=0.765 & \\
\hline\noalign{\smallskip}
A\tablefootmark{a} &  &  B\tablefootmark{b}\\
\hline\noalign{\smallskip}
 $P_{\rm A}$=1.8098\,d &  &  $P_{\rm B}$=0.5206\,d\\
 $a_{\rm A}$=1.38$\times$10$^6$\,km &  &  $a_{\rm B}$=1.62$\times$10$^6$\,km\\
 $i_{\rm A}$=78.4$^{\circ}$\tablefootmark{c} &  &  $i_{\rm B}$=89.17$^{\circ}$\\
 $e_{\rm A}$=0.0 &  &  $e_{\rm B}$=0.0\\
 \hline
\end{tabular}
\tablefoot{
\tablefoottext{a}{\citet{2005AJ....129.1001F};}
\tablefoottext{b}{this paper;}
\tablefoottext{c}{assuming $i_{\rm A}$=$i_{\rm AB}$.}
}
\end{table}

\subsection{Long-term photometric behavior using archival data}

\begin{figure}[thb]
\includegraphics[width=1.0\linewidth]{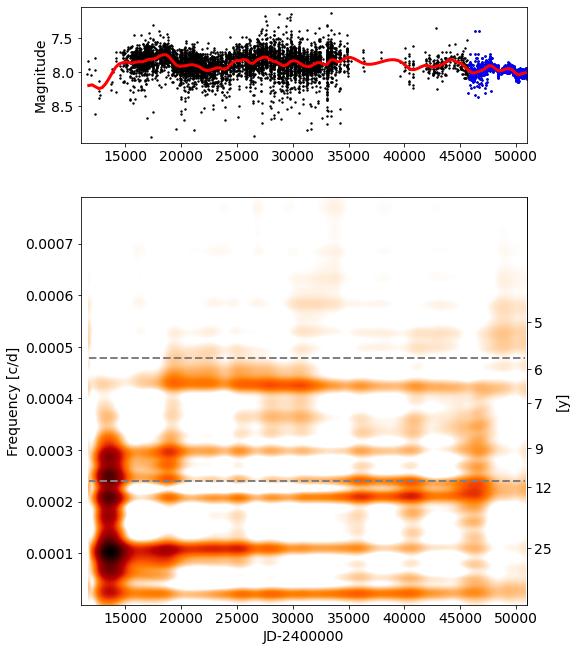}
      \caption{Long-term photometry and time-frequency analysis of V815\,Her. {\it Top:} DASCH light curve of $\approx$100 years for V815\,Her from scanned photographic plates (black dots) supplemented with photoelectric data (blue dots) spanning more than a century between 1890-1998 with 5 years' overlap. The spline smoothed data %averaged over 365 days 
      are indicated by a red line. {\it Bottom:} Time-frequency analysis for the available photographic+photoelectric data of V815\,Her using the \texttt{TiFrAn} code \citep{2009A&A...501..695K}. The plot indicates dominant cycles of different lengths and amplitudes on timescales corresponding to $\sim$6.5, $\sim$9.1, $\sim$11.5, $\sim$13 and $\sim$26 years, the latter two are probably multiples of the 6.5-year period. Dashed lines indicate the 2092\,d period of the wide orbit and its double at 4184\,d ($\approx$11.5yr).
      See the text for more.}
        \label{fig_tifran}
\end{figure}

V815\,Her has data from scanned photographic plates in the Digital Access to a Sky Century @ Harvard \citep[DASCH,][]{2009ASPC..410..101G} database for about hundred years between 1890-1990 supplemented with photoelectric data collected in \citet{2000A&A...362..223J} with an overlap of 5 years. We stitched the two data sets together by shifting the photoelectric B band light curve to match the median of the DASCH light curve in the 5 year overlap. These data cannot depict the rotational modulation of the G star, but can be used to search for long-term variability and trends. We combine the photographic and photoelectric data sets in Fig.~\ref{fig_tifran}.

Looking for possible activity cycles we used our Time Frequency Analyzer package \texttt{TiFrAn} \citep{2009A&A...501..695K} with Choi-Williams distribution kernel which gives good resolution in the frequency domain and somewhat less in the time-domain. The data were averaged in bins of 365 days. Although there is a huge difference between the observational noise of the photographic and photoelectric data, the magnitude of the noise does not alter the frequency pattern. The different kernels of \texttt{TiFrAn} and their features are discussed in detail in \citet{2009A&A...501..695K}.

The data and results are plotted in Fig.~\ref{fig_tifran}. We reveal a cycle with a slowly increasing period, which has a steady amplitude at $\sim$6.5 years. Two other cycles of $\sim$9.1 and $\sim$13 years with similarly strong amplitudes are present throughout the observations. Between them, we also find an $\sim$11.5-year cycle which is still clearly visible despite its weaker amplitude. This fits quite well to twice the period of the outer orbit, i.e. 11.45 years (=4184\,d); see the lower dashed line in Fig.~\ref{fig_tifran}. The longest persistent feature (ignoring the one corresponding to the total data length) points towards a timescale around $\sim$24-26 years with weakening amplitude. The cycle lengths of $\sim$13 and $\sim$26 years are perhaps harmonics of $\sim$6.5 years, although in contrast to the slow growth of the 6.5-year cycle over time, the 13-year cycle seems to decrease towards the end of the time series. At this point, it is difficult to interpret the possible physical background of each cycle that appears in the brightness change. In any case, we will attempt to list some ideas in Sect.~\ref{sect_disc}.

%-------------------------------------
\section{Spectroscopic data and astrophysical properties of V815\,Her~Aa}\label{sect_par}

\subsection{STELLA-SES spectra}\label{sect_stella_data}
The spectroscopic observations were carried out with the \mbox{1.2-m}
STELLA-II telescope of the STELLA robotic observatory
\citep{2010AdAst2010E..19S} located at Iza\~{n}a Observatory in
Tenerife, Spain. STELLA-II is equipped with the fibre-fed, fixed-format
STELLA Echelle Spectrograph (SES) providing an average
spectral resolution of $R=55,000$ along the covered 3900–8800\,\AA\ wavelength range. Further details on the performance
of the system and the data-reduction procedure can be found
in \citet{2008SPIE.7019E..0LW,2012SPIE.8451E..0KW}. 

Altogether 621 high-resolution spectra were recorded between 07 March 
and 11 November, 2018. With a default exposure time of 3000\,s an average signal-to-noise ratio (S/N) of $\approx$200:1 could be reached and only 18 spectra had to be discarded due to their poor quality. This data set were used for both the spectral synthesis detailed in the next section and the Doppler-inversions (see Sect.\,\ref{sect_di}).
The observing log is summarized in Table~\ref{tab_a_1}. Assuming synchronized rotation for the G star in the close binary subsystem V815\,Her~A, the rotational phases are calculated using the orbital period from \citet{2005AJ....129.1001F} according to the following equation:
\begin{equation}\label{phases}
    \mathrm{HJD}=2450204.5802 + 1.80983433\times E.
\end{equation}

The zero phase is the time of maximum radial velocity, i.e., the primary component is closest to us in $\phi$=0.75 phase, accordingly, the $\lambda$=270$^{\circ}$ longitude of the primary is facing the secondary.

\subsection{Spectral synthesis}\label{sme}

For the determination of some of the basic astrophysical parameters, we used the spectral synthesis code SME \citep{2017A&A...597A..16P}. Ten good quality (S/N$\sim$200) STELLA-SES echelle spectra were randomly selected to carry out spectral synthesis independently on each. During the computations, local thermodynamical equilibrium (LTE) was assumed and MARCS atmospheric models \citep{2008A&A...486..951G} were used. Atomic line data were taken from the VALD line database \citep{kupka_vald}. A global macroturbulence relationship was adopted from  \citet[][see their Eq.~1]{2005ApJS..159..141V}.
The parameters, $T_{\rm eff}$ effective temperature, $\log g$ surface gravity, $[\mathrm{Fe}/\mathrm{H}]$ metallicity, and $v_{\rm mic}$ microturbulence were determined iteratively; for details on the fitting method see \citet{2019A&A...627A..52K}. For the fits the equatorial rotational velocity $v\sin i$ was kept at 30\kms\ %\citep[cf.][]{2005AJ....129.1001F}
since the Doppler-inversion (see Sect.~\ref{sect_di} in this paper) proved to be artifact-free at this value, while using e.g. 27\kms\ and 33\kms\ caused severely %artificial 
artifact-ridden results having unrealistic equatorial brightening with polar darkening or polar brightening with equatorial darkening, respectively. We note that if $v\sin i$ was treated as a free parameter as well, the process yielded only slightly different results compared to the value of 30\kms.

\begin{figure}[thb]
    \includegraphics[width=\columnwidth]{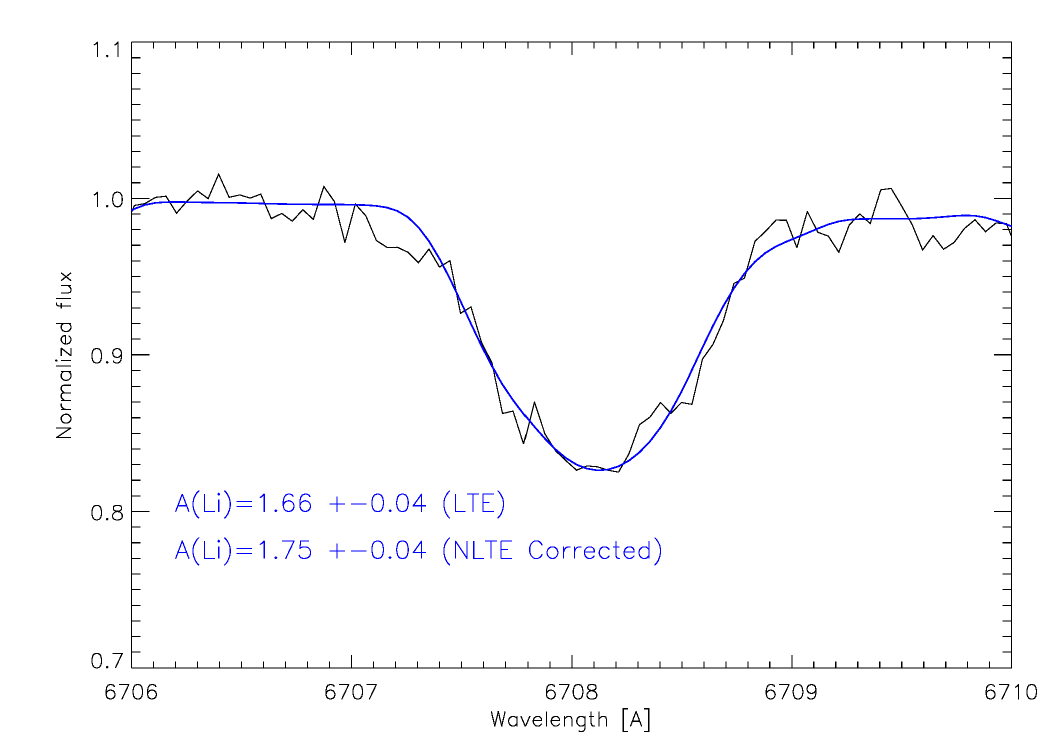}
      \caption{An example plot of the observed Li\,{\sc i} 6707\,\AA\ spectral region (black) fitted with a synthetic spectrum (blue).}
        \label{fig_liplot}
\end{figure}

Lithium 6707\AA\ abundance was determined also by SME using LTE synthesis together with 3D-NLTE departure coefficients taken from \citet{2018A&A...618A..16H}. We selected the same ten spectra of high signal-to-noise ratio as above, for which we determined the lithium abundance separately. Averaging these yielded logarithmic abundances of $A(\mathrm{Li})_\mathrm{LTE} = 1.66\pm0.04$ and $A(\mathrm{Li})_\mathrm{NLTE} = 1.75\pm0.04$.   Here, logarithmic abundance is defined as $A(\mathrm{Li})=\log(N_{\mathrm{Li}}/N_{\mathrm{H}}) + 12$ where $N_{\mathrm{Li}}$ and $N_{\mathrm{H}}$ are the number densities of lithium and hydrogen, respectively. Fig.~\ref{fig_liplot} shows an example fit to the observed Li\,{\sc i} 6707\,\AA\ spectral region. In a solar-like star like the G component in the V815\,Her system, Li-6707 is gradually depleting during the pre-MS evolution over a few tens of Myr \citep[e.g.][]{2009IAUS..258...81H}. According to \citet[][]{2015MNRAS.449.4131S}, in the Pleiades, $A(\mathrm{Li})$ typically decreases from the initial 3.3 to below $\sim$1.0 by the age of $\sim$100\,Myr, however, highly dependent on the evolutionary models applied. Nevertheless, the correspondingly estimated age of a few tens of Myr is also consistent with the result of our spectral energy distribution analysis; see Sect.~\ref{sect_vosa}.

% ---------------------------- T2
\begin{table}
 \centering%%%
\caption{Adopted/calculated astrophysical parameters for V815\,Her~Aa}
\label{tab_param}
%\begin{footnotesize}
\begin{tabular}{l l l}
\hline\noalign{\smallskip}
Parameter & Value \\
\hline\hline
\noalign{\smallskip}
Spectral type\tablefootmark{a}           & G6 V \\
Gaia distance [pc]    &  $32.087\pm0.127$ \\
$V_{\rm br}$    [mag]\tablefootmark{a}          & 7.56 \\
$(B-V)$     [mag]\tablefootmark{a}         & 0.71 \\
$M_{\rm bol}$     [mag]        & $4.90\pm0.03$ \\
Luminosity [${L_{\odot}}$]         & $0.87\pm0.03$ \\
$T_{\rm eff}$ [K] &      5582$\pm$63 \\
$\log g$ (in cgs) &      4.27$\pm$0.04 \\
$v\sin i$ [\kms]    &       $30.0\pm1.5$  \\
$P_{\rm rot}=P_{\rm orb}$ [d]\tablefootmark{a}   &           $1.80983433$  \\
Inclination of the rotation axis  [\degr]\tablefootmark{a}            &       $75\pm5$   \\
Radius      [$R_{\odot}$]           &      $1.1\pm0.1$    \\
Mass          [$M_{\odot}$]           & $\lesssim$1.0   \\
Microturbulence  [\kms] & $1.14\pm 0.19$ \\
Macroturbulence  [\kms]\tablefootmark{b} & 4.2  \\
Metallicity [Fe/H] &  0.06$\pm$ 0.03  \\
$A(\mathrm{Li})_\mathrm{LTE}$ (log) & 1.66$\pm$0.04  \\
$A(\mathrm{Li})_\mathrm{NLTE}$ (log) & 1.75$\pm$0.04  \\
Age [Myr]  & $\approx$30 \\
\hline
 \end{tabular}
 \tablefoot{
\tablefoottext{a}{\citet{2005AJ....129.1001F};}
\tablefoottext{b}{\citet{2005ApJS..159..141V}.}
}
\end{table}

\subsection{Analysis with VOSA}\label{sect_vosa}
\begin{figure}[h]
    \includegraphics[width=\columnwidth]{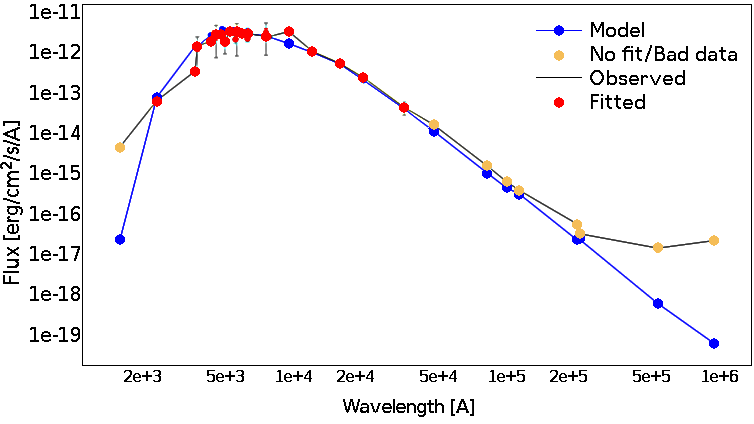}
      \caption{Spectral energy distribution for V815\,Her generated by the VOSA SED analyzer tool \citep{2008A&A...492..277B}. The synthetic spectrum (blue line) is fitted to the archival photometry (red dots) from the available VO catalogues and surveys using ATLAS9 Kurucz ODFNEW/NOVER
models with $T_{\rm eff}$=5500\,K effective temperature, $\log g$=4.5 surface gravity, and [Fe/H]=0.2 metallicity.
      }
        \label{fig_sed}
\end{figure}

We performed a spectral energy distribution (SED) synthesis using the virtual observatory (VO) SED analyzer tool VOSA \citep{2008A&A...492..277B} to build the SED of V815\,Her from the available VO catalogues
and surveys (\emph{GALEX}, \emph{Tycho}, \emph{SLOAN/SDSS}, \emph{Pan-Starrs}, \emph{Gaia}, \emph{2MASS}, \emph{AKARI/IRC}, \emph{WISE}, and \emph{IRAS}).
For fitting the resulting SED in Fig.~\ref{fig_sed} ATLAS9 Kurucz ODFNEW/NOVER models \citep{2003IAUS..210P.A20C} were used with the nearest grid values of $T_{\rm eff}$=5500\,K effective temperature, $\log g$=4.5 surface gravity, and $[\mathrm{Fe}/\mathrm{H}]$=0.2 metallicity.

It is noteworthy that the SED (see Fig.\,\ref{fig_sed}) shows an excess towards infrared (IR) wavelengths. This could be the characteristics of a very young age since pre-main sequence stars of $\lesssim$10\,Myr \citep{2011ARA&A..49...67W} usually possess primordial protoplanetary disc, which causes excess in near-IR (cf. Sect.\,\ref{sme}).

\begin{figure}[h]
    \includegraphics[width=1.00\columnwidth]{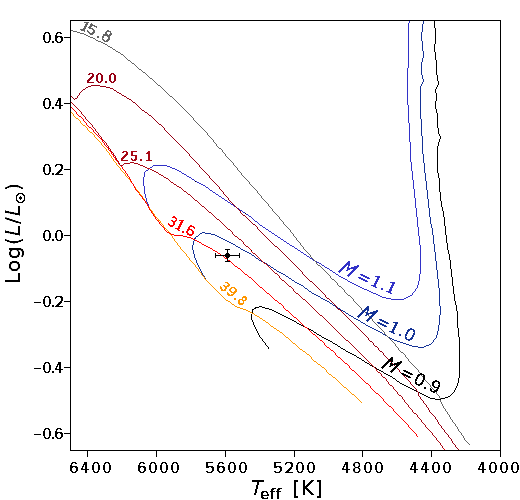}
      \caption{Position of V815\,Her~Aa (black dot) on the HRD plotted by using the VOSA SED analyzer tool \citep{2008A&A...492..277B}. Evolutionary tracks for the Geneva model \citep{2019A&A...624A.137H} indicate an age of 30\,Myr for a $M$$\lesssim 1$$M_{\odot}$ ZAMS star.
      }
        \label{fig_hrd}
\end{figure}

In Fig.~\ref{fig_hrd} we plotted the star on the Hertzsprung$-$Russell-diagram (HRD) along with theoretical isochrones and evolutionary tracks for the most suitable Geneva model \citep{2019A&A...624A.137H}. According to the plot, the age of the G star with $M/M_{\odot}\lesssim1.0$ mass ratio is $\approx$26 million years. We note, however, that errors of age determination increase towards the zero-age main sequence (ZAMS) as the isochrones line up more and more densely. Nevertheless, this age determination is in agreement with the measured Li-6707 abundance in Fig.~\ref{fig_liplot}.

%On the other hand, infrared excess towards the 

\begin{table*}[thb]
\centering
\caption{Temporal distribution of the subsequent data sets for each individual Doppler image.}
\label{disets}
\begin{tabular}{c c c c c c }
\hline
\hline\noalign{\smallskip}
Data & Mean-HJD & Mean-date & Number & Data range & Data range \\
subset &2\,450\,000+  & yyyy-mm-dd & of spectra &in days& in $P_{\rm rot}$  \\
\hline
\noalign{\smallskip}
S01 & 8202.0111 & 2018-03-24 & 12 & 6.069& 3.354 \\
S02 & 8212.6014 & 2018-04-04 & 15 & 8.094& 4.472 \\
S03 & 8223.9904 & 2018-04-15 & 17 & 8.100& 4.475 \\
S04 & 8238.1401 & 2018-04-29 & 29 & 8.161& 4.509 \\
S05 & 8248.3858 & 2018-05-09 & 33 & 8.147& 4.502 \\
S06 & 8255.8846 & 2018-05-17 & 27 & 6.175& 3.412 \\
S07 & 8263.4391 & 2018-05-24 & 30 & 8.175& 4.517 \\
S08 & 8272.1472 & 2018-06-02 & 38 & 7.234& 3.997 \\
S09 & 8280.1488 & 2018-06-10 & 31 & 8.168& 4.513 \\
S10 & 8289.2001 & 2018-06-19 & 47 & 7.255& 4.009 \\
S11 & 8296.4747 & 2018-06-26 & 39 & 6.219& 3.436 \\
S12 & 8303.9658 & 2018-07-04 & 38 & 7.204& 3.981 \\
S13 & 8315.0355 & 2018-07-15 & 29 & 6.124& 3.384 \\
S14 & 8324.4312 & 2018-07-24 & 31 & 6.220& 3.437 \\
S15 & 8357.6721 & 2018-08-27 & 38 & 8.128& 4.491 \\
S16 & 8367.9366 & 2018-09-06 & 35 & 10.001& 5.526 \\
S17 & 8381.8476 & 2018-09-20 & 28 & 8.075& 4.462 \\
S18 & 8390.1971 & 2018-09-28 & 18 & 7.115& 3.931 \\
S19 & 8402.4761 & 2018-10-10 & 10 & 6.015& 3.323 \\
\hline
\end{tabular}
\end{table*}

\subsection{Luminosity and size}
The \emph{Gaia} EDR3 parallax of $\pi=31.1649\pm0.1225\,\mathrm{mas}$ \citep{2016A&A...595A...1G,2021A&A...649A...1G}
yields a distance of $d=32.087\pm0.127\,\mathrm{pc}$ for V815\,Her.
Assuming $V_{\rm br}\approx7\fm56$
for the brightest $V$ magnitude with $B-V$ of 0$\fm$71 \citep[cf.][]{2005AJ....129.1001F}, and neglecting interstellar extinction
yields an absolute visual magnitude of $M_V=5\fm03\pm0\fm03$.
Taking the $BC=-0.132$ bolometric correction from \citet{1996ApJ...469..355F} gives a bolometric magnitude of $M_{\mathrm{bol}}=4\fm90 \pm0\fm03$. This gives $L/L_{\odot}=0.87\pm0.03$ when using a value of  $M_{\mathrm{bol},\odot}=4\fm74$ for the Sun.

Assuming an inclination of $i$=75$\pm5^{\circ}$ \citep[cf.][]{2005AJ....129.1001F}, the projected equatorial rotational velocity $v\sin i$ of 30\kms\ with $\pm$1.5\kms\ error (see Sect.~\ref{sme}) would yield a stellar radius of $R/R_{\odot}=1.1\pm0.1$. We note, that at a distance of $\approx$32\,pc the estimated limb-darkened angular diameter of the G star of 0.346($\pm$0.009)\,mas \citep[Mid-infrared stellar Diameters and Fluxes compilation Catalogue V10,][]{2019yCat.2361....0C} would yield a similar radius of  $\approx$1.2$R_{\odot}$.
Although the $\gtrsim$1.1$R_{\odot}$ size is a bit larger than predicted for a non-magnetic G5 ZAMS star, still, such inflation is indeed to be expected for a ZAMS star with enhanced surface magnetic activity \citep[cf.][and their references]{2016A&A...586A..52J}.
When combining this radius with the effective temperature, the Stefan$-$Boltzmann law would give $L/L_{\odot}$ ratio of 1.05$\pm$0.15, still in line with the value derived from the bolometric magnitude, if the estimated errors are taken into account.
Moreover, the 1.1$R_{\odot}$ radius, when combined with the most likely mass ratio of $M/M_{\odot}\lesssim1.0$ (cf. Fig.\,\ref{fig_hrd} and \cite{2005AJ....129.1001F}) would yield $\log g\approx4.31$ for the G component in a pretty good agreement with the one obtained from SME.

In Table~\ref{tab_param} we summarize the astrophysical parameters of the G star determined in our study, along with the others adopted from the literature.

\section{Surface temperature maps of the G star}\label{sect_di}
%\nopagebreak

\subsection{Imaging code \emph{iMAP}}
To reconstruct the surface temperature maps we use the state-of-the-art Doppler imaging code \emph{iMAP} \citep{2012A&A...548A..95C}. This code carries out multi-line inversion on a list of photospheric lines. We selected 38 non-blended absorption lines from the 5049-6750\,\AA\ region with suitable line-depth, temperature sensitivity and well-defined continuum. The stellar surface is modeled on a 5$^{\circ}\times$5$^{\circ}$ spherical grid. Each local line profile is computed with a full radiative solver \citep{carroll_solver}. The local line profiles are disk integrated, and the individually modeled disk-integrated lines are averaged. Atomic line data are taken from the Vienna Atomic Line Database \citep[VALD,][]{kupka_vald}. Model atmospheres from \citet{castelli_mod} are interpolated for the necessary temperature, gravity, or metallicity values. When solving the radiative transfer, local thermodynamical equilibrium (LTE) is assumed. Additional input parameters are micro- and macroturbulence, and the projected equatorial velocity (see  Sect.\,\ref{sect_par}). For the surface temperature reconstructions \emph{iMAP} uses an iterative regularization based on a Landweber algorithm \citep{2012A&A...548A..95C}. 
The iterative regularization has been proven to always converge on the same image solution, therefore, no additional constraints are imposed \citep[see Appendix\,A in][]{2012A&A...548A..95C}.

\begin{figure*}[thb!!!]
    \includegraphics[width=\linewidth]{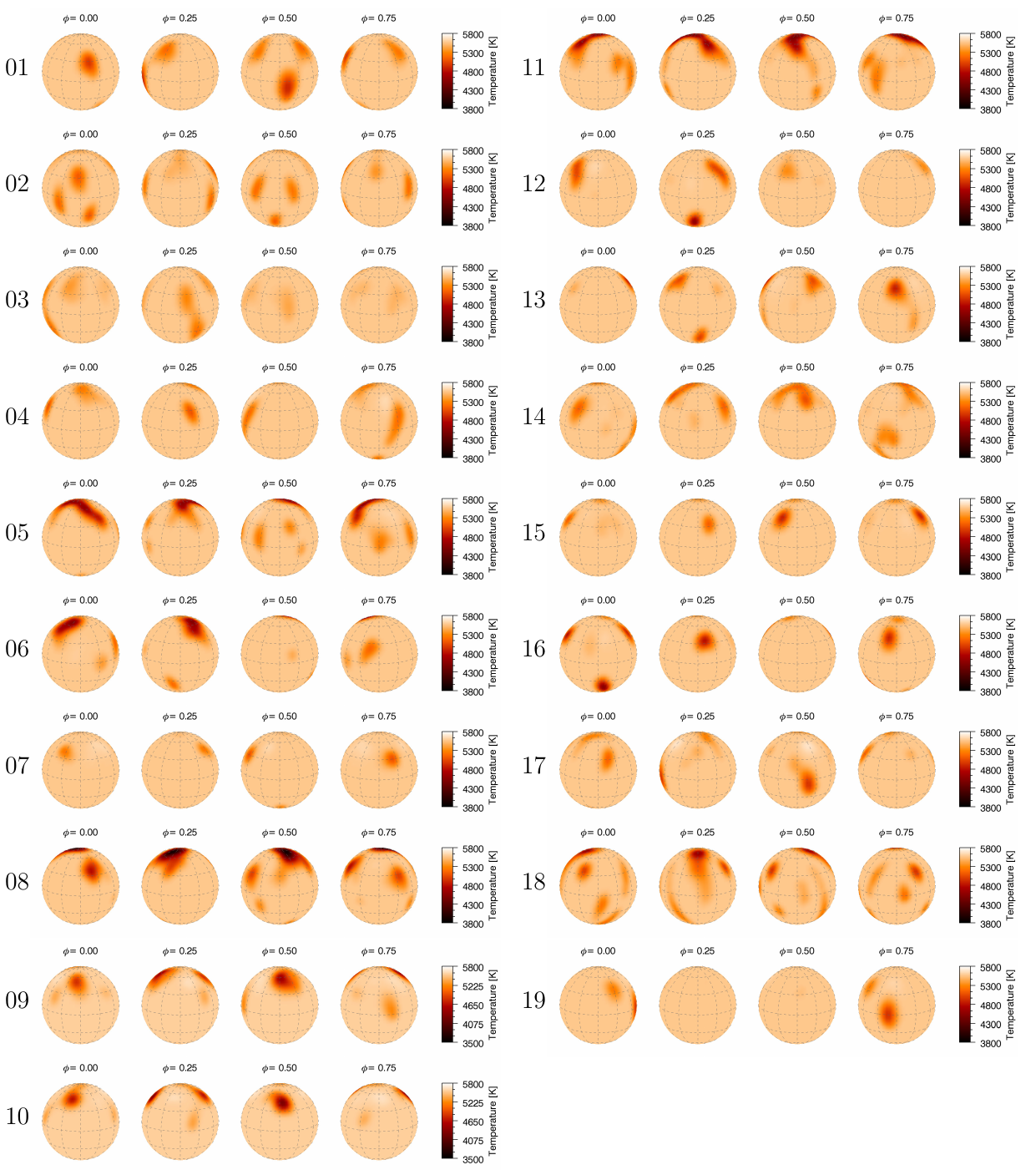}
\caption{Nineteen time-series Doppler images of V815\,Her for the 2018 observing season. From top to bottom, in chronological order, Doppler images S01-S10 are on the left, and Doppler images S11-S19 are on the right.}
\label{fig:DIs}
\end{figure*}
%\begin{multicols}{2}
%\end{multicols}
%\newpage
\nopagebreak
\subsection{Time-series Doppler images}\label{DIs}
From the STELLA-SES spectra described in Sect.\,\ref{sect_stella_data} we formed consecutive and independent data subsets in such a way that each of them covers the entire rotation phase in sufficient density. As a result, we created enough subsets for 19 time-series Doppler images. For the temporal distribution of the Doppler images see Table~\ref{disets}. The 19 time-series Doppler images of V815\,Her~Aa
are plotted in Fig.\,\ref{fig:DIs}, while the corresponding line-profile fits are
given in Figs.\,\ref{fig:di_profs1}-\ref{fig:di_profs3} in Appendix\,\ref{App:profiles}. In order to follow the change in the spot distribution even more spectacularly, in Appendix\,\ref{App:DIsMerc}, we also present our Doppler images in the Mercator projection. The surface temperature maps indicate a constantly changing surface structure on a time scale of a few weeks.
Even the coverage of the visible pole with a spot is not continuous: in images S05-S06 and S08-S10 and in S18 we see strong polar spots, which, for example, are almost absent in images S03, in S07 and S19. Spot temperature ranges are not constant either, the highest contrast (see e.g., S08) reaches 1800\,K, while the smallest temperature contrast (S07) is only a few hundred degrees. 
The lifespan of the longest-lived spots can typically be estimated at 2-3 weeks. However, with few exceptions, it is difficult to clearly point out the corresponding spots in successive images. The obvious main reason for this is rapid spot evolution, in which surface differential rotation is most likely involved. What is certain is that during the observed period we do not see a permanent polar spot, nor is there any sign of the formation of an active longitude. For further discussion of surface evolution see Sect.\,\ref{surfevol}.

\subsection{Surface differential rotation}\label{sect_dr}

The latitude dependent surface rotation can usually be inferred from the cross-correlation of two successive Doppler images \citep{1997MNRAS.291....1D}. However, random changes in spot distribution can easily mask the expected cross-correlation pattern of differential rotation. Therefore, 
for the detection of the surface shear operating in the background, we use {\tt ACCORD} \citep[e.g.,][]{2012A&A...539A..50K,2015A&A...573A..98K}, a technique which enables the averaging of latitudinal cross-correlations in the case of a sufficient number of pairs of Doppler images not too distant in time, this way suppressing the effect of randomness and amplifying the correlation pattern attributed to surface differential rotation.

\begin{figure}[thb]
    \centering
    \includegraphics[width=1.0\columnwidth]{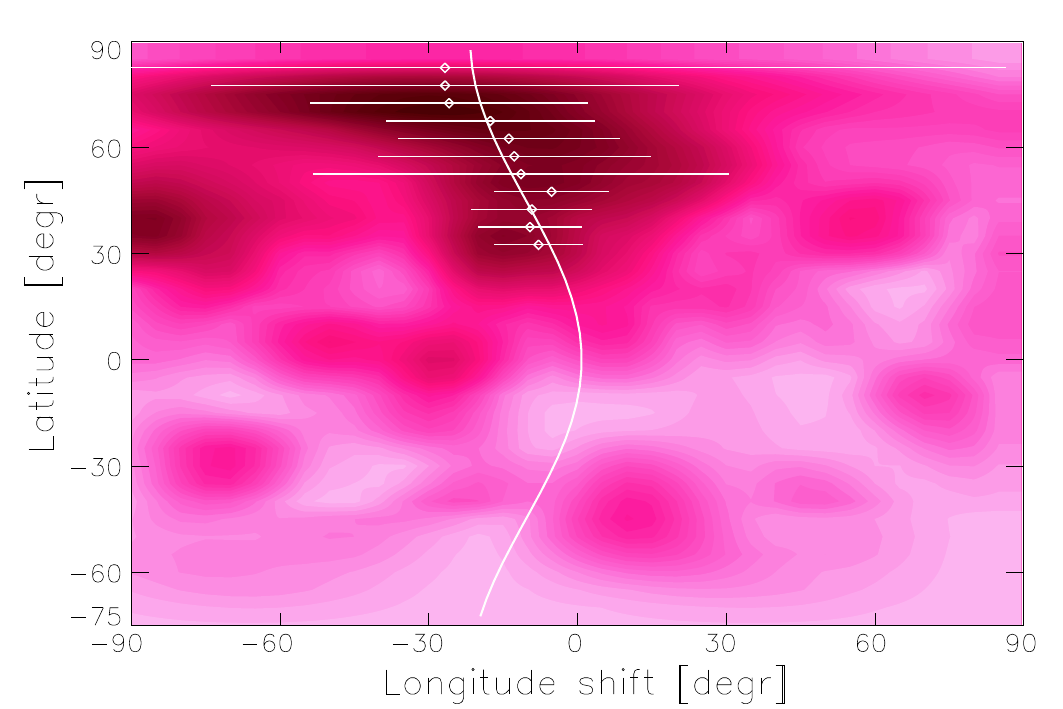}
     \caption{The average cross-correlation function map derived from 18 individual cross-correlations indicates how much longitude shift occurs at a given latitude due to surface shear during $\sim$6$P_{\rm rot}$ (which is the average time difference between the mean-HJDs of the consecutive maps; see Table\,\ref{disets}).}
   \label{fig_ccf}
\end{figure}

The overall correlation pattern in the resulting average cross-correlation map (Fig.\,\ref{fig_ccf}) is fitted by a rotational law in the form
\begin{equation}
    \Omega(\beta)=\Omega_{\mathrm{eq}}-\Delta\Omega\sin^2\beta,
\end{equation}
where $\Omega(\beta)$ is the latitude ($\beta$) dependent angular velocity, while $\Omega_{\mathrm{eq}}$ is the angular velocity at the equator, and $\Delta\Omega=\Omega_{\mathrm{eq}}-\Omega_{\mathrm{pole}}$ gives the difference between the equatorial and polar angular velocities. The dimensionless surface shear parameter $\alpha$ is defined as $\alpha_{\mathrm{DR}}=\Delta\Omega/\Omega_{\mathrm{eq}}$.
The resulting fit indicates a weak solar-type differential rotation, with $\Omega_{\mathrm{eq}}=199.06\pm0.40$\degr/day and $\Delta\Omega=1.99\pm0.60$\degr/day, yielding $\alpha_{\mathrm{DR}}=0.010\pm0.003$ shear parameter. In comparison, the solar pole-to-equator angular velocity difference is $\Delta\Omega_{\odot}=4.1\degr$/day, i.e. twice that of V815\,Her. In the average correlation map, the stronger correlations are observed at latitudes higher than $\approx$30\degr, this is probably due to the fact that at lower latitudes the spots have less contrast, and perhaps in connection with this, their lifetime is also shorter. According to the rotational law that best fits the average cross-correlation pattern, it is most likely that the equatorial regions of the G star are (most) synchronized to the orbit. We note that the errors of the fitted rotation law plotted in Fig.\,\ref{fig_ccf} 
were estimated based on the error bars of the locations with the best correlations at the different latitude bands. However, this is the error of the fitting process, which does not necessarily reflect the true errors resulting from the method.

\section{Discussions}\label{sect_disc}

\subsection{Multiplicity and dynamical evolution}
%,  circumbinary disc(s) and age}

The multiple nature of the V815\,Her system makes it very special, as the size of the wide (i.e. AB, in other words `outer' or `long') orbit of the 2+2 hierarchical quadruple is relatively small, only $a_{\rm AB}$$\approx$1.6AU (cf. Table\,\ref{Tab:twoplustwo}). The investigation of similar systems is also exciting because their secular gravitational evolution can have a significant impact on the future fate of the components. Long-term gravitational effects in such quadruple systems can have a serious impact on the evolution of the orbital eccentricities, since both binaries can act as a distant perturber on the other pair. Unfortunately, the evolution of such a 2+2 hierarchy is still an under-researched topic \citep[but see][]{2016MNRAS.461.3964V}. It is certain, however, that this interaction is not only dynamic (e.g. Kozai oscillations, also known as von Zeipel-Lidov-Kozai mechanism or Lidov-Kozai cycles; for a recent review see \citealt{2019MEEP....7....1I}), but due to the small size and large eccentricity of the wide orbit, as well as the compactness of the subsystems, tidal effects must also be taken into account \citep[cf.][]{2007ApJ...669.1298F}.

Due to the Kozai mechanism, in hierarchical triple systems a distant third body (as well as a distant close binary in a 2+2 system) may even decircularize the inner orbit \citep[e.g.][see also their references]{2021MNRAS.500.3481H}. The time scale of these Kozai cycles is in the order of $P^2_{\rm out}/P$, i.e. few times 10$^3$ years. As a matter of fact, this can be seen in action on HD\,74438 \citep[][]{2022NatAs...6..681M}, a 2+2 hierarchical system in a highly eccentric wide orbit, where the eccentricity of one of the inner orbits was probably pumped by gravitational effects. According to this, the circularized close binaries of V815\,Her will not necessarily remain so either. However, Kozai oscillations can be suppressed drastically when tidal friction enters the picture, shrinking and circularizing the inner orbit(s) on the tidal dissipation timescale \citep{2007ApJ...669.1298F}. For both inner binaries, this timescale should be on the order of 10$^5$ years \citep{2016ApJ...824...15V}, i.e. much shorter than $\sim$10$^7$-year age of the Aa component and so the assumed age of the four-star system (in case the components are coeval). For this reason, we do not expect a large-amplitude Kozai oscillation in the case of V815\,Her.

We note that in \citet{2022NatAs...6..681M} HD\,74438 is reported to be one of the shortest period 2+2 quadruple systems with $P\approx$5.7yr for the wide orbit, coincidentally almost exactly the same as the AB orbital period of V815\,Her. Indeed, along with a few compact and coplanar (or even doubly eclipsing) 2+2 four-star system \citep{2018ApJS..235....6T,2021MNRAS.503.3759B,2023MNRAS.522...90K,2023A&A...675A.113Z}, V815\,Her is among the shortest period 2+2 hierarhical systems. However, considering the size of the wide orbit, the V815\,Her system is much more compact compared to HD\,74438, which can be a significant difference in terms of the strength of secondary (i.e. tidal) effects; see also Sect.~\ref{ltbrightness} below in this paper.

\subsection{The V815\,Her B close binary subsystem}

\begin{figure}[thb]
    \centering
    \includegraphics[width=1\columnwidth]{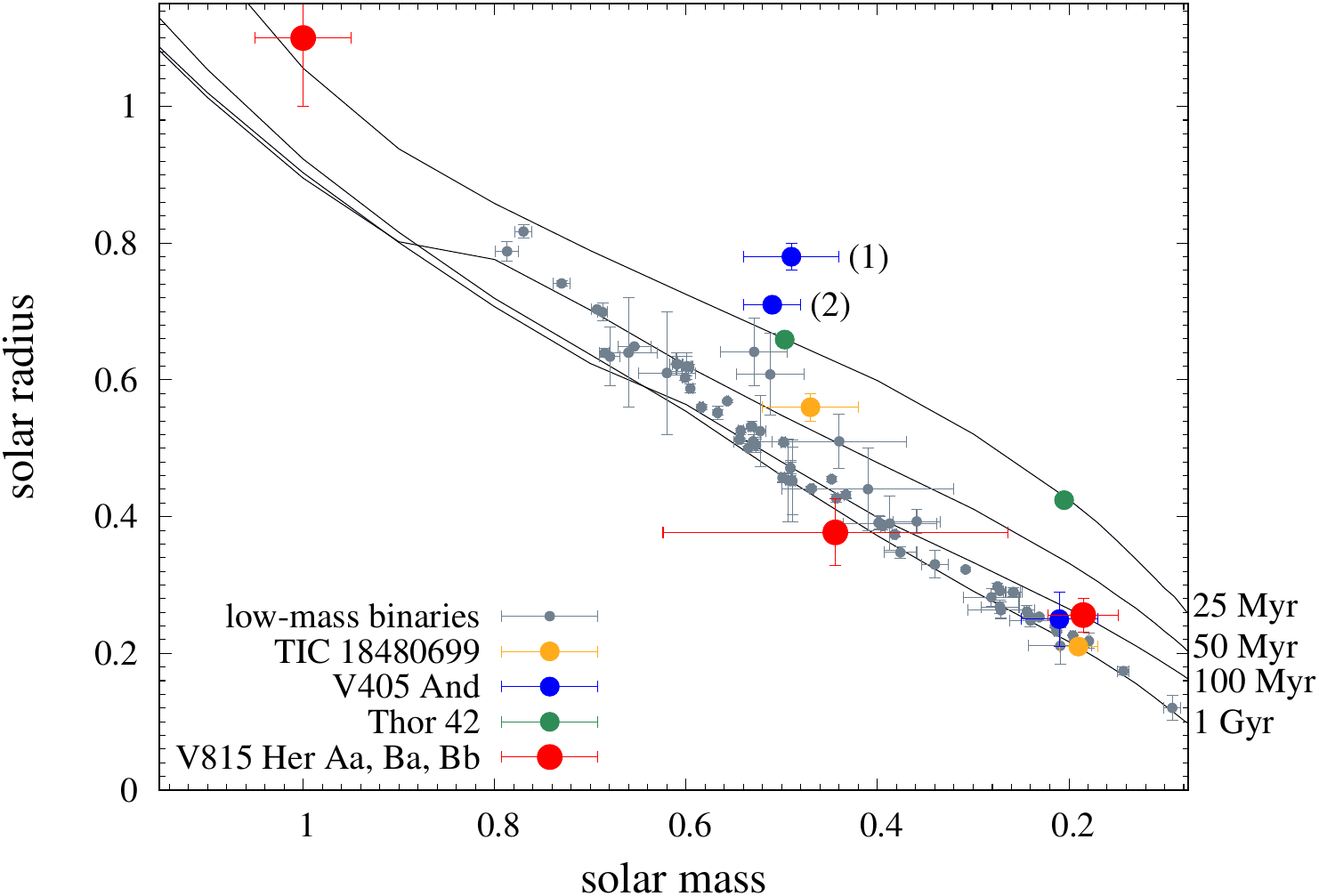}
     \caption{Mass-radius relation of low mass eclipsing binary stars with well-determined parameters from the compilation of Table B.2. from \cite{2019A&A...625A..68S} plotted with grey dots. Coloured symbols (see legend) show the position of TIC\,18480699 = 2MASS J04463285+1901432: \cite{2006AJ....131..555H}, V405\,And (1): \cite{2009A&A...504.1021V} and (2): \cite{2011AJ....142..106R}, Thor\,42 = CRTS J055255.7-004426: \cite{2020MNRAS.491.4902M}, and V815\,Her Aa, Ba and Bb (present paper). Non-magnetic, solar metallicity isochrones are plotted from \cite{2015A&A...577A..42B}. }
   \label{fig_mass-radius}
\end{figure} 

The components of the low-mass eclipsing binary V815\,Her\,B of the quadruple system are plotted on the mass-radius diagram on Fig.\ref{fig_mass-radius} together with other eclipsing binary components from the literature. Note that the oversize of the primary of V405\,And were determined by two different methods using independent data (cf. caption of Fig.\ref{fig_mass-radius}). In the figure the non-eclipsing component V815\,Her~Aa is also given for reference to the age determination. Together with V815\,Her~Ba+Bb three other systems are plotted in color (see the legends in Fig.\ref{fig_mass-radius}). While the masses of both the primaries and secondaries of these four systems are very similar to each other, laying on the two sides of the full-convection limit of $\approx$0.35\,$M_\odot$ \citep{1997A&A...327.1039C}, the primaries show about twofold difference in radius. The fully convective secondaries are quite similar to each other in size, except for  Thor-42 \citep{2020MNRAS.491.4902M}. This detailed study of Thor-42 was not able to simultaneously model the mass, radius, temperature and luminosity of the components.

The relative oversize of low-mass magnetically active stars compared to inactive stars is a known phenomenon \citep[cf.][]{2001ApJ...559..353M}. \citet{2011ApJ...728...48K} suggested rotation as a third parameter of the mass-radius relation. Fast rotation, maintained by the binarity influence the magnetic field and through this, the strength of the magnetic activity of the late-type dwarfs. \citet{2021ApJ...907...27M} modelled Thor-42 using both magnetoconvection and starspots and through this they were able to reconcile all stellar parameters of the system, but at a different age that \citet{2020MNRAS.491.4902M} found. This result underlines the importance of magnetic field in estimating fundamental parameters of active stars. We may say that the apparent discrepancy between the ages of the components of the V815\,Her quadruple system shown in Fig.\,\ref{fig_mass-radius} could be largely due to the magnetic activity of the components which, in the same time, could also be different star-by-star.

\subsection{Brightness change along the wide orbit}\label{ltbrightness}

\begin{figure}[thb]
	\centering
	\includegraphics[width=\linewidth]{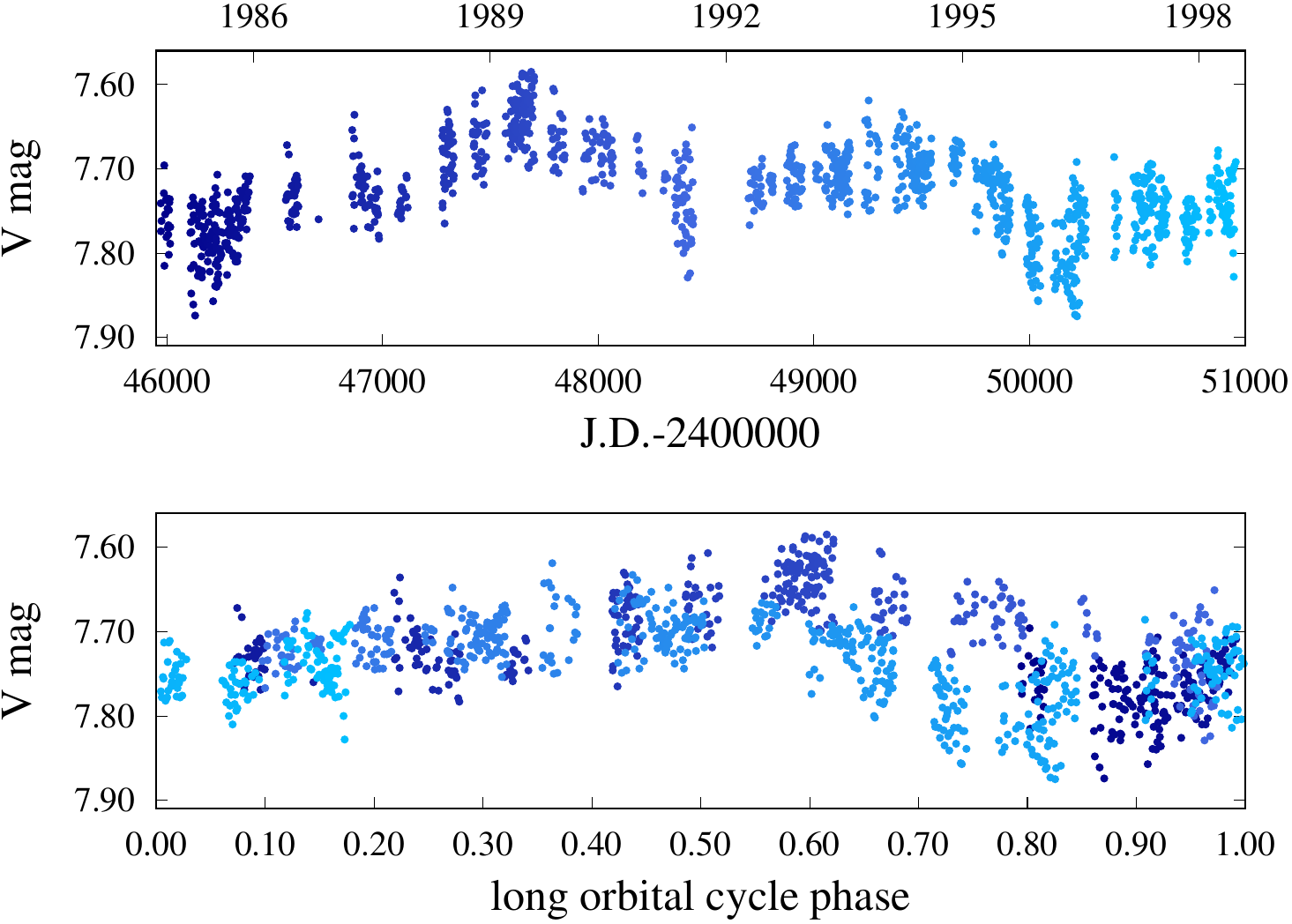}
	\caption{{\it Top}: $V$ light curve of V815\,Her from \cite{2000A&A...362..223J} between 1984-98. {\it Bottom}: Folded light curve using the 5.73\,yr (=2092.2\,d) period of the wide orbit. As zero phase the time of the periastron passage was set. The continuous color gradient helps to distinguish successive phases.}
	\label{fig_long_phot}
\end{figure}

In the top panel of Fig.~\ref{fig_long_phot} we plot the $V$ observations of V815\,Her between 1984-98 from \cite{2000A&A...362..223J}. In these data we found a long-term brightness modulation, which showed an exciting coincidence with the approximate period of  5.73\,yr of the outer (AB) orbit. Therefore, in the bottom panel of Fig.~\ref{fig_long_phot} we replot the $V$ data along with the folded light curve phased with the period of the wide orbit taken from \cite{2005AJ....129.1001F}. It is interesting that near zero phase (which is the periastron of the wide orbit), the brightness decreases, while away from it the brightness gradually increases.
The question arises as to whether this is just a coincidence or really a change related to the wide orbit, and if it is the latter, then what is causing it.

On the other hand, the time-frequency analysis of the photometric data on a much longer term plotted in Fig.~\ref{fig_tifran} does not show a strong cycle corresponding \emph{exactly} to the period of the outer orbit, instead we see a strong signal of a 6.5-year cycle with gradually increasing period, and strangely, a signal that appears almost exactly at twice the orbital period. Nevertheless, it cannot be ruled out that the slowly changing, roughly 6.5-year photometric period is the changing activity cycle of the Aa component, which is perhaps triggered along the wide orbit.
According to \citet{2005AJ....129.1001F} the eccentricity of the wide orbit is quite large ($\approx$0.77), therefore gravitational effects exerted on the G star by V815\,Her~B change significantly along the outer orbit. Tidal interactions have indeed a potential impact on the internal structure and the turbulent/mixing processes \citep[e.g.][]{2007A&A...461.1057T,2021A&A...653A.127K}, ultimately, therefore, for the operation of the magnetic dynamo. This may also result in a change of the strength of spot activity of the G star, which ultimately causes the long-term change in brightness. We note here that such a possible binary-induced, i.e. tidally triggered activity was reported in \citet{2011A&A...535A..98S}.

An alternative explanation could be that the brightness change along the wide orbit is somehow related to the dust-gas material component of any kind (circumstellar or circumbinary) present in the quadruple star system. Assuming that there is a circumbinary disk around the Ba+Bb binary (consistently with the infrared surplus indicated by Fig.~\ref{fig_sed}), such a disc if tilted to the plane of the wide orbit, can reflect the radiation of the G star differently, depending on their mutual position, i.e. the orbital phase. In our case, the reflected light can only be estimated cautiously in the case of a disc of unknown composition, size and geometry. However, if we estimate the contribution of the reflection from the disc to $\lesssim$1\% (which is a rough upper estimate, cf. \citealt{2021AJ....162...98B}), then we can only explain a few percent of the long-term change in the mean brightness  with reflection (say, $\lesssim$0.02 mag). Therefore, we conclude that changing spot activity results in the modulated brightness along the outer orbit, even though light reflected from any disc contributes little, if at all.

\subsection{Surface evolution of V815\,Her~Aa}\label{surfevol}
\subsubsection{Spot decay and turbulent-driven magnetic diffusion}

According to observations, the lifetime of sunspots depends on their area, based on which linear and nonlinear spot decay theories were born \citep[e.g.,][etc.]{1963BAICz..14...91B,1975SoPh...42..107K,1997SoPh..176..249P,2008SoPh..250..269H,2015ApJ...800..130L,2021ApJ...908..133M,2021A&A...653A..50F}. The underlying theories assume different mechanisms that also depend on the morphology of the sunspots/sunspot groups.
When starspot lifetimes and starspot decay are discussed based on the solar analogy, the question arises as to whether it is a matter of scaling or other considerations are necessary \citep[see e.g.,][and their references]{2019ApJ...871..187N}. The average spot area on V815\,Her is about a factor of $\sim$100 larger compared to long-lived big sunspots (see Fig.\,\ref{fig:DIs}). We note however, that the spatial resolution of our Doppler images does not reach the size range of sunspots, but in principle they are able to resolve surface structures corresponding to the largest sunspot groups. In this context, it is perhaps more reasonable to compare the starspots of our Doppler images with groups of sunspots instead of large individual sunspots.

As for the spot lifetime, based on our time-series Doppler images, the spotted surface of V815\,Her is completely rearranged in a few steps, typically within $\sim$25-50 days (cf. Fig.\,\ref{fig:DIs} and Table\,\ref{disets}). This is therefore the approximate upper limit of spot lifetime. At the same time, the lifespan of individual spots can vary widely. The shortest-lived spots can only be observed in one or at most two consecutive images, so their lifetime is $\sim$10-15 days.

We emphasize here, that it is difficult to discuss on the starspot decay in the context of sunspot theory without knowledge of the true morphology of starspots, i.e., do we see monolithic spots on Doppler images or clusters of smaller spots? It is even possible that one does not exclude the other, nevertheless, usually the limited spatial resolution of observations does not allow us to make a clear distinction \citep[but see][]{2018A&A...620A.162J}.
Moreover, spot decay may strongly be influenced by surface and subsurface flows. 

\citet[][]{2007A&A...464.1049I} modeled the temporal decay of the surface magnetic flux when assuming different spot configurations while large-scale flows (differential rotation, meridional circulation) are switched on and off. Their results indicate that spot conglomerates decay faster compared to large monolithic spots, and flows also play an important role in speeding up the decay. We note that the $\Delta\Omega\approx2^{\circ}$/d surface shear on V815\,Her is much weaker than on the Sun (see Sect.\,\ref{sect_dr}). If we take into account the $t_{\rm l}$ lap time, i.e. the time taken for the equator to lap the poles, then in the case of V815\,Her we get $360^{\circ}/\Delta\Omega$=180\,d (i.e. about twice the solar lap time of $\sim$88\,d). This is much longer than the observed spot lifetimes on V815\,Her, therefore, differential rotation is likely to play a minor role (if at all) in spot decay. However, even with flows, the simulations in \citet[][]{2007A&A...464.1049I} generally predict longer spot lifetimes than what we see on V815\,Her.

\citet[][]{2014ApJ...795...79B} argued that, assuming the same physical background and starting from the Gnevyshev$-$Waldmeier (hereafter GW) rule, the decay of sunspots and starspots can be scaled by the value of magnetic diffusivity through the turbulent scale length (which is connected to the supergranulation on the Sun). They suggested an anomalous turbulent-driven magnetic diffusion, which can account for the decay of either sunspots or starspots by scaling the characteristic supergranule/diffusion lengths.

By following the empirical GW rule, i.e. assuming a linear relationship between the maximum
$A_{\rm s}$ spot area and its $t_{\rm s}$ lifetime, based on our time-series Doppler images, taking $t_{\rm s}$$\sim$20\,d and $R_{\rm s}$$\sim$10$^{\circ}$, equivalent of 1.3\,10$^5$\,km, as average spot lifetime and corresponding characteristic maximum size, respectively, we can estimate a characteristic value of the spot decay rate $\Delta A_{\rm s}/\Delta t_{\rm s}$ of about $-$4.0\,10$^4$\,km$^{2}$\,s$^{-1}$. In Appendix\,\ref{app_decay}, we also present a more detailed study to measure the spot decay rate. Although the method is not mature enough yet, it is certainly more quantitative than the above estimate. We note that this detailed study yields even faster decay of %$dA_{\rm s}/dt_{\rm s}$$\approx$
about $-$7.0\,10$^4$\,km$^{2}$\,s$^{-1}$.

Now, using $t_{\rm s}$=20\,d and $R_{\rm s}$=1.3\,10$^5$\,km (see above), we get from $t_{\rm s}\sim R_{\rm s}^2/\eta_{\rm T}$ \citep[][their Eq.\,3]{2014ApJ...795...79B} a value of $\eta_{\rm T}\approx10^4$\,km$^{2}$\,s$^{-1}$ for the turbulent diffusivity.
According to \citet[][their Eq.\,19]{2008ApJ...683.1153C}, from 
\begin{equation}
    \eta_{\rm T}= \bigg( \frac{l}{260\,{\rm km}}\bigg)^{5/4}
\end{equation}
this value would yield a turbulent scale length of $l$=4.2\,10$^5$\,km or $\approx$0.55\,$R_*$, i.e., in the order of the stellar radius. Indeed, if we assumed an even faster spot decay from Appendix\,\ref{app_decay}, we would get about twice this amount. This large turbulent scale length is probably an overestimate, which may result from an overestimation of the maximum spot size. However, if, for example, several smaller spots were assumed instead of a monolithic spot, the turbulent scale length could also be reduced to a more plausible value \citep[cf.][]{2014ApJ...795...79B}. Finally, we should also note that our measurement of the spot decay rate can be significantly distorted if, in addition to the spot shrinking, a rapid flux emergence occurs simultaneously in the same region. This would limit the validity of the interpretation that the observed spot decay rate and turbulent diffusion are directly related.

\subsubsection{Latitudinal and longitudinal distribution of spots}
In order to make it easier to assess the distribution of spots on the time-series Doppler maps, for each temperature reconstruction we calculated the sums of the $f$ filling factor values along the $\lambda$ longitude and $\beta$ latitude (we have 72 longitude and 36 latitude pixels on a 5$^{\circ}\times$5$^{\circ}$ spherical grid). When calculating $f_\lambda$ and $f_\beta$, we took into account the geometric distortion on the spherical surface according to the following formula:
\begin{equation}
f_{\lambda,\beta} = 1 - \frac{\sum_{i=1}^{n_{\lambda,\beta}} \cos(\beta_i) T_i^4 }{T_{\rm eff}^4 \sum_{i=1}^{n_{\lambda,\beta}} \cos(\beta_i)},
\end{equation}
where $T_i$ and $\beta_i$ are the temperature and the latitude of the $i$th pixel, while the limit of the summation is the image pixel number over either the longitude ($n_\lambda$=72), or the latitude ($n_\beta$=33, since we do not see the star below $\beta$=$-$75$^{\circ}$). The resulting spot filling factors for the nineteen Doppler images are shown in Appendix\,\ref{App:FF}. Based on the latitudinal distribution of the spots, we see that especially in the case of low spot coverage, the spots tend to appear at mid-latitudes. In these cases (e.g. S01-S02-S03-S04, S07, S12-S13, S18) a somewhat equatorially symmetric distribution is seen. On the other hand, at other times spots tend to appear at high latitudes, closer to the visible pole. We note however, that, since we cannot see the other pole and its immediate surroundings, we cannot say whether the appearance of the polar spots shows any symmetry to the equator.

\citet{2000A&A...355.1087G} studied the dynamics of magnetic flux tubes in young stars by numerical simulations and found that at a rotation rate of V815\,Her~Aa the spot emergence latitude is most likely between 25$^{\circ}$-50$^{\circ}$ (see their ZAMS star model of 1 solar mass with $\sim$10$\Omega/\Omega_{\odot}$ rotation). However, similarly to V815\,Her~Aa, the coexistence of low-to-mid latitude spots and high latitude/polar spots on G-K type main sequence stars has already been supported by numerous Doppler imaging studies \citep[e.g.,][and references therein]{2009A&ARv..17..251S}. Moreover, EK\,Dra, the rapidly rotating young Sun, showed a very similar starspot distribution in the recent Doppler imaging study by \citet{2021MNRAS.502.3343S}, in which the polar and mid-latitude (>20$^{\circ}$) spots are also consistent with flux emergence and transport simulations. According to \citet{2000A&A...355.1087G}, after flux emergence at mid latitudes a poleward flux transport by meridional circulation and/or magnetic stresses could explain the formation of polar spots \citep[for the formation of polar spots on rapid rotators see also][]{2001ApJ...551.1099S,2018A&A...620A.177I}. Maybe a weak sign of this can be seen in some of our subsequent Doppler image pairs (S04-S05, S13-S14, S16-S17), where the formation of the polar spot is preceded by a nearby lower latitude spot, although the available time resolution is not  enough for a strict conclusion. Here, we note that the internal structure of V815\,Her~Aa is indeed affected by its close companion Ab, and this may be important also from the point of view of magnetic flux tube dynamics \citep[e.g.,][]{1996A&A...314..503S,2003A&A...405..291H,2003A&A...405..303H}.

The longitudinal spot distribution, in contrast, is more random, we see no definite indication of the existence of active longitudes. Although the changes can still mostly be followed in successive images, however, no systematic pattern can be seen in the evolution of the spotted surface over the entire period.

\subsubsection{Surface differential rotation}

From our cross-correlation study, we infer that the active G star has a weak solar-type differential rotation. Actually, this finding is in a fair agreement with the recent results obtained for two similar close binary systems V471\,Tau \citep[][]{2021A&A...650A.158K} and EI\,Eri \citep[][]{2023A&A...674A.143K} in which the active component preserves its rapid rotation as a result of synchronization, but at the same time the degree of surface shear due to differential rotation is presumably confined by tidal interaction. These observations support the idea that differential rotation is suppressed in low-mass stars tidally locked in close binary systems \citep[see also][and their references]{2007AN....328.1030C}.
From a theoretical point of view, this kind of behavior is not unexpected, since a marked initial differential rotation structure would converge to rigid body-like rotation due to tidal shear energy dissipation \citep{2021A&A...653A.127K}. Our Fig.~\ref{fig_ccf} also suggests that the equatorial region of V815\,Her~Aa is synchronized the most with the orbital period of V815\,Her~A. We note that this is consistent with tidal perturbations being expected to be strongest around equatorial latitudes.

\section{Summary}\label{sect_sum}

 The V815\,Her system undoubtedly evolves dynamically and most likely magnetically, while the two subsystems, V815\,Her~A and B, interact gravitationally.
Our main conclusions are summarized as follows:
   \begin{itemize}
      \item The wide component in V815\,Her previously apostrophized as a `third body' is actually an eclipsing close binary subsystem of two M dwarfs, i.e., V815\,Her is a 2+2 hierarchical quadruple system.
      \item V815\,Her is presumably a young system, supported by the infrared surplus indicated by the spectral energy distribution in Fig.~\ref{fig_sed}. The young age is also supported by the lithium abundance that has not yet been depleted. In accordance with this, the evolutionary state of V815\,Her~Aa is a zero-age main sequence star, with an age of $\sim$30M years.
      \item A rapid spot evolution can be observed on the surface of the active G component V815\,Her~Aa, the spot distribution can change completely on a time range of few times ten days.
      \item We measured a weak solar-type surface differential rotation on the G star, which is probably suppressed by the tidal force of the close companion.%in agreement with \citet{2017AN....338..903K}
      \item Based on the long-term photometric data, we found a slowly increasing cycle length of about 6.5 years on average, which we interpret as a spot cycle. The length of the spot cycle is not necessarily closely related to the 5.73-year period of the wide orbit, but we would not rule out the possibility of this at this point. We suspect that the reason for the slow change may be the modulation due to the eccentricity of the wide orbit.
   \end{itemize}

The case of V815\,Her is a good example of the particular importance of tidal effects for the dynamo mechanism. For this reason, our quartet is certainly interesting for further study of how gravitational evolution of the system affects activity.

\begin{acknowledgements}
We thank our referee, Dr. Emre I\c{s}{\i}k, for his very useful and constructive feedback, in which he made several important comments that contributed to the further development of the paper.
Authors gratefully thank Dr. Attila Moór from Konkoly Observatory for the discussion about protoplanetary and debris disks and the light reflected from them. This work was supported by the Hungarian National Research, Development and Innovation Office grants OTKA K131508, KKP-14398 and by the Lend\"ulet Program of the Hungarian Academy of Sciences, project No. LP2018-7/2019. Authors acknowledge the financial support of the Austrian-Hungarian Action Foundation (98\"ou5, 101\"ou13, 112\"ou1). LK acknowldeges the support of the Hungarian National Research, Development and Innovation Office grant PD-134784. 
LK and KV are Bolyai J\'anos research Fellows.
KV is supported by the Bolyai+ grant \'UNKP-22-5-ELTE-1093, BS is supported by the \'UNKP-22-3 New National Excellence Program of the Ministry for Culture and
Innovation from the source of the National Research, Development and Innovation Fund. 
STELLA was made possible by funding through the State of Brandenburg (MWFK) and the German Federal Ministry of Education and Research (BMBF). The facility is a collaboration of the AIP in Brandenburg with the IAC in Tenerife. 
This publication makes use of VOSA, developed under the Spanish Virtual Observatory (https://svo.cab.inta-csic.es) project funded by MCIN/AEI/10.13039/501100011033/ through grant PID2020-112949GB-I00.
VOSA has been partially updated by using funding from the European Union's Horizon 2020 Research and Innovation Programme, under Grant Agreement no 776403 (EXOPLANETS-A).
This work has made use of data from the European Space Agency (ESA) mission
{\it Gaia} (\url{https://www.cosmos.esa.int/gaia}), processed by the {\it Gaia}
Data Processing and Analysis Consortium (DPAC,
\url{https://www.cosmos.esa.int/web/gaia/dpac/consortium}). Funding for the DPAC
has been provided by national institutions, in particular the institutions
participating in the {\it Gaia} Multilateral Agreement. This publication makes use of the Digital Access to a Sky Century at Harvard (DASCH), a project to digitize the photographic plate collection held at Harvard College Observatory. DASCH has been supported by NSF grants AST-0407380, AST-0909073, and AST-1313370, for which we are grateful.
\end{acknowledgements}

\bibliography{v815her}

\begin{appendix}
%\twocolumn
\onecolumn

\section{Times of eclipsing minima from \emph{TESS} light curves}

\begin{table*}[h]
\caption{Times of minima of V815\,Her~B from Sector 26 (\emph{left}), Sector 40 (\emph{middle}), and Sector 53 (\emph{right}) light curves.}
 \label{Tab:V815_Her_B_ToM}
\begin{tabular}{@{}crc||crc||crc}
\hline
BJD & Cycle  & std. dev. & BJD & Cycle  & std. dev. & BJD & Cycle  & std. dev. \\ 
$-2\,400\,000$ & no. &   [d] & $-2\,400\,000$ & no. &   [d] & $-2\,400\,000$ & no. &   [d] \\
%\hline\smallskip
\hline
59010.396759 & $-$1.0 & 0.005419 & 59390.969044 & 730.0 & 0.000579 & 59750.207340 & 1420.0 & 0.001752 \\ 
59010.655434 & $-$0.5 & 0.000611 & 59391.228904 & 730.5 & 0.001037 & 59750.467594 & 1420.5 & 0.001804 \\ 
59010.917386 &	0.0  &	0.000308 & 59391.490881 & 731.0 & 0.000731 & 59750.729019 & 1421.0 & 0.000407 \\ 
59011.177977 &	0.5  &	0.000852 & 59391.750806 & 731.5 & 0.000616 & 59750.986819 & 1421.5 & 0.000808 \\ 
59011.436903 &	1.0  &	0.000463 & 59392.010439 & 732.0 & 0.000859 & 59751.249397 & 1422.0 & 0.000430 \\ 
59011.703819 &	1.5  &	0.000760 & 59392.270268 & 732.5 & 0.000889 & 59751.509992 & 1422.5 & 0.002570 \\ 
59011.958773 &	2.0  &	0.000525 & 59392.530909 & 733.0 & 0.000864 & 59751.770279 & 1423.0 & 0.000575 \\ 
59012.216689 &	2.5  &	0.000328 & 59392.792132 & 733.5 & 0.000830 & 59752.028581 & 1423.5 & 0.001020 \\ 
59012.478749 &	3.0  &	0.000290 & 59393.052663 & 734.0 & 0.000633 & 59752.287769 & 1424.0 & 0.007730 \\ 
59012.730444 &	3.5  &	0.002701 & 59393.311377 & 734.5 & 0.001224 & 59752.550820 & 1424.5 & 0.002143 \\ 
59012.997801 &	4.0  &	0.000545 & 59393.571995 & 735.0 & 0.000457 & 59752.810861 & 1425.0 & 0.000600 \\ 
59013.256687 &	4.5  &	0.000690 & 59393.835471 & 735.5 & 0.002979 & 59753.071759 & 1425.5 & 0.001207 \\ 
59013.523129 &	5.0  &	0.000308 & 59394.093038 & 736.0 & 0.000608 & 59753.330231 & 1426.0 & 0.002246 \\ 
59013.781077 &	5.5  &	0.000576 & 59394.353760 & 736.5 & 0.000651 & 59753.593639 & 1426.5 & 0.001204 \\ 
59014.041462 &	6.0  &	0.000656 & 59394.611848 & 737.0 & 0.000606 & 59753.853110 & 1427.0 & 0.000520 \\ 
59014.298518 &	6.5  &	0.000725 & 59394.874266 & 737.5 & 0.001242 & 59754.112817 & 1427.5 & 0.002296 \\ 
59014.558817 &	7.0  &	0.000488 & 59395.134036 & 738.0 & 0.000886 & 59754.372525 & 1428.0 & 0.000551 \\ 
59014.822105 &	7.5  &	0.000913 & 59395.395363 & 738.5 & 0.000718 & 59754.633838 & 1428.5 & 0.002193 \\ 
59015.089820 &	8.0  &	0.000879 & 59395.655174 & 739.0 & 0.001446 & 59754.893547 & 1429.0 & 0.001495 \\ 
59015.345284 &	8.5  &	0.000737 & 59395.914606 & 739.5 & 0.000931 & 59755.155476 & 1429.5 & 0.003858 \\ 
59015.602487 &	9.0  &	0.000425 & 59396.175866 & 740.0 & 0.000544 & 59755.414605 & 1430.0 & 0.002845 \\ 
59015.866312 &	9.5  &	0.000881 & 59396.435910 & 740.5 & 0.001447 & 59755.675277 & 1430.5 & 0.001056 \\ 
59016.127331 &	10.0 &	0.000456 & 59396.696363 & 741.0 & 0.000695 & 59755.934886 & 1431.0 & 0.000762 \\ 
59016.381487 &	10.5  &	0.000440 & 59396.956394 & 741.5 & 0.007025 & 59763.483106 & 1445.5 & 0.002752 \\ 
59016.643627 &	11.0  &	0.000205 & 59397.216360 & 742.0 & 0.000450 & 59763.746312 & 1446.0 & 0.000439 \\ 
59016.907289 &	11.5  &	0.000494 & 59397.475688 & 742.5 & 0.029455 & 59764.004660 & 1446.5 & 0.001141 \\ 
59017.165940 &	12.0  &	0.000268 & 59397.736504 & 743.0 & 0.000448 & 59764.264120 & 1447.0 & 0.000456 \\ 
59017.421941 &	12.5  &	0.000440 & 59397.997957 & 743.5 & 0.018673 & 59764.523280 & 1447.5 & 0.001219 \\ 
59017.685807 &	13.0  &	0.000299 & 59398.258231 & 744.0 & 0.000392 & 59764.785534 & 1448.0 & 0.000492 \\ 
59017.949074 &	13.5  &	0.000974 & 59398.519739 & 744.5 & 0.000633 & 59765.045838 & 1448.5 & 0.001699 \\ 
59018.206886 &	14.0  &	0.000320 & 59398.778495 & 745.0 & 0.000447 & 59765.306575 & 1449.0 & 0.000420 \\ 
59018.467208 &	14.5  &	0.017412 & 59399.038859 & 745.5 & 0.002356 & 59765.566480 & 1449.5 & 0.002744 \\ 
59018.724460 &  15.0  &	0.000314 & 59399.299451 & 746.0 & 0.000442 & 59765.827445 & 1450.0 & 0.001626 \\ 
59018.984292 &	15.5  &	0.021397 & 59399.559764 & 746.5 & 0.001107 & 59766.087395 & 1450.5 & 0.001636 \\ 
59019.246745 &	16.0  &	0.000258 & 59399.820430 & 747.0 & 0.000690 & 59766.347915 & 1451.0 & 0.000397 \\ 
59019.507475 &	16.5  &	0.000424 & 59400.080220 & 747.5 & 0.000908 & 59766.608793 & 1451.5 & 0.001091 \\ 
59019.765580 &	17.0  &	0.000468 & 59400.340231 & 748.0 & 0.000553 & 59766.868377 & 1452.0 & 0.000841 \\ 
59020.025038 &	17.5  &	0.000752 & 59400.599451 & 748.5 & 0.002930 & 59767.128870 & 1452.5 & 0.004383 \\ 
	           &	  &           &	 59400.861610 & 749.0 & 0.000455 & 59767.388947 & 1453.0 & 0.000701 \\ 
	           &	  &           &	 59401.121024 & 749.5 & 0.006293 & 59767.648351 & 1453.5 & 0.008222 \\ 
	           &	  &           &	 59401.382010 & 750.0 & 0.000480 & 59767.907794 & 1454.0 & 0.000645 \\ 
	           &	  &           &	 59401.641153 & 750.5 & 0.001030 & 59768.170459 & 1454.5 & 0.005906 \\ 
	           &	  &           &	 59401.902890 & 751.0 & 0.000336 & 59768.429865 & 1455.0 & 0.000545 \\ 
	           &	  &           &	 59402.159590 & 751.5 & 0.008119 &                &       &           \\
	           &	  &           &	 59402.422958 & 752.0 & 0.000671 &                &       &           \\
	           &	  &           &	 59402.683010 & 752.5 & 0.009989 &                &       &           \\
	           &	  &           &	 59402.942458 & 753.0 & 0.001320 &                &       &           \\
	           &	  &           &	 59403.202713 & 753.5 & 0.000738 &                &       &           \\
	           &	  &           &	 59403.464449 & 754.0 & 0.000372 &                &       &           \\
	           &	  &           &	 59403.724330 & 754.5 & 0.000899 &                &       &           \\
	           &	  &           &	 59403.984932 & 755.0 & 0.000486 &                &       &           \\
	           &	  &           &	 59404.245716 & 755.5 & 0.001709 &                &       &           \\
%\hline
\end{tabular}
\end{table*}

\twocolumn

\section{Log of spectroscopic data}

\vspace{0.6cm}
%---------------------------- TA1
\begin{center}
\tablehead
{\hline\hline\noalign{\smallskip}
\setcounter{footnote}{0}HJD\footnote{a}&  Date &Phase\footnote{b}\setcounter{footnote}{-1} & S/N & Subset\\
\hline\noalign{\smallskip}}
\topcaption{Observing log of STELLA-SES spectra of V815\,Her from 2018 used for the individual Doppler reconstructions shown in Fig.~\ref{fig:DIs}.}\label{tab_a_1}
\begin{supertabular}{c c c r c}
\tabletail{
\hline\hline\noalign{\smallskip}
\multicolumn{5}{l}{\setcounter{footnote}{0}\footnote{a}{2,450,000+}}\\
\multicolumn{5}{l}{\footnote{b}{Phases computed using Eq.~\ref{phases}.}}\\}
\tablelasttail{
\hline\hline\noalign{\smallskip}
\multicolumn{5}{l}{\setcounter{footnote}{0}\footnote{a}{2,450,000+.}}\\
\multicolumn{5}{l}{\footnote{b}{Phases computed using Eq.~\ref{phases}.}}\\}
8198.641 & 21.03.2018 & 0.011 &          197 & S01  \\
8199.672 & 22.03.2018 & 0.581 &          225 & S01  \\
8199.718 & 22.03.2018 & 0.607 &          221 & S01  \\
8200.641 & 23.03.2018 & 0.116 &          213 & S01  \\
8200.679 & 23.03.2018 & 0.138 &          228 & S01  \\
8201.646 & 24.03.2018 & 0.671 &          233 & S01  \\%SME
8201.720 & 24.03.2018 & 0.712 &          211 & S01  \\
8203.679 & 26.03.2018 & 0.795 &          170 & S01  \\
8203.717 & 26.03.2018 & 0.817 &          171 & S01  \\
8204.636 & 27.03.2018 & 0.324 &          207 & S01  \\
8204.673 & 27.03.2018 & 0.344 &          209 & S01  \\
8204.711 & 27.03.2018 & 0.365 &          201 & S01  \\
8208.630 & 31.03.2018 & 0.530 &          213 & S02  \\
8208.676 & 31.03.2018 & 0.556 &          200 & S02  \\
8209.691 & 01.04.2018 & 0.117 &          178 & S02  \\
8211.617 & 03.04.2018 & 0.181 &          214 & S02  \\
8211.662 & 03.04.2018 & 0.205 &          218 & S02  \\
8211.700 & 03.04.2018 & 0.227 &          217 & S02  \\
8212.598 & 04.04.2018 & 0.723 &          211 & S02  \\
8212.681 & 04.04.2018 & 0.768 &          196 & S02  \\
8212.718 & 04.04.2018 & 0.790 &          179 & S02  \\
8213.620 & 05.04.2018 & 0.289 &          224 & S02  \\
8213.666 & 05.04.2018 & 0.312 &          218 & S02  \\
8213.703 & 05.04.2018 & 0.334 &          223 & S02  \\
8214.648 & 06.04.2018 & 0.856 &          223 & S02  \\
8216.687 & 08.04.2018 & 0.983 &          173 & S02  \\
8216.724 & 08.04.2018 & 0.002 &          189 & S02  \\
8219.577 & 11.04.2018 & 0.580 &          204 & S03  \\
8219.615 & 11.04.2018 & 0.599 &          216 & S03  \\
8219.652 & 11.04.2018 & 0.621 &          238 & S03  \\
8219.697 & 11.04.2018 & 0.645 &          225 & S03  \\
8222.655 & 14.04.2018 & 0.281 &          183 & S03  \\
8222.700 & 14.04.2018 & 0.304 &          207 & S03  \\
8223.624 & 15.04.2018 & 0.816 &          233 & S03  \\
8223.670 & 15.04.2018 & 0.842 &          225 & S03  \\
8224.606 & 16.04.2018 & 0.358 &          225 & S03  \\
8224.669 & 16.04.2018 & 0.392 &          227 & S03  \\
8225.586 & 17.04.2018 & 0.899 &          216 & S03  \\
8225.660 & 17.04.2018 & 0.940 &          230 & S03  \\
8226.584 & 18.04.2018 & 0.452 &          201 & S03  \\
8226.621 & 18.04.2018 & 0.471 &          225 & S03  \\
8227.602 & 19.04.2018 & 0.013 &          186 & S03  \\
8227.640 & 19.04.2018 & 0.035 &          177 & S03  \\
8227.677 & 19.04.2018 & 0.054 &          183 & S03  \\
8234.538 & 26.04.2018 & 0.846 &          199 & S04  \\
8234.586 & 26.04.2018 & 0.872 &          216 & S04  \\
8234.623 & 26.04.2018 & 0.892 &          212 & S04  \\
8234.668 & 26.04.2018 & 0.917 &          210 & S04  \\
8234.706 & 26.04.2018 & 0.939 &          190 & S04  \\
8235.575 & 27.04.2018 & 0.418 &          221 & S04  \\
8235.613 & 27.04.2018 & 0.440 &          218 & S04  \\
8235.650 & 27.04.2018 & 0.459 &          211 & S04  \\
8235.695 & 27.04.2018 & 0.485 &          212 & S04  \\
8236.567 & 28.04.2018 & 0.967 &          218 & S04  \\
8236.605 & 28.04.2018 & 0.988 &          216 & S04  \\
8236.642 & 28.04.2018 & 0.008 &          218 & S04  \\
8236.687 & 28.04.2018 & 0.034 &          223 & S04  \\
8237.571 & 29.04.2018 & 0.521 &          230 & S04  \\
8237.608 & 29.04.2018 & 0.543 &          227 & S04  \\
8237.692 & 29.04.2018 & 0.588 &          200 & S04  \\
8238.578 & 30.04.2018 & 0.078 &          196 & S04  \\
8238.615 & 30.04.2018 & 0.098 &          201 & S04  \\
8239.532 & 01.05.2018 & 0.605 &          209 & S04  \\
8239.578 & 01.05.2018 & 0.631 &          215 & S04  \\
8239.615 & 01.05.2018 & 0.650 &          221 & S04  \\
8239.653 & 01.05.2018 & 0.672 &          239 & S04  \\
8241.601 & 03.05.2018 & 0.749 &          208 & S04  \\
8241.639 & 03.05.2018 & 0.770 &          211 & S04  \\
8241.676 & 03.05.2018 & 0.790 &          223 & S04  \\
8242.579 & 04.05.2018 & 0.288 &          209 & S04  \\
8242.616 & 04.05.2018 & 0.310 &          199 & S04  \\
8242.654 & 04.05.2018 & 0.329 &          210 & S04  \\
8242.699 & 04.05.2018 & 0.355 &          168 & S04  \\
8243.535 & 05.05.2018 & 0.817 &          194 & S05  \\
8243.631 & 05.05.2018 & 0.871 &          195 & S05  \\
8245.508 & 06.05.2018 & 0.907 &          189 & S05  \\
8245.546 & 07.05.2018 & 0.929 &          213 & S05  \\
8245.583 & 07.05.2018 & 0.948 &          209 & S05  \\
8245.629 & 07.05.2018 & 0.974 &          210 & S05  \\
8245.666 & 07.05.2018 & 0.993 &          216 & S05  \\
8246.530 & 08.05.2018 & 0.473 &          214 & S05  \\
8246.567 & 08.05.2018 & 0.492 &          224 & S05  \\
8246.605 & 08.05.2018 & 0.514 &          233 & S05  \\
8246.650 & 08.05.2018 & 0.537 &          233 & S05  \\
8247.512 & 08.05.2018 & 0.014 &          214 & S05  \\
8247.632 & 09.05.2018 & 0.081 &          220 & S05  \\
8247.669 & 09.05.2018 & 0.101 &          202 & S05  \\
8248.535 & 10.05.2018 & 0.580 &          232 & S05  \\
8248.580 & 10.05.2018 & 0.604 &          228 & S05  \\
8248.617 & 10.05.2018 & 0.625 &          228 & S05  \\
8248.655 & 10.05.2018 & 0.646 &          246 & S05  \\
8248.700 & 10.05.2018 & 0.670 &          232 & S05  \\
8249.524 & 11.05.2018 & 0.126 &          232 & S05  \\
8249.569 & 11.05.2018 & 0.152 &          227 & S05  \\
8249.607 & 11.05.2018 & 0.171 &          229 & S05  \\
8249.644 & 11.05.2018 & 0.193 &          250 & S05  \\
8250.526 & 12.05.2018 & 0.681 &          227 & S05  \\
8250.571 & 12.05.2018 & 0.704 &          229 & S05  \\
8250.609 & 12.05.2018 & 0.726 &          222 & S05  \\
8250.646 & 12.05.2018 & 0.745 &          245 & S05  \\
8250.691 & 12.05.2018 & 0.771 &          232 & S05  \\
8251.516 & 12.05.2018 & 0.227 &          204 & S05  \\
8251.562 & 13.05.2018 & 0.252 &          201 & S05  \\
8251.599 & 13.05.2018 & 0.272 &          210 & S05  \\
8251.637 & 13.05.2018 & 0.293 &          219 & S05  \\
8251.682 & 13.05.2018 & 0.319 &          204 & S05  \\
8252.496 & 13.05.2018 & 0.768 &          223 & S06  \\
8252.533 & 14.05.2018 & 0.788 &          241 & S06  \\
8252.690 & 14.05.2018 & 0.876 &          233 & S06  \\
8253.502 & 14.05.2018 & 0.325 &          161 & S06  \\
8253.585 & 15.05.2018 & 0.371 &          172 & S06  \\
8253.694 & 15.05.2018 & 0.431 &          182 & S06  \\
8254.482 & 15.05.2018 & 0.865 &          209 & S06  \\
8254.520 & 15.05.2018 & 0.886 &          209 & S06  \\
8254.557 & 16.05.2018 & 0.908 &          215 & S06  \\
8254.603 & 16.05.2018 & 0.932 &          214 & S06  \\
8254.640 & 16.05.2018 & 0.953 &          206 & S06  \\
8254.678 & 16.05.2018 & 0.975 &          208 & S06  \\
8256.493 & 17.05.2018 & 0.976 &          221 & S06  \\
8256.531 & 18.05.2018 & 0.998 &          225 & S06  \\
8256.568 & 18.05.2018 & 0.017 &          225 & S06  \\
8256.614 & 18.05.2018 & 0.043 &          235 & S06  \\
8256.652 & 18.05.2018 & 0.065 &          214 & S06  \\
8256.697 & 18.05.2018 & 0.088 &          229 & S06  \\
8257.529 & 19.05.2018 & 0.548 &          226 & S06  \\
8257.566 & 19.05.2018 & 0.570 &          222 & S06  \\
8257.604 & 19.05.2018 & 0.591 &          235 & S06  \\
8257.687 & 19.05.2018 & 0.637 &          233 & S06  \\
8258.513 & 19.05.2018 & 0.092 &          182 & S06  \\
8258.551 & 20.05.2018 & 0.114 &          211 & S06  \\
8258.596 & 20.05.2018 & 0.140 &          215 & S06  \\
8258.633 & 20.05.2018 & 0.159 &          208 & S06  \\
8258.671 & 20.05.2018 & 0.181 &          232 & S06  \\
8259.482 & 20.05.2018 & 0.627 &          218 & S07  \\
8259.519 & 20.05.2018 & 0.649 &          226 & S07  \\
8259.557 & 21.05.2018 & 0.670 &          228 & S07  \\
8259.602 & 21.05.2018 & 0.694 &          228 & S07  \\
8259.677 & 21.05.2018 & 0.735 &          224 & S07  \\
8260.566 & 22.05.2018 & 0.228 &          215 & S07  \\
8260.604 & 22.05.2018 & 0.249 &          227 & S07  \\
8260.691 & 22.05.2018 & 0.296 &          211 & S07  \\
8261.573 & 23.05.2018 & 0.784 &          210 & S07  \\
8261.611 & 23.05.2018 & 0.804 &          224 & S07  \\
8261.686 & 23.05.2018 & 0.847 &          215 & S07  \\
8262.513 & 23.05.2018 & 0.302 &          225 & S07  \\
8262.551 & 24.05.2018 & 0.324 &          214 & S07  \\
8262.588 & 24.05.2018 & 0.346 &          224 & S07  \\
8262.691 & 24.05.2018 & 0.401 &          206 & S07  \\
8263.497 & 24.05.2018 & 0.846 &          237 & S07  \\
8263.535 & 25.05.2018 & 0.868 &          230 & S07  \\
8263.572 & 25.05.2018 & 0.887 &          226 & S07  \\
8265.483 & 26.05.2018 & 0.945 &          209 & S07  \\
8265.521 & 27.05.2018 & 0.964 &          232 & S07  \\
8265.558 & 27.05.2018 & 0.986 &          224 & S07  \\
8265.692 & 27.05.2018 & 0.059 &          215 & S07  \\
8266.453 & 27.05.2018 & 0.480 &          213 & S07  \\
8266.500 & 27.05.2018 & 0.506 &          213 & S07  \\
8266.537 & 28.05.2018 & 0.525 &          222 & S07  \\
8266.582 & 28.05.2018 & 0.551 &          240 & S07  \\
8266.685 & 28.05.2018 & 0.607 &          230 & S07  \\
8267.459 & 28.05.2018 & 0.037 &          212 & S07  \\
8267.531 & 29.05.2018 & 0.076 &          209 & S07  \\
8267.657 & 29.05.2018 & 0.145 &          215 & S07  \\
8268.450 & 29.05.2018 & 0.583 &          167 & S08  \\
8268.491 & 29.05.2018 & 0.607 &          201 & S08  \\
8268.529 & 30.05.2018 & 0.626 &          234 & S08  \\
8268.566 & 30.05.2018 & 0.648 &          211 & S08  \\
8269.451 & 30.05.2018 & 0.136 &          190 & S08  \\
8269.492 & 30.05.2018 & 0.159 &          217 & S08  \\
8269.530 & 31.05.2018 & 0.181 &          229 & S08  \\
8269.567 & 31.05.2018 & 0.200 &          219 & S08  \\
8269.654 & 31.05.2018 & 0.248 &          220 & S08  \\
8270.448 & 31.05.2018 & 0.688 &          215 & S08  \\
8270.494 & 31.05.2018 & 0.712 &          214 & S08  \\
8270.531 & 01.06.2018 & 0.733 &          204 & S08  \\
8270.569 & 01.06.2018 & 0.755 &          201 & S08  \\
8270.655 & 01.06.2018 & 0.802 &          183 & S08  \\
8271.476 & 01.06.2018 & 0.256 &          221 & S08  \\
8271.513 & 01.06.2018 & 0.275 &          220 & S08  \\
8271.550 & 02.06.2018 & 0.297 &          231 & S08  \\
8271.681 & 02.06.2018 & 0.368 &          222 & S08  \\
8272.452 & 02.06.2018 & 0.795 &          212 & S08  \\
8272.493 & 02.06.2018 & 0.817 &          216 & S08  \\
8272.530 & 03.06.2018 & 0.838 &          223 & S08  \\
8272.568 & 03.06.2018 & 0.858 &          236 & S08  \\
8272.658 & 03.06.2018 & 0.908 &          225 & S08  \\
8273.452 & 03.06.2018 & 0.348 &          223 & S08  \\
8273.494 & 03.06.2018 & 0.370 &          222 & S08  \\
8273.531 & 04.06.2018 & 0.391 &          219 & S08  \\%SME
8273.568 & 04.06.2018 & 0.411 &          242 & S08  \\
8273.648 & 04.06.2018 & 0.456 &          223 & S08  \\
8274.450 & 04.06.2018 & 0.898 &          202 & S08  \\
8274.494 & 04.06.2018 & 0.922 &          222 & S08  \\
8274.532 & 05.06.2018 & 0.943 &          221 & S08  \\
8274.569 & 05.06.2018 & 0.965 &          232 & S08  \\
8274.680 & 05.06.2018 & 0.025 &          209 & S08  \\
8275.457 & 05.06.2018 & 0.455 &          239 & S08  \\
8275.502 & 05.06.2018 & 0.481 &          238 & S08  \\
8275.539 & 06.06.2018 & 0.500 &          241 & S08  \\
8275.646 & 06.06.2018 & 0.559 &          249 & S08  \\
8275.684 & 06.06.2018 & 0.580 &          239 & S08  \\
8276.473 & 06.06.2018 & 0.016 &          219 & S09  \\
8276.518 & 06.06.2018 & 0.042 &          220 & S09  \\
8276.555 & 07.06.2018 & 0.062 &          236 & S09  \\
8276.633 & 07.06.2018 & 0.104 &          239 & S09  \\
8276.670 & 07.06.2018 & 0.126 &          225 & S09  \\
8277.425 & 07.06.2018 & 0.543 &          218 & S09  \\
8277.462 & 07.06.2018 & 0.563 &          225 & S09  \\
8277.507 & 07.06.2018 & 0.588 &          230 & S09  \\
8277.545 & 08.06.2018 & 0.610 &          237 & S09  \\
8277.649 & 08.06.2018 & 0.666 &          236 & S09  \\
8278.446 & 08.06.2018 & 0.106 &          227 & S09  \\
8278.491 & 08.06.2018 & 0.132 &          216 & S09  \\
8279.451 & 09.06.2018 & 0.661 &          232 & S09  \\
8279.496 & 09.06.2018 & 0.687 &          239 & S09  \\
8279.534 & 10.06.2018 & 0.708 &          237 & S09  \\
8279.652 & 10.06.2018 & 0.773 &          239 & S09  \\
8279.690 & 10.06.2018 & 0.795 &          233 & S09  \\
8280.472 & 10.06.2018 & 0.227 &          219 & S09  \\
8280.517 & 10.06.2018 & 0.250 &          232 & S09  \\
8280.555 & 11.06.2018 & 0.271 &          247 & S09  \\
8280.644 & 11.06.2018 & 0.321 &          222 & S09  \\
8280.688 & 11.06.2018 & 0.345 &          217 & S09  \\
8283.411 & 13.06.2018 & 0.850 &          188 & S09  \\
8283.448 & 13.06.2018 & 0.871 &          237 & S09  \\
8283.493 & 13.06.2018 & 0.895 &          228 & S09  \\
8283.661 & 14.06.2018 & 0.988 &          223 & S09  \\
8284.411 & 14.06.2018 & 0.402 &          178 & S09  \\
8284.449 & 14.06.2018 & 0.423 &          191 & S09  \\
8284.493 & 14.06.2018 & 0.447 &          209 & S09  \\
8284.531 & 15.06.2018 & 0.469 &          227 & S09  \\
8284.641 & 15.06.2018 & 0.529 &          200 & S09  \\
8285.411 & 15.06.2018 & 0.955 &          205 & S10  \\
8285.487 & 15.06.2018 & 0.998 &          200 & S10  \\
8285.524 & 16.06.2018 & 0.017 &          215 & S10  \\
8285.665 & 16.06.2018 & 0.095 &          204 & S10  \\
8286.411 & 16.06.2018 & 0.507 &          186 & S10  \\
8286.449 & 16.06.2018 & 0.529 &          181 & S10  \\
8286.528 & 17.06.2018 & 0.572 &          240 & S10  \\
8286.624 & 17.06.2018 & 0.626 &          232 & S10  \\
8286.669 & 17.06.2018 & 0.649 &          210 & S10  \\
8287.412 & 17.06.2018 & 0.060 &          208 & S10  \\
8287.449 & 17.06.2018 & 0.081 &          216 & S10  \\
8287.494 & 17.06.2018 & 0.105 &          228 & S10  \\
8287.532 & 18.06.2018 & 0.126 &          245 & S10  \\
8287.642 & 18.06.2018 & 0.187 &          231 & S10  \\
8287.687 & 18.06.2018 & 0.213 &          219 & S10  \\
8288.410 & 18.06.2018 & 0.612 &          215 & S10  \\
8288.447 & 18.06.2018 & 0.631 &          208 & S10  \\
8288.492 & 18.06.2018 & 0.657 &          218 & S10  \\
8288.529 & 19.06.2018 & 0.677 &          232 & S10  \\
8288.601 & 19.06.2018 & 0.718 &          225 & S10  \\
8288.639 & 19.06.2018 & 0.739 &          206 & S10  \\
8288.684 & 19.06.2018 & 0.763 &          204 & S10  \\
8289.412 & 19.06.2018 & 0.165 &          205 & S10  \\
8289.450 & 19.06.2018 & 0.186 &          212 & S10  \\
8289.494 & 19.06.2018 & 0.210 &          207 & S10  \\
8289.532 & 20.06.2018 & 0.231 &          238 & S10  \\
8289.600 & 20.06.2018 & 0.271 &          243 & S10  \\
8289.638 & 20.06.2018 & 0.290 &          217 & S10  \\
8289.675 & 20.06.2018 & 0.312 &          220 & S10  \\
8290.413 & 20.06.2018 & 0.719 &          219 & S10  \\
8290.450 & 20.06.2018 & 0.739 &          225 & S10  \\
8290.495 & 20.06.2018 & 0.765 &          227 & S10  \\
8290.533 & 21.06.2018 & 0.784 &          232 & S10  \\
8290.601 & 21.06.2018 & 0.823 &          249 & S10  \\
8290.639 & 21.06.2018 & 0.845 &          234 & S10  \\
8290.676 & 21.06.2018 & 0.864 &          228 & S10  \\
8291.413 & 21.06.2018 & 0.272 &          219 & S10  \\
8291.450 & 21.06.2018 & 0.292 &          230 & S10  \\
8291.495 & 21.06.2018 & 0.317 &          232 & S10  \\
8291.597 & 22.06.2018 & 0.374 &          248 & S10  \\
8291.635 & 22.06.2018 & 0.395 &          228 & S10  \\
8291.672 & 22.06.2018 & 0.415 &          226 & S10  \\
8292.413 & 22.06.2018 & 0.824 &          227 & S10  \\
8292.450 & 22.06.2018 & 0.844 &          231 & S10  \\
8292.592 & 23.06.2018 & 0.924 &          240 & S10  \\
8292.629 & 23.06.2018 & 0.943 &          217 & S10  \\
8292.667 & 23.06.2018 & 0.965 &          208 & S10  \\
8293.451 & 23.06.2018 & 0.396 &          222 & S11  \\
8293.495 & 23.06.2018 & 0.422 &          213 & S11  \\
8293.590 & 24.06.2018 & 0.474 &          213 & S11  \\
8293.628 & 24.06.2018 & 0.496 &          203 & S11  \\
8293.665 & 24.06.2018 & 0.515 &          178 & S11  \\
8294.413 & 24.06.2018 & 0.930 &          203 & S11  \\
8294.451 & 24.06.2018 & 0.949 &          208 & S11  \\
8294.496 & 24.06.2018 & 0.975 &          206 & S11  \\
8294.592 & 25.06.2018 & 0.029 &          220 & S11  \\
8294.629 & 25.06.2018 & 0.048 &          197 & S11  \\
8294.666 & 25.06.2018 & 0.068 &          189 & S11  \\
8295.442 & 25.06.2018 & 0.497 &          209 & S11  \\
8295.479 & 25.06.2018 & 0.519 &          214 & S11  \\
8295.517 & 25.06.2018 & 0.538 &          187 & S11  \\
8295.589 & 26.06.2018 & 0.579 &          225 & S11  \\
8295.627 & 26.06.2018 & 0.601 &          200 & S11  \\
8295.672 & 26.06.2018 & 0.625 &          198 & S11  \\
8296.413 & 26.06.2018 & 0.035 &          199 & S11  \\
8296.451 & 26.06.2018 & 0.054 &          208 & S11  \\
8296.495 & 26.06.2018 & 0.080 &          197 & S11  \\%SME
8296.587 & 27.06.2018 & 0.129 &          224 & S11  \\
8296.625 & 27.06.2018 & 0.151 &          201 & S11  \\
8296.662 & 27.06.2018 & 0.170 &          183 & S11  \\
8297.413 & 27.06.2018 & 0.587 &          217 & S11  \\
8297.450 & 27.06.2018 & 0.606 &          216 & S11  \\%SME
8297.495 & 27.06.2018 & 0.632 &          239 & S11  \\
8297.590 & 28.06.2018 & 0.684 &          229 & S11  \\
8297.627 & 28.06.2018 & 0.706 &          228 & S11  \\
8297.664 & 28.06.2018 & 0.725 &          216 & S11  \\
8298.413 & 28.06.2018 & 0.140 &          199 & S11  \\
8298.451 & 28.06.2018 & 0.159 &          217 & S11  \\
8298.495 & 28.06.2018 & 0.185 &          240 & S11  \\
8298.591 & 29.06.2018 & 0.237 &          214 & S11  \\
8298.629 & 29.06.2018 & 0.258 &          213 & S11  \\
8298.666 & 29.06.2018 & 0.278 &          190 & S11  \\
8299.499 & 29.06.2018 & 0.740 &          234 & S11  \\
8299.594 & 30.06.2018 & 0.792 &          225 & S11  \\
8299.632 & 30.06.2018 & 0.813 &          219 & S11  \\
8299.669 & 30.06.2018 & 0.833 &          199 & S11  \\
8300.413 & 30.06.2018 & 0.245 &          226 & S12  \\
8300.496 & 30.06.2018 & 0.290 &          244 & S12  \\
8300.592 & 01.07.2018 & 0.344 &          217 & S12  \\
8300.629 & 01.07.2018 & 0.363 &          228 & S12  \\
8300.667 & 01.07.2018 & 0.385 &          217 & S12  \\
8301.591 & 02.07.2018 & 0.895 &          221 & S12  \\
8301.628 & 02.07.2018 & 0.916 &          219 & S12  \\
8301.665 & 02.07.2018 & 0.936 &          196 & S12  \\
8302.413 & 02.07.2018 & 0.350 &          172 & S12  \\
8302.451 & 02.07.2018 & 0.369 &          168 & S12  \\
8302.495 & 02.07.2018 & 0.395 &          193 & S12  \\
8302.594 & 03.07.2018 & 0.449 &          173 & S12  \\
8302.631 & 03.07.2018 & 0.471 &          164 & S12  \\
8303.415 & 03.07.2018 & 0.902 &          136 & S12  \\
8303.452 & 03.07.2018 & 0.924 &          152 & S12  \\
8303.497 & 03.07.2018 & 0.948 &          146 & S12  \\
8303.558 & 04.07.2018 & 0.982 &          172 & S12  \\
8303.602 & 04.07.2018 & 0.006 &          151 & S12  \\
8303.640 & 04.07.2018 & 0.027 &          150 & S12  \\
8303.677 & 04.07.2018 & 0.047 &          131 & S12  \\
8304.413 & 04.07.2018 & 0.455 &          165 & S12  \\
8304.450 & 04.07.2018 & 0.474 &          161 & S12  \\
8304.556 & 05.07.2018 & 0.533 &          180 & S12  \\
8304.601 & 05.07.2018 & 0.559 &          173 & S12  \\
8304.638 & 05.07.2018 & 0.578 &          162 & S12  \\
8305.412 & 05.07.2018 & 0.005 &          171 & S12  \\
8305.450 & 05.07.2018 & 0.027 &          169 & S12  \\
8305.567 & 06.07.2018 & 0.092 &          183 & S12  \\
8305.611 & 06.07.2018 & 0.115 &          183 & S12  \\
8305.649 & 06.07.2018 & 0.137 &          177 & S12  \\
8306.413 & 06.07.2018 & 0.560 &          183 & S12  \\
8306.450 & 06.07.2018 & 0.580 &          214 & S12  \\
8306.598 & 07.07.2018 & 0.662 &          198 & S12  \\
8306.635 & 07.07.2018 & 0.683 &          188 & S12  \\
8306.672 & 07.07.2018 & 0.703 &          165 & S12  \\
8307.413 & 07.07.2018 & 0.112 &          186 & S12  \\
8307.450 & 07.07.2018 & 0.132 &          179 & S12  \\
8307.618 & 08.07.2018 & 0.225 &          185 & S12  \\
8311.536 & 12.07.2018 & 0.390 &          174 & S13  \\%SME
8311.581 & 12.07.2018 & 0.416 &          163 & S13  \\
8311.618 & 12.07.2018 & 0.435 &          152 & S13  \\
8312.449 & 12.07.2018 & 0.895 &          156 & S13  \\
8312.540 & 13.07.2018 & 0.944 &          188 & S13  \\
8312.585 & 13.07.2018 & 0.970 &          176 & S13  \\
8312.622 & 13.07.2018 & 0.990 &          163 & S13  \\
8313.541 & 14.07.2018 & 0.497 &          189 & S13  \\
8313.586 & 14.07.2018 & 0.522 &          196 & S13  \\
8313.623 & 14.07.2018 & 0.542 &          161 & S13  \\
8313.661 & 14.07.2018 & 0.563 &          154 & S13  \\
8314.541 & 15.07.2018 & 0.049 &          142 & S13  \\
8315.411 & 15.07.2018 & 0.531 &          211 & S13  \\
8315.448 & 15.07.2018 & 0.552 &          216 & S13  \\
8315.541 & 16.07.2018 & 0.602 &          206 & S13  \\
8315.586 & 16.07.2018 & 0.628 &          216 & S13  \\
8315.623 & 16.07.2018 & 0.647 &          205 & S13  \\
8316.410 & 16.07.2018 & 0.083 &          223 & S13  \\%SME
8316.448 & 16.07.2018 & 0.105 &          231 & S13  \\
8316.541 & 17.07.2018 & 0.154 &          213 & S13  \\
8316.586 & 17.07.2018 & 0.180 &          220 & S13  \\
8316.623 & 17.07.2018 & 0.200 &          204 & S13  \\
8316.661 & 17.07.2018 & 0.221 &          191 & S13  \\
8317.411 & 17.07.2018 & 0.636 &          214 & S13  \\
8317.448 & 17.07.2018 & 0.657 &          215 & S13  \\
8317.541 & 18.07.2018 & 0.707 &          196 & S13  \\
8317.585 & 18.07.2018 & 0.733 &          190 & S13  \\
8317.622 & 18.07.2018 & 0.752 &          188 & S13  \\
8317.660 & 18.07.2018 & 0.774 &          171 & S13  \\
8321.408 & 21.07.2018 & 0.844 &          196 & S14  \\
8321.446 & 21.07.2018 & 0.865 &          217 & S14  \\
8321.514 & 21.07.2018 & 0.904 &          203 & S14  \\
8321.551 & 22.07.2018 & 0.923 &          181 & S14  \\
8321.596 & 22.07.2018 & 0.949 &          177 & S14  \\
8321.633 & 22.07.2018 & 0.969 &          164 & S14  \\
8321.670 & 22.07.2018 & 0.990 &          137 & S14  \\
8322.408 & 22.07.2018 & 0.396 &          198 & S14  \\
8322.526 & 23.07.2018 & 0.463 &          202 & S14  \\
8322.608 & 23.07.2018 & 0.508 &          154 & S14  \\%SME
8323.408 & 23.07.2018 & 0.949 &          177 & S14  \\
8323.544 & 24.07.2018 & 0.024 &          175 & S14  \\
8323.589 & 24.07.2018 & 0.050 &          158 & S14  \\
8324.408 & 24.07.2018 & 0.501 &          171 & S14  \\
8324.502 & 24.07.2018 & 0.555 &          179 & S14  \\
8324.539 & 25.07.2018 & 0.575 &          175 & S14  \\
8324.584 & 25.07.2018 & 0.601 &          163 & S14  \\
8324.622 & 25.07.2018 & 0.620 &          149 & S14  \\
8325.407 & 25.07.2018 & 0.054 &          179 & S14  \\
8325.499 & 25.07.2018 & 0.105 &          196 & S14  \\
8325.537 & 26.07.2018 & 0.125 &          178 & S14  \\
8325.574 & 26.07.2018 & 0.146 &          175 & S14  \\
8326.403 & 26.07.2018 & 0.604 &          200 & S14  \\
8326.532 & 27.07.2018 & 0.675 &          219 & S14  \\
8326.570 & 27.07.2018 & 0.697 &          201 & S14  \\
8326.615 & 27.07.2018 & 0.721 &          186 & S14  \\
8327.406 & 27.07.2018 & 0.159 &          206 & S14  \\
8327.508 & 27.07.2018 & 0.215 &          189 & S14  \\
8327.546 & 28.07.2018 & 0.237 &          187 & S14  \\
8327.583 & 28.07.2018 & 0.256 &          177 & S14  \\
8327.628 & 28.07.2018 & 0.282 &          158 & S14  \\
8353.432 & 22.08.2018 & 0.540 &          207 & S15  \\
8353.476 & 22.08.2018 & 0.563 &          193 & S15  \\
8353.513 & 22.08.2018 & 0.583 &          195 & S15  \\
8353.551 & 23.08.2018 & 0.604 &          184 & S15  \\
8353.595 & 23.08.2018 & 0.628 &          146 & S15  \\
8354.431 & 23.08.2018 & 0.090 &          203 & S15  \\
8354.476 & 23.08.2018 & 0.116 &          199 & S15  \\
8354.550 & 24.08.2018 & 0.157 &          178 & S15  \\
8355.430 & 24.08.2018 & 0.643 &          208 & S15  \\
8355.475 & 24.08.2018 & 0.669 &          210 & S15  \\
8355.512 & 24.08.2018 & 0.688 &          201 & S15  \\
8355.550 & 25.08.2018 & 0.710 &          193 & S15  \\
8356.429 & 25.08.2018 & 0.195 &          182 & S15  \\
8356.474 & 25.08.2018 & 0.219 &          171 & S15  \\
8356.512 & 25.08.2018 & 0.241 &          189 & S15  \\
8356.549 & 26.08.2018 & 0.262 &          169 & S15  \\
8357.428 & 26.08.2018 & 0.748 &          190 & S15  \\
8357.473 & 26.08.2018 & 0.771 &          192 & S15  \\
8357.510 & 26.08.2018 & 0.793 &          169 & S15  \\
8357.548 & 27.08.2018 & 0.813 &          155 & S15  \\
8358.427 & 27.08.2018 & 0.298 &          188 & S15  \\
8358.472 & 27.08.2018 & 0.324 &          195 & S15  \\
8358.509 & 27.08.2018 & 0.344 &          182 & S15  \\
8359.403 & 28.08.2018 & 0.838 &          206 & S15  \\
8359.440 & 28.08.2018 & 0.859 &          199 & S15  \\
8359.484 & 28.08.2018 & 0.883 &          187 & S15  \\
8359.522 & 29.08.2018 & 0.905 &          182 & S15  \\
8359.559 & 29.08.2018 & 0.924 &          166 & S15  \\
8360.402 & 29.08.2018 & 0.391 &          218 & S15  \\
8360.439 & 29.08.2018 & 0.410 &          208 & S15  \\
8360.484 & 29.08.2018 & 0.436 &          198 & S15  \\
8360.522 & 30.08.2018 & 0.458 &          170 & S15  \\
8360.559 & 30.08.2018 & 0.477 &          159 & S15  \\
8361.401 & 30.08.2018 & 0.943 &          200 & S15  \\
8361.439 & 30.08.2018 & 0.962 &          194 & S15  \\
8361.484 & 30.08.2018 & 0.988 &          179 & S15  \\
8361.522 & 31.08.2018 & 0.010 &          172 & S15  \\
8361.559 & 31.08.2018 & 0.029 &          142 & S15  \\
8362.488 & 31.08.2018 & 0.543 &          183 & S16  \\
8363.400 & 01.09.2018 & 0.046 &          193 & S16  \\
8363.437 & 01.09.2018 & 0.067 &          178 & S16  \\
8363.482 & 01.09.2018 & 0.091 &          159 & S16  \\
8363.520 & 02.09.2018 & 0.113 &          149 & S16  \\
8364.400 & 02.09.2018 & 0.599 &          214 & S16  \\
8364.437 & 02.09.2018 & 0.620 &          183 & S16  \\
8364.482 & 02.09.2018 & 0.644 &          184 & S16  \\
8365.399 & 03.09.2018 & 0.151 &          208 & S16  \\
8365.436 & 03.09.2018 & 0.172 &          187 & S16  \\
8365.481 & 03.09.2018 & 0.196 &          171 & S16  \\
8365.519 & 04.09.2018 & 0.218 &          163 & S16  \\%SME
8366.419 & 04.09.2018 & 0.714 &          189 & S16  \\
8366.464 & 04.09.2018 & 0.740 &          189 & S16  \\
8366.501 & 04.09.2018 & 0.760 &          175 & S16  \\
8366.538 & 05.09.2018 & 0.781 &          166 & S16  \\
8368.395 & 06.09.2018 & 0.807 &          186 & S16  \\
8368.432 & 06.09.2018 & 0.828 &          166 & S16  \\
8368.477 & 06.09.2018 & 0.852 &          139 & S16  \\
8369.395 & 07.09.2018 & 0.359 &          201 & S16  \\
8369.432 & 07.09.2018 & 0.380 &          204 & S16  \\
8369.477 & 07.09.2018 & 0.404 &          191 & S16  \\
8369.515 & 07.09.2018 & 0.426 &          180 & S16  \\
8370.394 & 08.09.2018 & 0.912 &          209 & S16  \\
8370.432 & 08.09.2018 & 0.933 &          206 & S16  \\
8370.477 & 08.09.2018 & 0.957 &          204 & S16  \\
8370.515 & 08.09.2018 & 0.979 &          173 & S16  \\
8371.369 & 09.09.2018 & 0.449 &          220 & S16  \\
8371.407 & 09.09.2018 & 0.471 &          197 & S16  \\
8371.452 & 09.09.2018 & 0.497 &          151 & S16  \\
8371.490 & 09.09.2018 & 0.516 &          160 & S16  \\
8372.369 & 10.09.2018 & 0.001 &          227 & S16  \\
8372.406 & 10.09.2018 & 0.023 &          212 & S16  \\
8372.452 & 10.09.2018 & 0.049 &          199 & S16  \\
8372.489 & 10.09.2018 & 0.068 &          187 & S16  \\
8378.364 & 16.09.2018 & 0.314 &          211 & S17  \\
8378.402 & 16.09.2018 & 0.336 &          209 & S17  \\
8378.447 & 16.09.2018 & 0.360 &          207 & S17  \\
8378.485 & 16.09.2018 & 0.381 &          186 & S17  \\
8378.522 & 17.09.2018 & 0.403 &          172 & S17  \\
8379.363 & 17.09.2018 & 0.867 &          207 & S17  \\
8379.400 & 17.09.2018 & 0.887 &          195 & S17  \\
8379.446 & 17.09.2018 & 0.913 &          210 & S17  \\
8379.507 & 17.09.2018 & 0.947 &          169 & S17  \\
8380.365 & 18.09.2018 & 0.419 &          203 & S17  \\
8380.403 & 18.09.2018 & 0.441 &          209 & S17  \\
8380.448 & 18.09.2018 & 0.467 &          197 & S17  \\
8380.486 & 18.09.2018 & 0.486 &          178 & S17  \\
8381.353 & 19.09.2018 & 0.966 &          147 & S17  \\
8382.360 & 20.09.2018 & 0.522 &          212 & S17  \\
8382.397 & 20.09.2018 & 0.544 &          206 & S17  \\
8382.435 & 20.09.2018 & 0.563 &          211 & S17  \\
8382.480 & 20.09.2018 & 0.589 &          179 & S17  \\
8383.357 & 21.09.2018 & 0.073 &          207 & S17  \\
8383.394 & 21.09.2018 & 0.095 &          197 & S17  \\
8383.440 & 21.09.2018 & 0.121 &          191 & S17  \\
8383.477 & 21.09.2018 & 0.140 &          171 & S17  \\
8385.364 & 23.09.2018 & 0.182 &          187 & S17  \\
8385.402 & 23.09.2018 & 0.204 &          187 & S17  \\
8385.447 & 23.09.2018 & 0.228 &          163 & S17  \\
8386.357 & 24.09.2018 & 0.730 &          186 & S17  \\
8386.394 & 24.09.2018 & 0.752 &          170 & S17  \\
8386.439 & 24.09.2018 & 0.776 &          139 & S17  \\
8387.355 & 25.09.2018 & 0.283 &          204 & S18  \\
8387.393 & 25.09.2018 & 0.305 &          201 & S18  \\
8387.438 & 25.09.2018 & 0.329 &          187 & S18  \\
8387.499 & 25.09.2018 & 0.363 &          158 & S18  \\
8388.354 & 26.09.2018 & 0.835 &          170 & S18  \\
8388.392 & 26.09.2018 & 0.855 &          157 & S18  \\
8388.437 & 26.09.2018 & 0.881 &          172 & S18  \\
8388.474 & 26.09.2018 & 0.900 &          145 & S18  \\
8389.353 & 27.09.2018 & 0.386 &          186 & S18  \\
8389.391 & 27.09.2018 & 0.408 &          190 & S18  \\
8389.436 & 27.09.2018 & 0.434 &          190 & S18  \\
8389.473 & 27.09.2018 & 0.453 &          160 & S18  \\
8392.426 & 30.09.2018 & 0.084 &          181 & S18  \\
8392.487 & 30.09.2018 & 0.119 &          145 & S18  \\
8394.350 & 02.10.2018 & 0.148 &          203 & S18  \\
8394.388 & 02.10.2018 & 0.168 &          205 & S18  \\
8394.433 & 02.10.2018 & 0.194 &          187 & S18  \\
8394.470 & 02.10.2018 & 0.213 &          167 & S18  \\
8399.345 & 07.10.2018 & 0.907 &          148 & S19  \\
8399.382 & 07.10.2018 & 0.929 &          148 & S19  \\
8400.367 & 08.10.2018 & 0.473 &          123 & S19  \\
8400.405 & 08.10.2018 & 0.494 &          123 & S19  \\
8403.360 & 11.10.2018 & 0.126 &          192 & S19  \\
8403.398 & 11.10.2018 & 0.147 &          198 & S19  \\
8404.342 & 12.10.2018 & 0.670 &          197 & S19  \\%SME
8404.379 & 12.10.2018 & 0.689 &          198 & S19  \\
8404.424 & 12.10.2018 & 0.715 &          150 & S19  \\
8405.359 & 13.10.2018 & 0.231 &          201 & S19  \\
\end{supertabular}
\end{center}

\clearpage
\onecolumn
\section{Fitted line profiles}\label{App:profiles}

\begin{figure*}[h!!!!]
    \begin{multicols}{4}  
    \includegraphics[width=\linewidth]{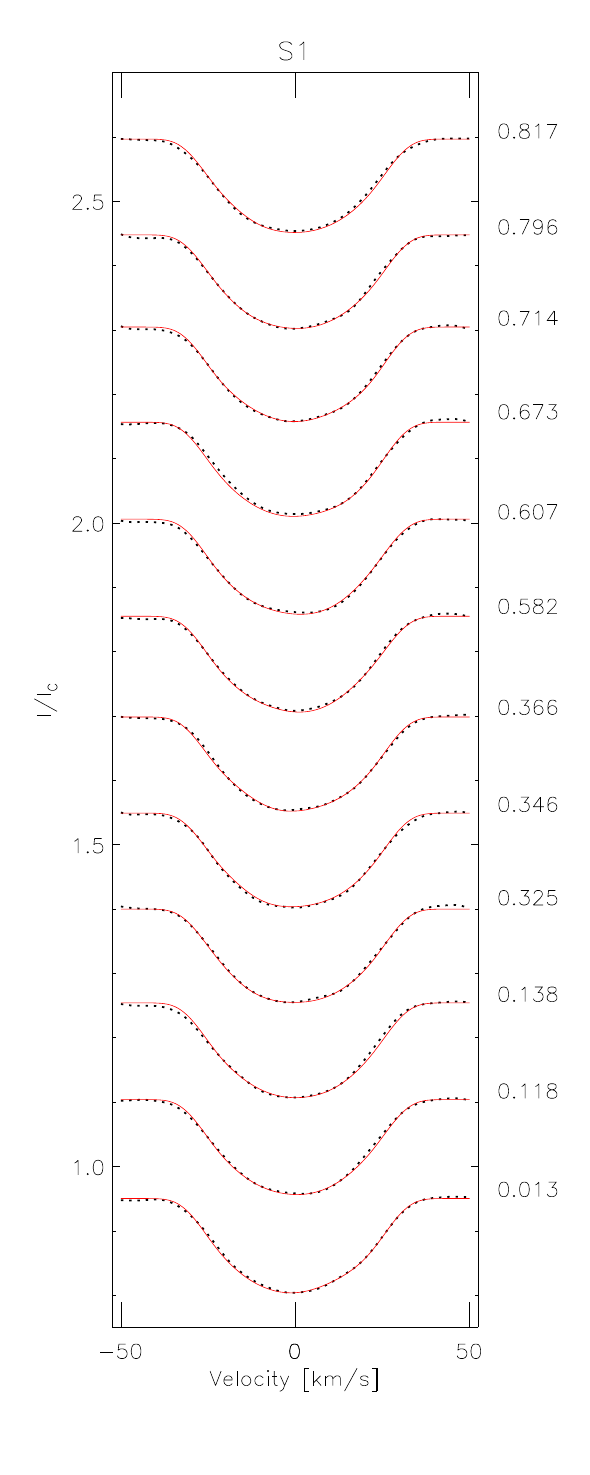}\par 
    \includegraphics[width=\linewidth]{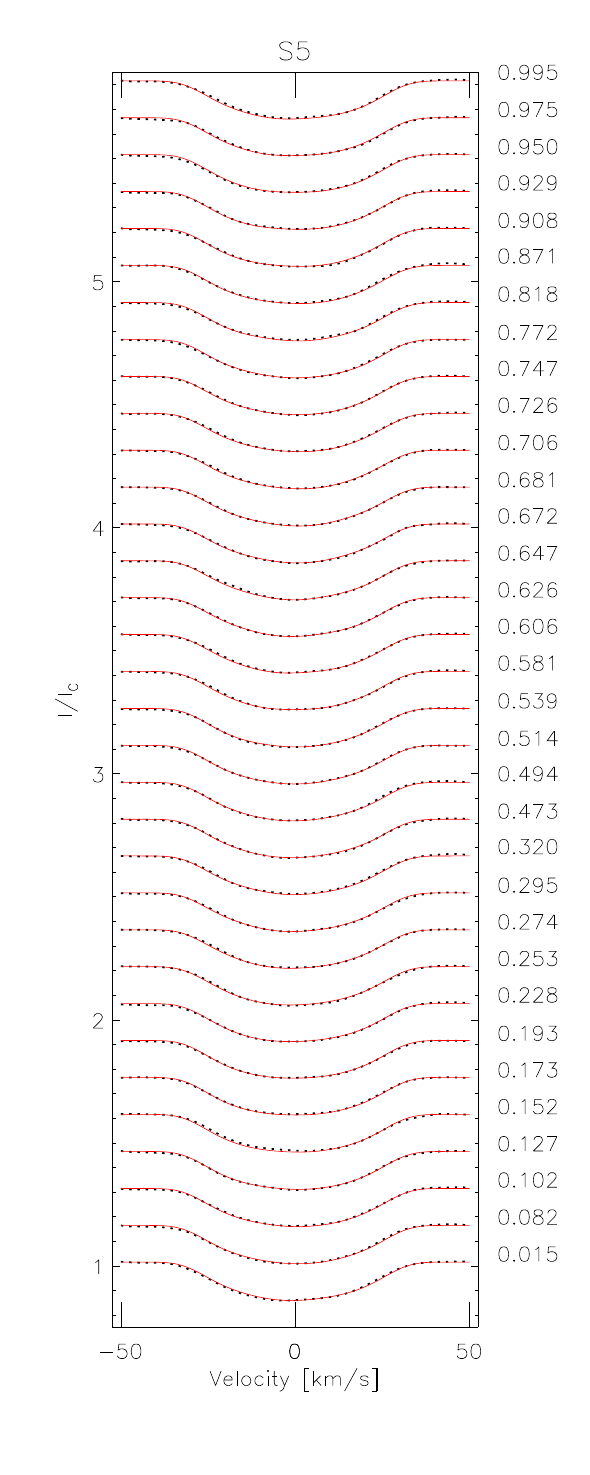}\par
    \includegraphics[width=\linewidth]{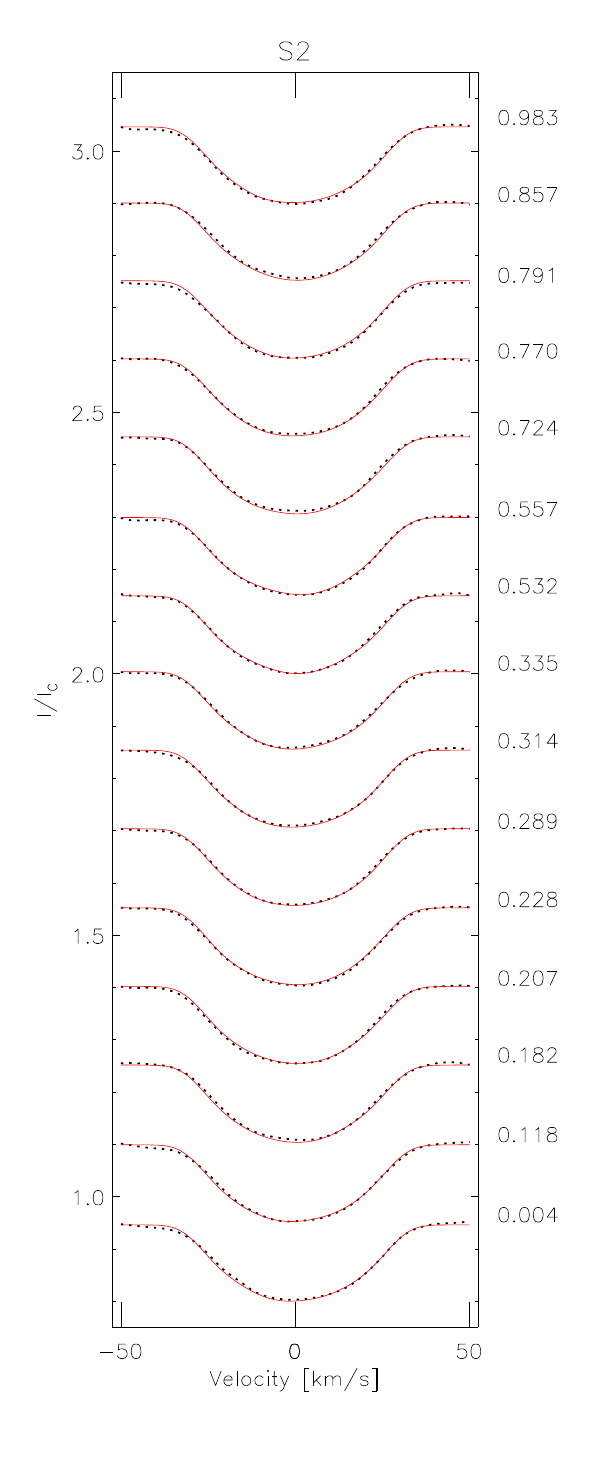}\par 
    \includegraphics[width=\linewidth]{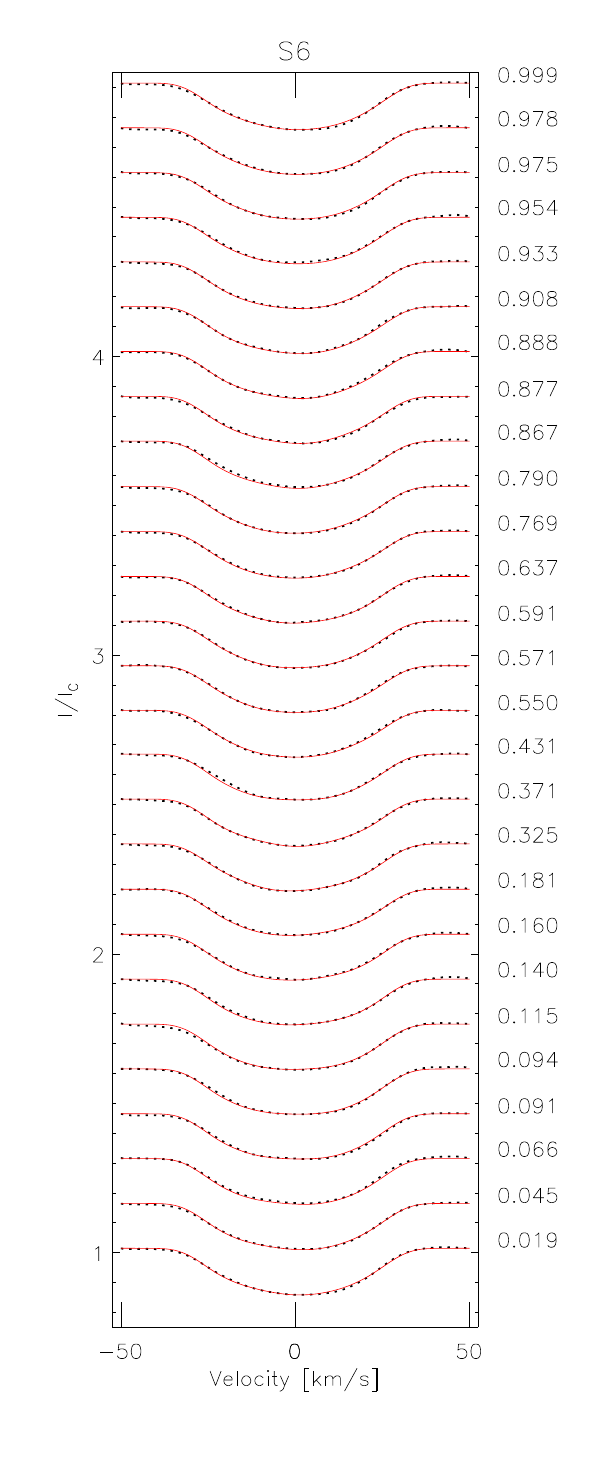}\par
     \includegraphics[width=\linewidth]{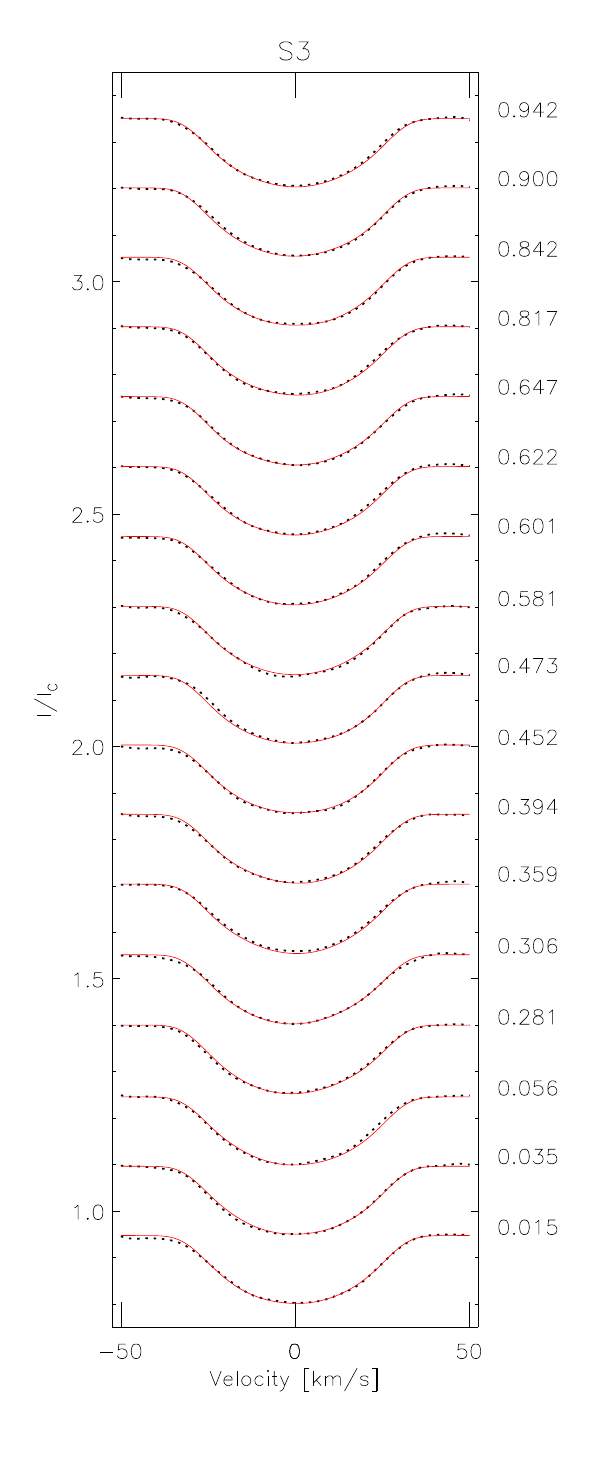}\par  
     \includegraphics[width=\linewidth]{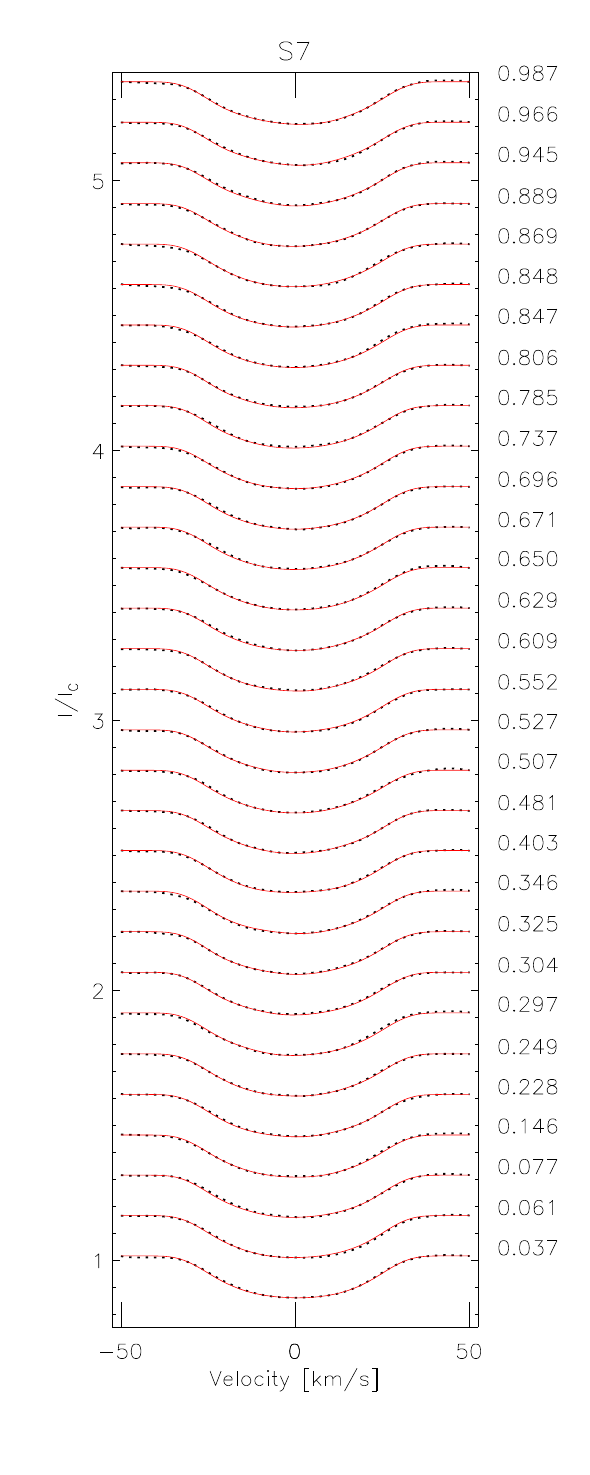}\par
    \includegraphics[width=\linewidth]{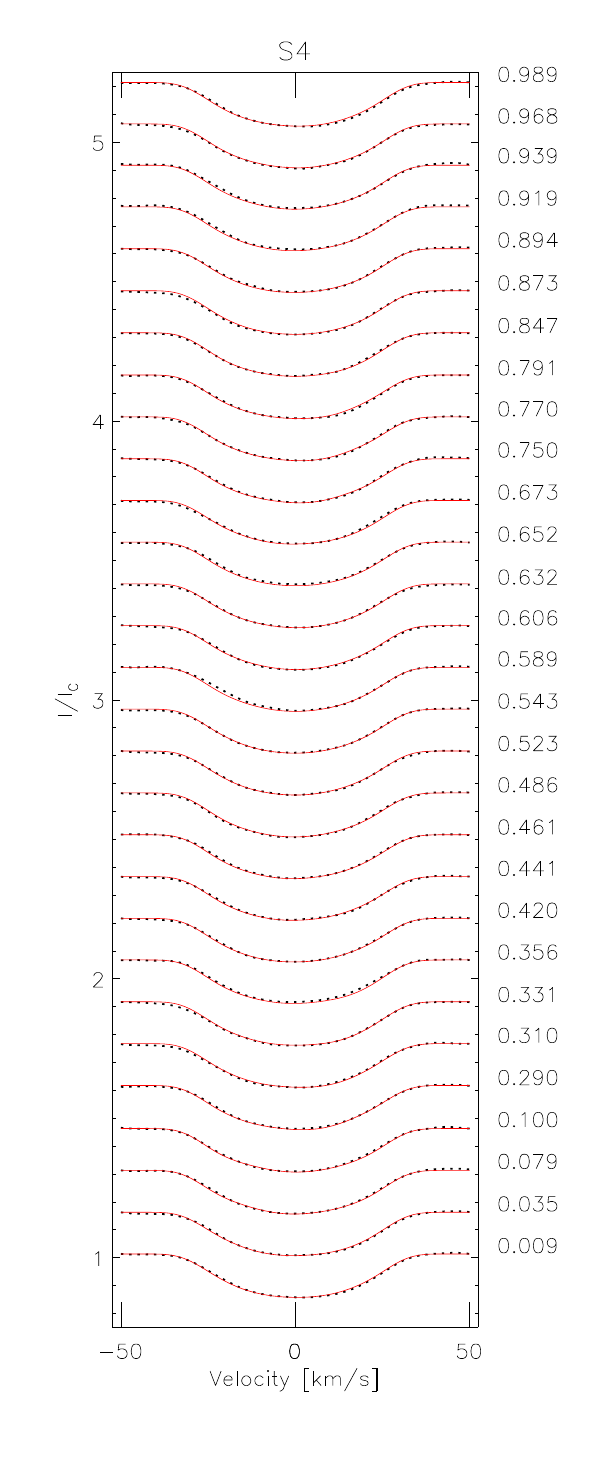}\par   
    \includegraphics[width=\linewidth]{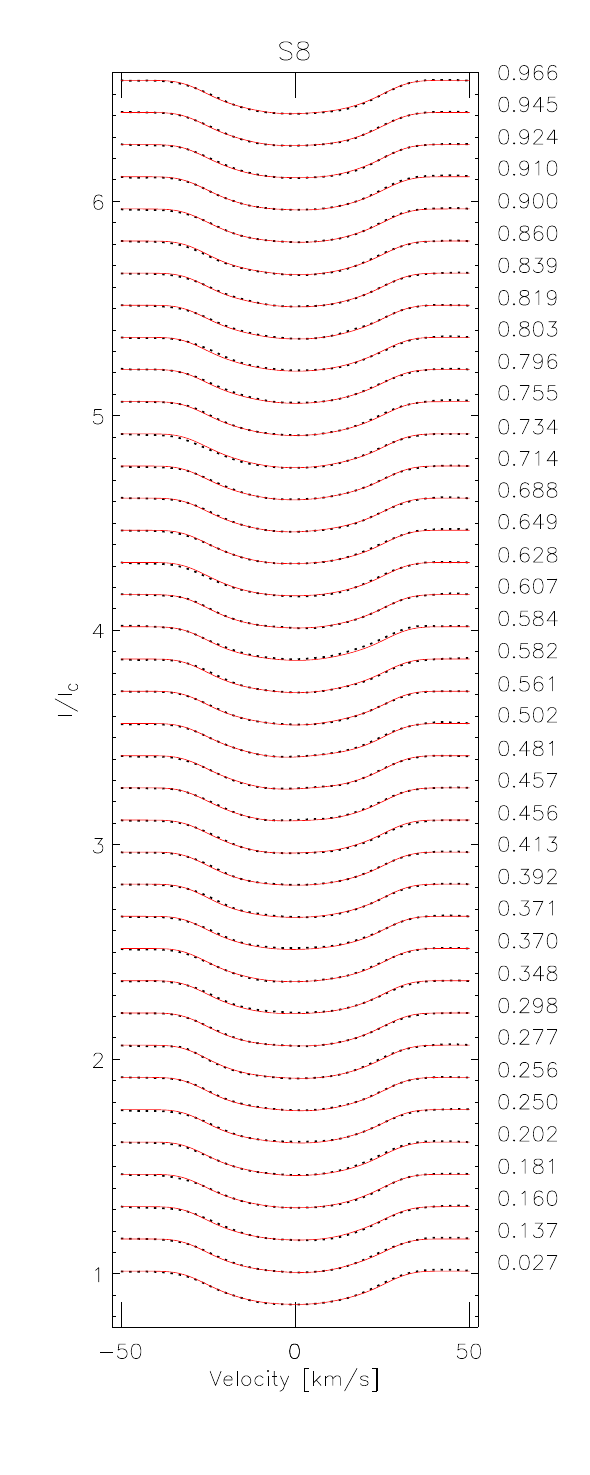}\par
\end{multicols}
\caption{Fitted line profile subsets corresponding to the Doppler images S01-S08 presented in Fig.~\ref{fig:DIs}. The phase values of each profile are shown on the right side of the panels.}
\label{fig:di_profs1}
\end{figure*}
%--
\begin{figure*}[h!!!!!]
%\ContinuedFloat
    \begin{multicols}{4}
    \includegraphics[width=\linewidth]{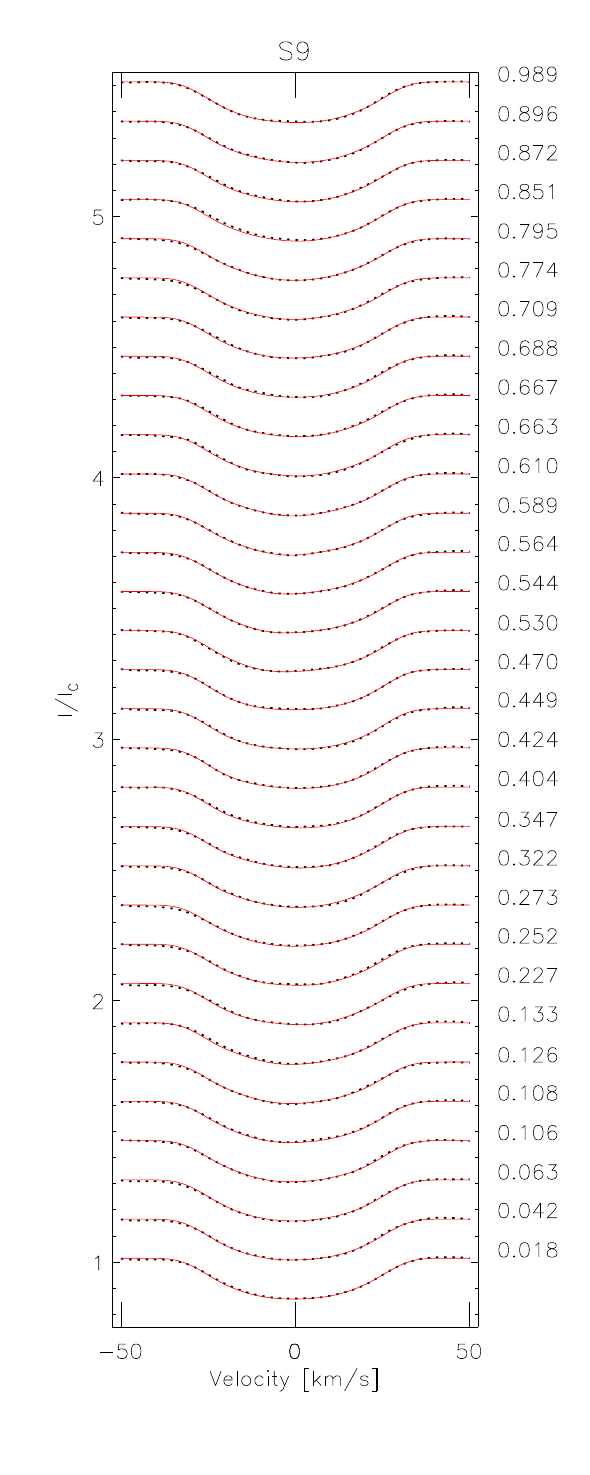}\par 
    \includegraphics[width=\linewidth]{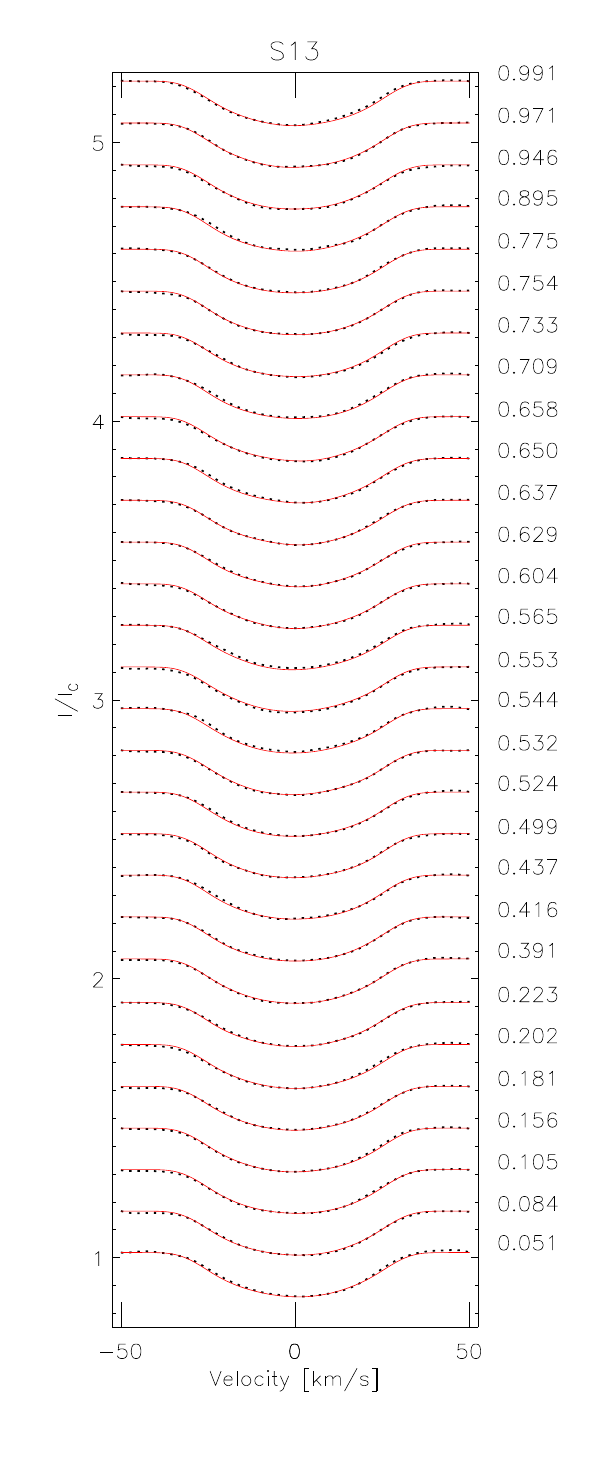}\par
     \includegraphics[width=\linewidth]{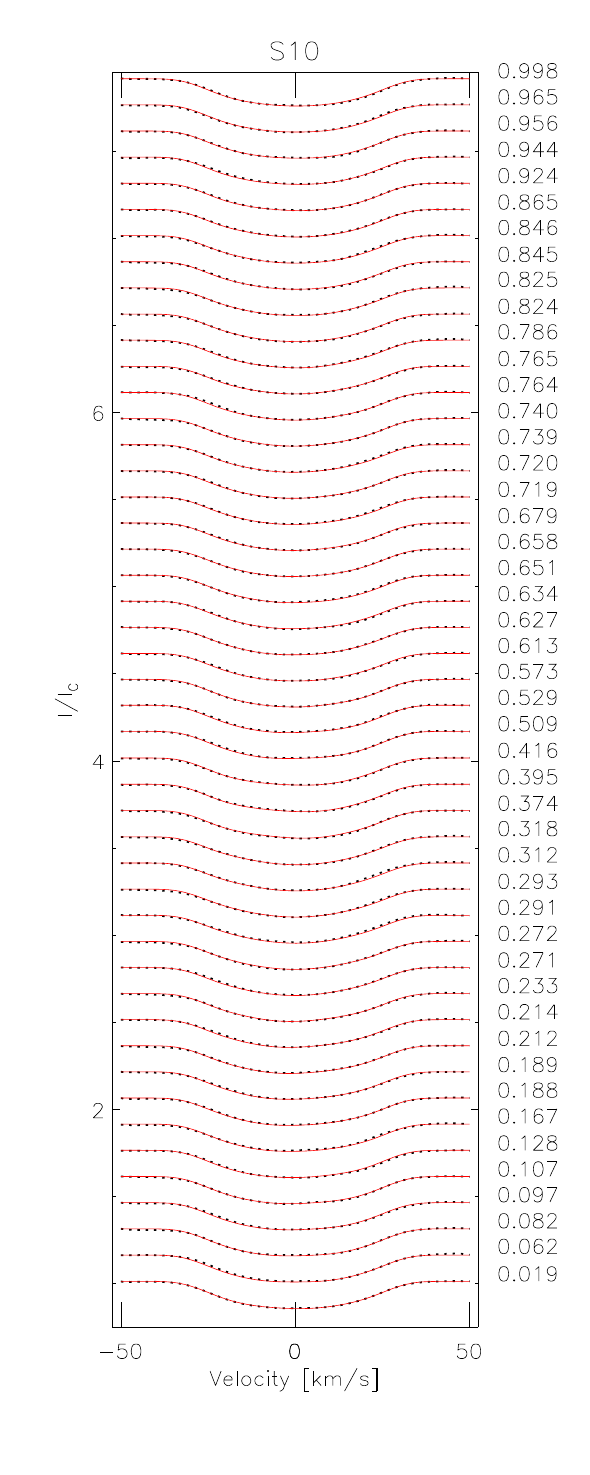}\par  
     \includegraphics[width=\linewidth]{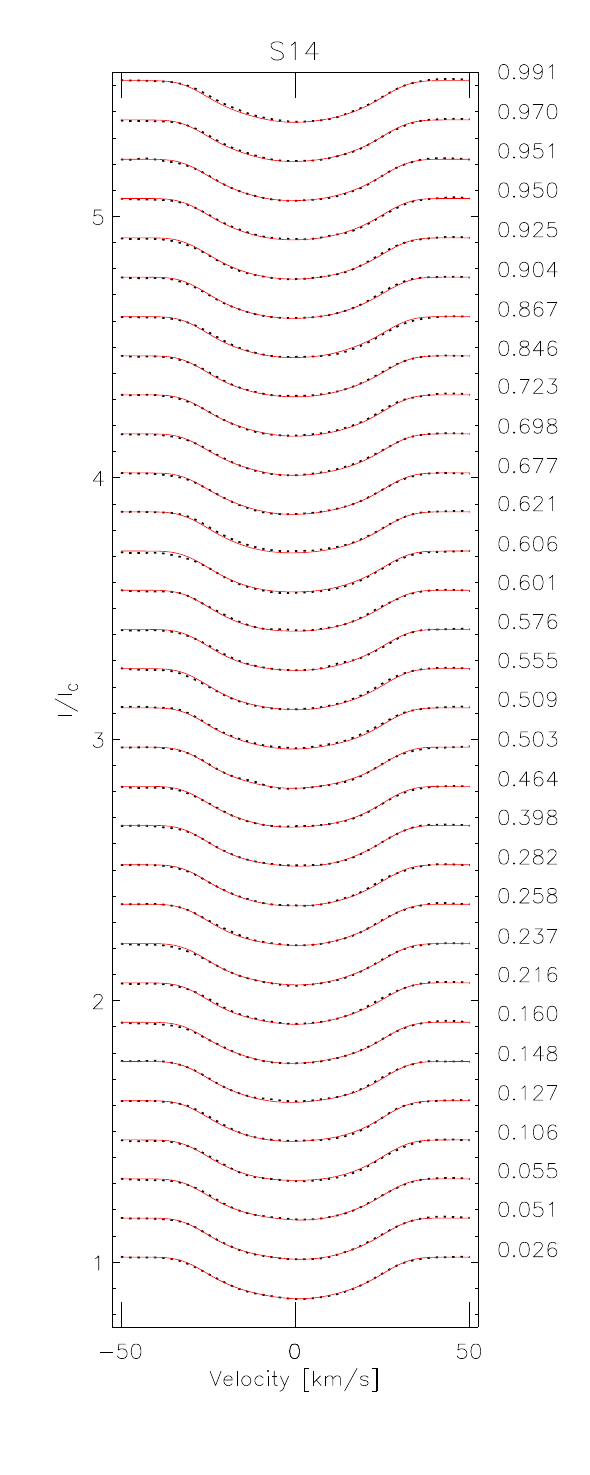}\par
    \includegraphics[width=\linewidth]{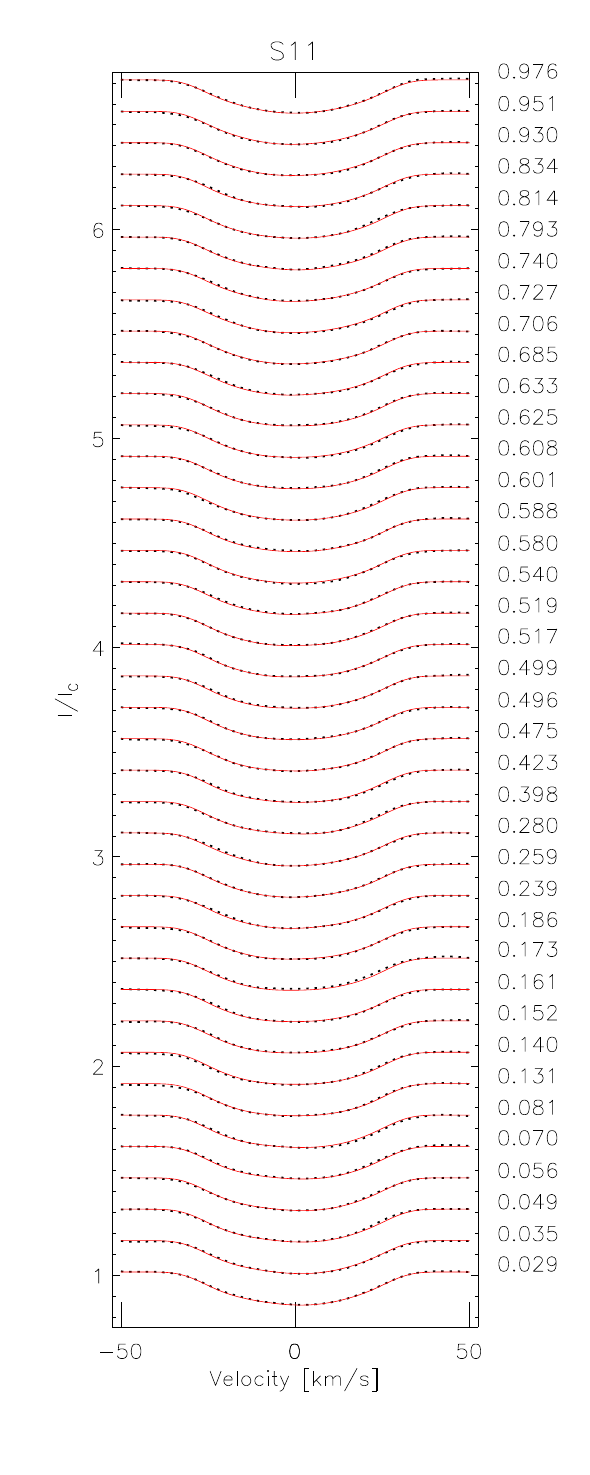}\par 
    \includegraphics[width=\linewidth]{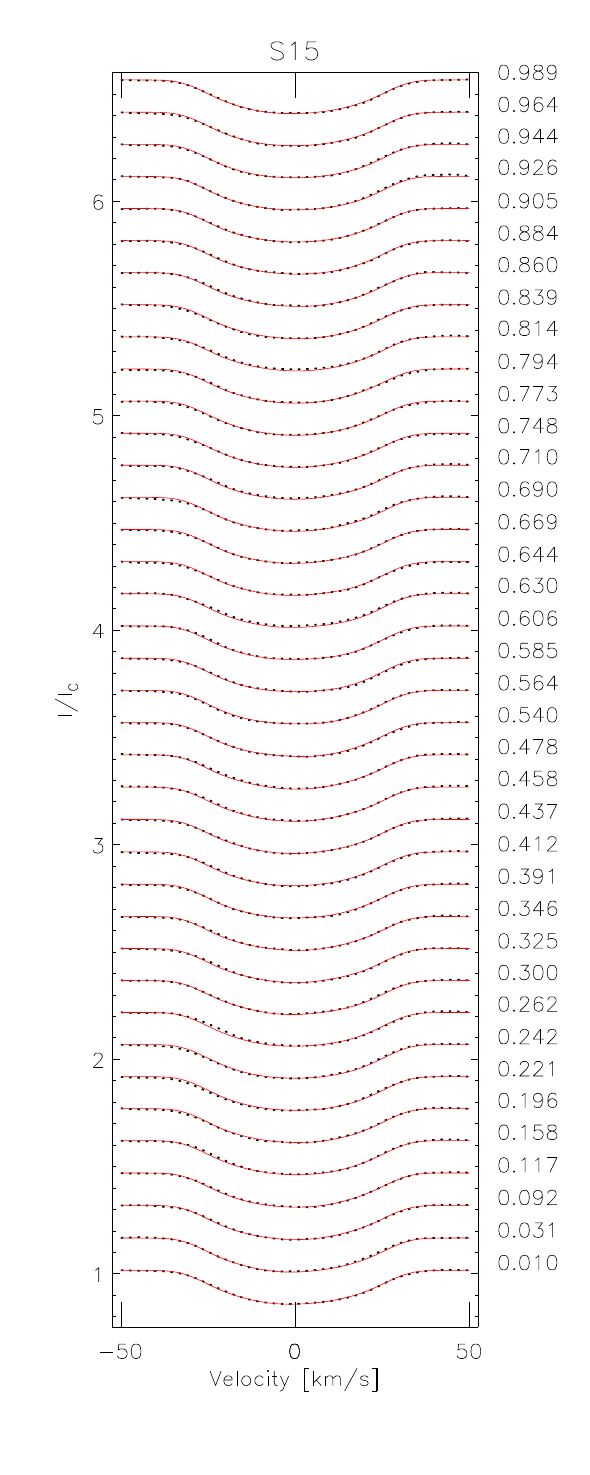}\par
    \includegraphics[width=\linewidth]{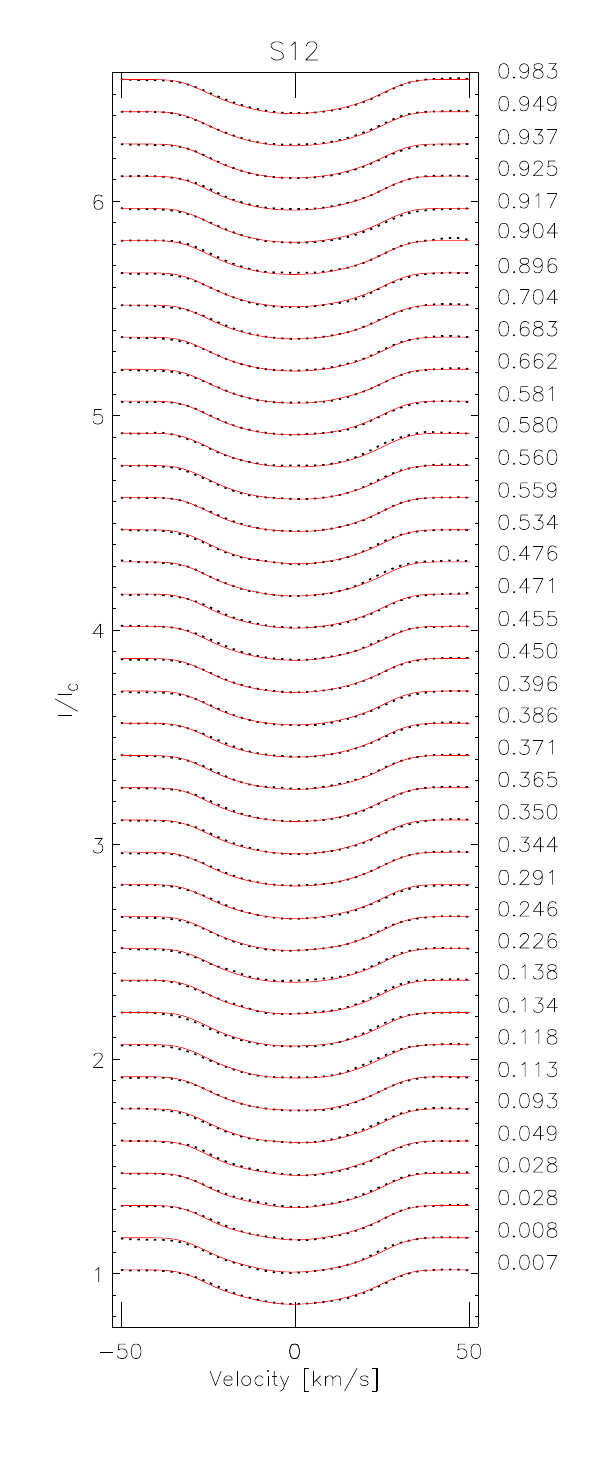}\par 
    \includegraphics[width=\linewidth]{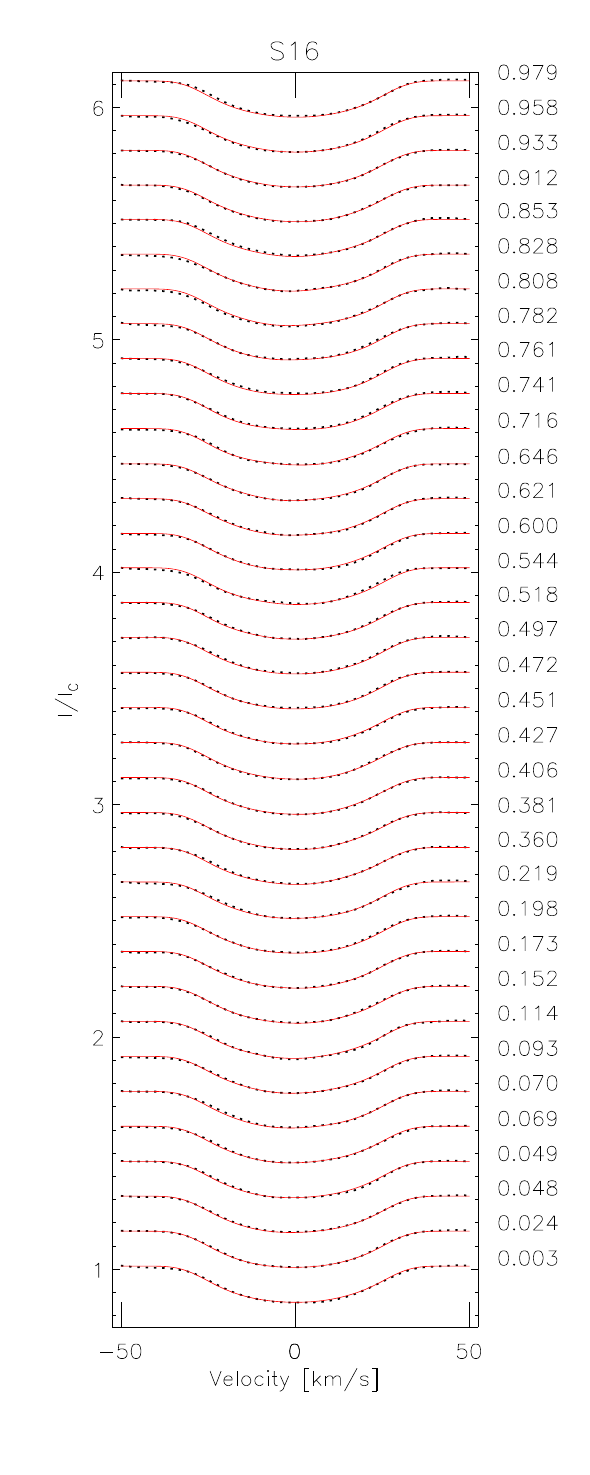}\par
\end{multicols}
\caption{Fitted line profile subsets corresponding to the Doppler images S09-S16 presented in Fig.~\ref{fig:DIs}. The phase values of each profile are shown on the right side of the panels.}
\label{fig:di_profs2}
\end{figure*}
%--
\begin{figure*}[h!!!!!]
%\ContinuedFloat
    \begin{multicols}{4}
     \includegraphics[width=\linewidth]{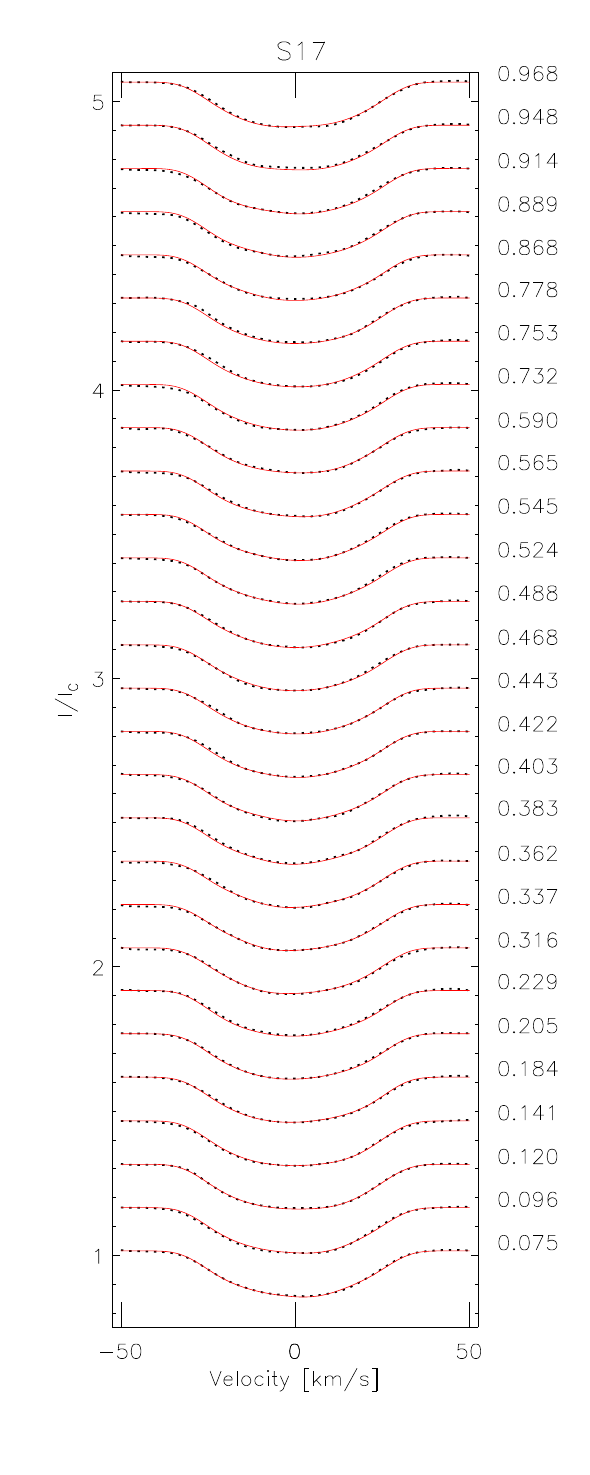}\par  
     \includegraphics[width=\linewidth]{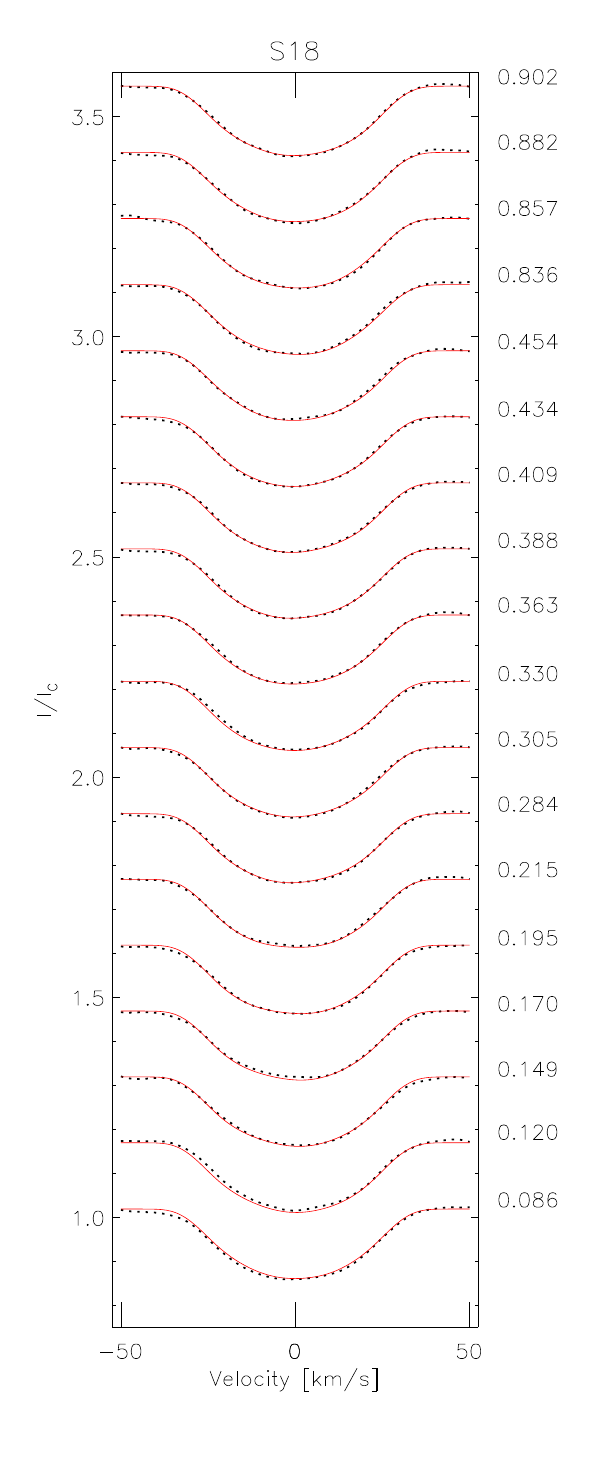}\par
     \includegraphics[width=\linewidth]{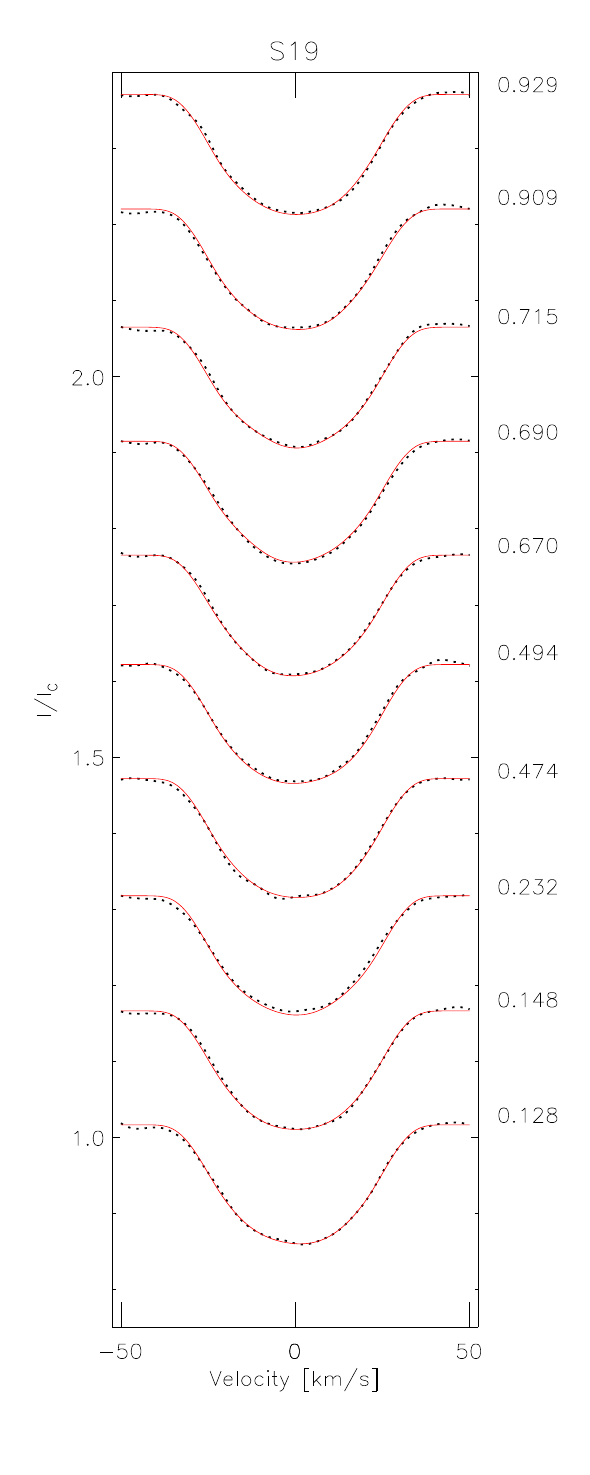}\par  
    \end{multicols}
\caption{Fitted line profile subsets corresponding to the Doppler images S17-S19 presented in Fig.~\ref{fig:DIs}. The phase values of each profile are shown on the right side of the panels.}
\label{fig:di_profs3}
\end{figure*}

\newpage

\twocolumn
\section{Doppler images in Mercator projection}\label{App:DIsMerc}
%\vspace*{-2mm}
\begin{figure}[h!!!!!]
     \vspace*{-1mm}\includegraphics[width=\linewidth]{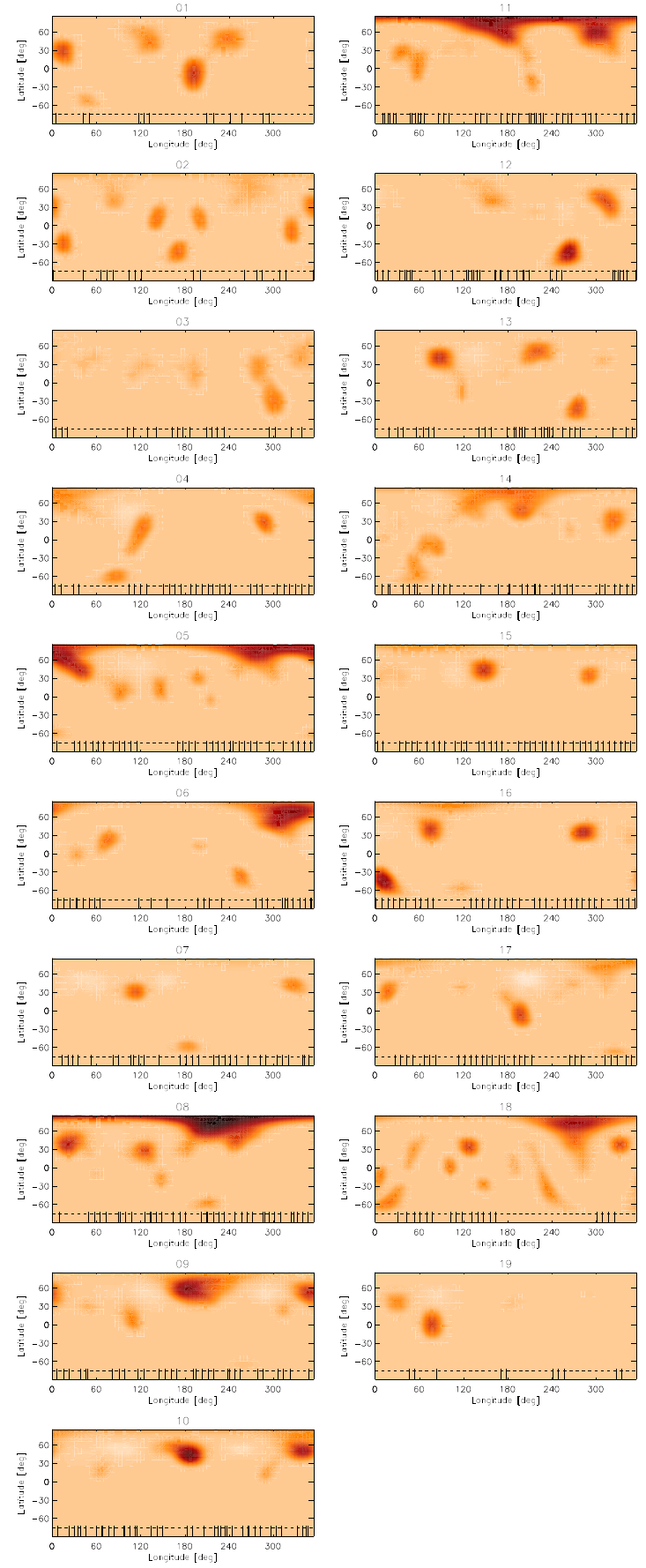}
\caption{Doppler images S01-S10 (left, top to bottom) and S11-S19 (right, top to bottom) plotted in Mercator projection. The maps and the temperature scale correspond to those shown in Fig.~\ref{fig:DIs}. The ticks on the bottom edge indicate the phases of the observations used for that map.}
\label{fig:mercall}
\end{figure}

\section{Spot filling factors}\label{App:FF}
%\vspace*{-2mm}
\begin{figure}[h!!!!!]
     \vspace*{-2mm}\includegraphics[width=0.994\linewidth]{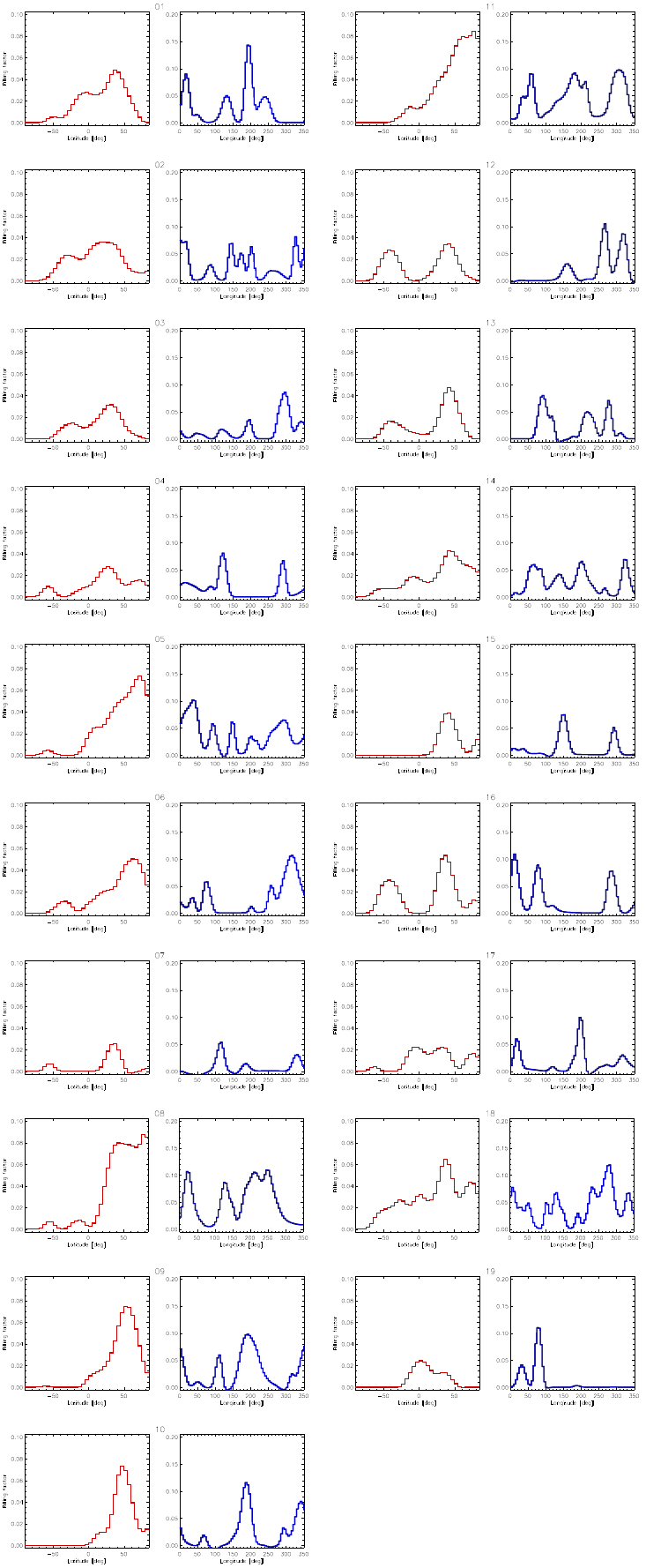}
\caption{The spot filling factors for the nineteen Doppler images in the arrangement corresponding to Fig.\,\ref{fig:mercall}. For each temperature map, the histogram of the spot filling factor is shown in two panels: on the left side (in red), summed along the latitude, and on the right side (in blue), summed up along the longitude.}
\label{fig:ff}
\end{figure}

\section{Measuring spot decay rate}\label{app_decay}

Relying strictly on our time-series Doppler images, below we attempt to quantitatively measure the spot decay rate $\Delta A_{\rm s}/\Delta t_{\rm s}$ for V815\,Her. %based on our time-series Doppler images.
For this we set up a simple statistic, in which we tracked the evolution of well-separated and easy-to-follow spots on our series of surface snapshots.

First, we identified spots on all 19 Doppler maps with the \texttt{Photutils} \citep{larry_bradley_2021_5796924} image segmentation function called \texttt{detect\_sources}
on the 72$\times$36 pixel images to detect spots below a specific threshold value. We required each spot to have at least ten pixels at least 50\,K below the quiescent photosperic temperature. As a next step, we used the \texttt{deblend\_sources} function to deblend the overlapping spots, with the minimum size of ten pixels. We ended up with 99 individual spots on all of the images. For each spot, we calculated the centroid position by weighting with temperature, and the area, taking into account the distortion of pixels with the cosine of the latitude. Then, we manually identified matching spots on consecutive image pairs. However, tracking decaying spots is made more difficult by interactions between neighboring spots (e.g. merging) or the emergence of new fluxes. Therefore, we only matched unambigous spot pairs, where it was reasonable to assume that we see the evolution of the same feature. The maximum allowed angular separation between the centroids of the subsequent images of the evolving spot was set to be 40$^{\circ}$.

\begin{figure}[h!!!]
	\centering
	\includegraphics[width=\linewidth]{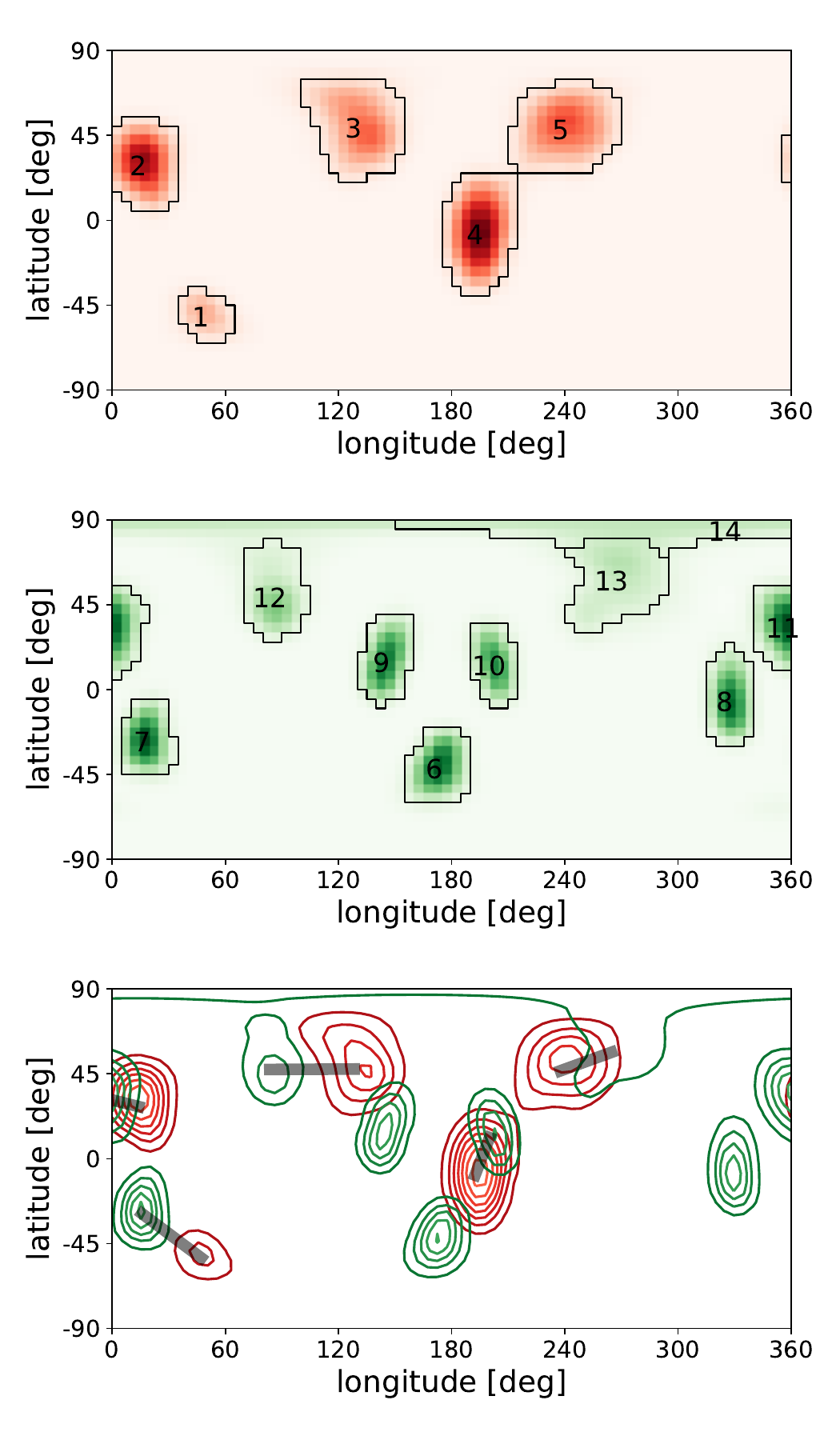}
	\caption{Demonstrating how spot segmentation and identification works. As an example, the upper panel shows the identified spots with their ID numbers (see also Table\,\ref{tab:spotid}) of the S01 Doppler image, and the middle one shows the identified spots of the S02 Doppler image. In the bottom panel, the spots of S01 are represented by red contours, the spots of S02 by green contour lines, and the paired spots are connected by gray lines.}
	\label{fig:spot_segmentation}
\end{figure}

As an example, in Fig.\,\ref{fig:spot_segmentation} we show the identification of the individual spots for S01 and S02 Doppler images (top and middle panels, respectively) and the spot pairing (see the bottom panel). For each spot pair, we calculated the decay rate as $\Delta A/\Delta t$, where $\Delta A$ is the change in spot area, and $\Delta t$ is the time difference between the two images. The average decay rate for the 18 spot pairs showing negative $\Delta A$ (i.e. shrinking) is $\Delta A/\Delta t=-$7.0$\pm$1.3\,10$^4$\,km$^{2}$\,s$^{-1}$. For identification, we provide a list of these spots in Table\,\ref{tab:spotid}, while the list of spot pairs found on consecutive Doppler images that we considered to represent successive evolutionary phases (i.e. spot decay) of the same spot is as follows:
\#2$-$\#11;
\#3$-$\#12;
\#4$-$\#10;
\#5$-$\#13;
\#9$-$\#16;
\#12$-$\#18;
\#20$-$\#23;
\#21$-$\#25;
\#29$-$\#34;
\#34$-$\#37;
\#36$-$\#40;
\#42$-$\#45;
\#43$-$\#48;
\#47$-$\#52;
\#48$-$\#53;
\#60$-$\#63;
\#69$-$\#75;
\#78$-$\#83.

Finally, we note that the method is not mature, due to the aforementioned sources of error (e.g. interacting spots, arbitrarily taken threshold values, etc.), we consider it only suitable for order of magnitude estimation for the time being.

\clearpage
\begin{center}
\tablehead
{\hline\hline\noalign{\smallskip}
\# &  Image & \setcounter{footnote}{0}$\lambda_{\rm s}$\footnote{a} & \setcounter{footnote}{0}$\beta_{\rm s}$\footnote{a} & Area\footnote{b} &  \setcounter{footnote}{0}$\overline{\gamma}_{\rm s}$\footnote{a} & \setcounter{footnote}{2}$\overline{T}_{\rm s}$\footnote{c} \\
\hline\noalign{\smallskip}}
\topcaption{Identified spots with their longitude and latitude coordinates and their calculated area. In the last two columns, the equivalent spot radius and average temperature values are provided for information.}\label{tab:spotid}
\begin{supertabular}{r r r r r r r}
\tabletail{
\hline\hline\noalign{\smallskip}
\multicolumn{5}{l}{\setcounter{footnote}{0}\footnote{a}{ degree}}\\
\multicolumn{5}{l}{\footnote{b}{ square degree}}\\
\multicolumn{5}{l}{\footnote{c}{ Kelvin}}\\}
\tablelasttail{
\hline\hline\noalign{\smallskip}
\multicolumn{5}{l}{\setcounter{footnote}{0}\footnote{a}{ degree}}\\
\multicolumn{5}{l}{\footnote{b}{ square degree}}\\
\multicolumn{5}{l}{\footnote{c}{ Kelvin}}\\}
1  &  S01  &  47  &  $-$53  &  413  &  11.5  &  5469 \\
2  &  S01  &  14  &  28  &  1414  &  21.2  &  5332 \\
3  &  S01  &  128  &  48  &  1398  &  21.1  &  5420 \\
4  &  S01  &  192  &  $-$9  &  2088  &  25.8  &  5303 \\
5  &  S01  &  238  &  47  &  1542  &  22.2  &  5404 \\
6  &  S02  &  171  &  $-$43  &  885  &  16.8  &  5387 \\
7  &  S02  &  16  &  $-$29  &  895  &  16.9  &  5376 \\
8  &  S02  &  325  &  $-$8  &  1135  &  19.0  &  5394 \\
9  &  S02  &  143  &  13  &  1070  &  18.5  &  5394 \\
10  &  S02  &  200  &  11  &  900  &  16.9  &  5405 \\
11  &  S02  &  356  &  31  &  1230  &  19.8  &  5374 \\
12  &  S02  &  84  &  47  &  854  &  16.5  &  5477 \\
13  &  S02  &  265  &  56  &  1036  &  18.2  &  5488 \\
14  &  S02  &  325  &  83  &  177  &  7.5  &  5486 \\
15  &  S03  &  192  &  15  &  1016  &  18.0  &  5489 \\
16  &  S03  &  120  &  24  &  766  &  15.6  &  5507 \\
17  &  S03  &  342  &  45  &  1256  &  20.0  &  5473 \\
18  &  S03  &  52  &  36  &  401  &  11.3  &  5520 \\
19  &  S03  &  299  &  $-$27  &  1484  &  21.7  &  5394 \\
20  &  S03  &  281  &  24  &  1245  &  19.9  &  5435 \\
21  &  S04  &  85  &  $-$62  &  547  &  13.2  &  5420 \\
22  &  S04  &  119  &  10  &  1728  &  23.5  &  5393 \\
23  &  S04  &  288  &  28  &  959  &  17.5  &  5363 \\
24  &  S04  &  14  &  70  &  1463  &  21.6  &  5435 \\
25  &  S05  &  1  &  $-$61  &  305  &  9.9  &  5480 \\
26  &  S05  &  215  &  $-$5  &  273  &  9.3  &  5474 \\
27  &  S05  &  146  &  14  &  956  &  17.4  &  5417 \\
28  &  S05  &  198  &  30  &  541  &  13.1  &  5436 \\
29  &  S05  &  92  &  11  &  1296  &  20.3  &  5429 \\
30  &  S05  &  32  &  64  &  2095  &  25.8  &  5132 \\
31  &  S05  &  276  &  73  &  2074  &  25.7  &  5119 \\
32  &  S06  &  257  &  $-$39  &  780  &  15.8  &  5399 \\
33  &  S06  &  34  &  $-$1  &  570  &  13.5  &  5471 \\
34  &  S06  &  75  &  20  &  1102  &  18.7  &  5390 \\
35  &  S06  &  317  &  66  &  2505  &  28.2  &  5203 \\
36  &  S07  &  184  &  $-$59  &  374  &  10.9  &  5444 \\
37  &  S07  &  112  &  31  &  1020  &  18.0  &  5380 \\
38  &  S07  &  328  &  41  &  756  &  15.5  &  5418 \\
39  &  S07  &  252  &  80  &  22  &  2.7  &  5526 \\
40  &  S08  &  210  &  $-$59  &  369  &  10.8  &  5443 \\
41  &  S08  &  147  &  $-$18  &  478  &  12.3  &  5440 \\
42  &  S08  &  125  &  26  &  1531  &  22.1  &  5354 \\
43  &  S08  &  24  &  41  &  1715  &  23.4  &  5291 \\
44  &  S08  &  234  &  72  &  3766  &  34.6  &  4917 \\
45  &  S09  &  108  &  11  &  1153  &  19.2  &  5416 \\
46  &  S09  &  48  &  29  &  252  &  9.0  &  5502 \\
47  &  S09  &  314  &  25  &  480  &  12.4  &  5476 \\
48  &  S09  &  351  &  52  &  1556  &  22.3  &  5308 \\
49  &  S09  &  201  &  63  &  2411  &  27.7  &  5270 \\
50  &  S10  &  261  &  $-$64  &  156  &  7.0  &  5502 \\
51  &  S10  &  108  &  10  &  1158  &  19.2  &  5412 \\
52  &  S10  &  316  &  25  &  458  &  12.1  &  5487 \\
53  &  S10  &  353  &  51  &  1352  &  20.7  &  5357 \\
54  &  S10  &  200  &  63  &  2531  &  28.4  &  5258 \\
55  &  S11  &  213  &  $-$22  &  579  &  13.6  &  5430 \\
56  &  S11  &  59  &  6  &  1444  &  21.4  &  5398 \\
57  &  S11  &  36  &  26  &  883  &  16.8  &  5406 \\
58  &  S11  &  201  &  74  &  3140  &  31.6  &  5020 \\
59  &  S11  &  303  &  53  &  1877  &  24.4  &  5168 \\
60  &  S12  &  262  &  $-$46  &  1487  &  21.8  &  5250 \\
61  &  S12  &  313  &  39  &  1600  &  22.6  &  5331 \\
62  &  S12  &  156  &  47  &  1138  &  19.0  &  5461 \\
63  &  S13  &  275  &  $-$43  &  1083  &  18.6  &  5348 \\
64  &  S13  &  219  &  49  &  1305  &  20.4  &  5366 \\
65  &  S13  &  309  &  36  &  230  &  8.5  &  5497 \\
66  &  S13  &  117  &  $-$10  &  775  &  15.7  &  5476 \\
67  &  S13  &  89  &  38  &  1571  &  22.4  &  5328 \\
68  &  S14  &  265  &  16  &  282  &  9.5  &  5509 \\
69  &  S14  &  324  &  32  &  1451  &  21.5  &  5399 \\
70  &  S14  &  54  &  $-$50  &  1093  &  18.6  &  5431 \\
71  &  S14  &  76  &  $-$10  &  1460  &  21.6  &  5409 \\
72  &  S14  &  201  &  49  &  1098  &  18.7  &  5325 \\
73  &  S14  &  179  &  72  &  2054  &  25.6  &  5366 \\
74  &  S15  &  148  &  44  &  1286  &  20.2  &  5325 \\
75  &  S15  &  291  &  35  &  784  &  15.8  &  5384 \\
76  &  S15  &  18  &  33  &  378  &  11.0  &  5521 \\
77  &  S15  &  68  &  82  &  304  &  9.8  &  5439 \\
78  &  S16  &  12  &  $-$46  &  1475  &  21.7  &  5223 \\
79  &  S16  &  117  &  $-$57  &  263  &  9.2  &  5488 \\
80  &  S16  &  284  &  35  &  1286  &  20.2  &  5294 \\
81  &  S16  &  75  &  41  &  1343  &  20.7  &  5332 \\
82  &  S16  &  99  &  79  &  411  &  11.4  &  5448 \\
83  &  S17  &  326  &  $-$67  &  254  &  9.0  &  5479 \\
84  &  S17  &  194  &  $-$0  &  2010  &  25.3  &  5367 \\
85  &  S17  &  18  &  32  &  917  &  17.1  &  5392 \\
86  &  S17  &  269  &  43  &  290  &  9.6  &  5503 \\
87  &  S17  &  343  &  75  &  1067  &  18.4  &  5439 \\
88  &  S18  &  234  &  $-$29  &  1830  &  24.1  &  5446 \\
89  &  S18  &  147  &  $-$28  &  446  &  11.9  &  5430 \\
90  &  S18  &  102  &  3  &  590  &  13.7  &  5395 \\
91  &  S18  &  190  &  4  &  513  &  12.8  &  5486 \\
92  &  S18  &  52  &  25  &  1011  &  17.9  &  5456 \\
93  &  S18  &  128  &  34  &  954  &  17.4  &  5339 \\
94  &  S18  &  19  &  $-$54  &  1180  &  19.4  &  5412 \\
95  &  S18  &  4  &  $-$19  &  977  &  17.6  &  5415 \\
96  &  S18  &  278  &  68  &  3152  &  31.7  &  5254 \\
97  &  S18  &  333  &  37  &  923  &  17.1  &  5362 \\
98  &  S19  &  78  &  1  &  1669  &  23.0  &  5329 \\
99  &  S19  &  30  &  37  &  908  &  17.0  &  5422 \\
\end{supertabular}
\end{center}

\end{appendix}

\end{document}